\documentclass[ twoside,openright,titlepage,numbers=noenddot,headinclude,a4paper, %1headlines,% letterpaper a4paper
                footinclude=true,cleardoublepage=empty,abstractoff, % <--- obsolete, remove (todo)
                BCOR=7mm,paper=a4,fontsize=11pt,%11pt,a4paper,%
                ngerman,listof=totocnumbered,listof=leveldown,%
                toc=bibliographynumbered, %bibliography=leveldown
                ]{scrreprt}

\PassOptionsToPackage{eulerchapternumbers,listings,%drafting,%
				 pdfspacing,dottedtoc,manychapters,%floatperchapter,%linedheaders,%
				 subfig,beramono,parts}{classicthesis}										
\usepackage{ifthen}
\newboolean{enable-backrefs} % enable backrefs in the bibliography
\setboolean{enable-backrefs}{true} % true false
\PassOptionsToPackage{utf8}{inputenc}	% latin9 (ISO-8859-9) = latin1+"Euro sign"
 \usepackage{inputenc}
\newcommand{\myTitle}{Charge Carrier Dynamics of Methylammonium Lead-Iodide Perovskite Solar Cells\xspace}
\newcommand{\mySubtitle}{From Microseconds to Minutes\xspace}

\newcommand{\submissiondate}{January 2016}

\newcommand{\myName}{Martin Neukom\xspace}

\newcommand{\myFaculty}{Albert-Ludwigs-University Freiburg\xspace}

\newcommand{\myUni}{Albert-Ludwigs-University Freiburg\xspace}

\newcommand{\myYear}{2016\xspace}

\newcommand{\myVersionNumber}{1.0\xspace}
\newcommand{\myVersion}{Version \myVersionNumber\xspace}

\newcounter{dummy} % necessary for correct hyperlinks (to index, bib, etc.)
\providecommand{\mLyX}{L\kern-.1667em\lower.25em\hbox{Y}\kern-.125emX\@}

\newcommand{\paios}{\textit{Paios} }
\newcommand{\setfos}{\textit{Setfos} }

\PassOptionsToPackage{english}{babel}   % change this to your language(s)

\usepackage{babel}					

\usepackage[numbers, compress]{natbib}				
\PassOptionsToPackage{fleqn}{amsmath}		% math environments and more by the AMS 
 \usepackage{amsmath}

\usepackage{color,soul}
\PassOptionsToPackage{T1}{fontenc} % T2A for cyrillics
	\usepackage{fontenc}     
\usepackage{textcomp} % fix warning with missing font shapes
\usepackage{scrhack} % fix warnings when using KOMA with listings package          
\usepackage{xspace} % to get the spacing after macros right  
\usepackage{mparhack} % get marginpar right
\usepackage{fixltx2e} % fixes some LaTeX stuff 
\usepackage{rotating} % To rotate tables and stuff
\PassOptionsToPackage{printonlyused}{acronym} %option smaller makes nicer acronym list, but all acronyms are capitalized
	\usepackage{acronym} % nice macros for handling all acronyms in the thesis
\usepackage{tabularx} % better tables
\usepackage{caption}
\usepackage{subfig}
\usepackage{supertabular} % mulitpage tables
\usepackage{listings} 
\PassOptionsToPackage{hyphens}{url} % break to long URLs. Must be here, because hyperref loads the url package!
\PassOptionsToPackage{hyperfootnotes=false,pdfpagelabels}{hyperref}
	\usepackage{hyperref}  % backref linktocpage pagebackref
\PassOptionsToPackage{pdftex}{graphicx}
	\usepackage{graphicx} 

\newcommand{\backrefnotcitedstring}{\relax}%(Not cited.)
\newcommand{\backrefcitedsinglestring}[1]{(Cited on page~#1.)}
\newcommand{\backrefcitedmultistring}[1]{(Cited on pages~#1.)}
\ifthenelse{\boolean{enable-backrefs}}%
{
		\PassOptionsToPackage{hyperpageref}{backref}
		\usepackage{backref} % to be loaded after hyperref package 
		   \renewcommand*{\backref}[1]{}  % disable standard
		   \renewcommand*{\backrefalt}[4]{% detailed backref
		      \ifcase #1 %
		         \backrefnotcitedstring%
		      \or%
		         \backrefcitedsinglestring{#2}%
		      \else%
		         \backrefcitedmultistring{#2}%
		      \fi}%
}{\relax}

\hypersetup{%
    colorlinks=true, linktocpage=true, pdfstartpage=3, pdfstartview=FitV,%
    breaklinks=true, pdfpagemode=UseNone, pageanchor=true, pdfpagemode=UseOutlines,%
    plainpages=false, bookmarksnumbered, bookmarksopen=true, bookmarksopenlevel=1,%
    hypertexnames=true, pdfhighlight=/O,%nesting=true,%frenchlinks,%
    urlcolor=RoyalBlue, linkcolor=RoyalBlue, citecolor=RoyalBlue, %pagecolor=RoyalBlue,%
    pdftitle={\myTitle},%
    pdfauthor={\textcopyright\ \myName, \myUni, \myFaculty},%
    pdfsubject={},%
    pdfkeywords={},%
    pdfcreator={pdfLaTeX},%
    pdfproducer={LaTeX with hyperref and classicthesis}%
}   

\makeatletter
\@ifpackageloaded{babel}%
    {%
       \addto\extrasamerican{%
				}%
       \addto\extrasngerman{% 
				}%	
    }{\relax}
\makeatother

\usepackage{classicthesis} 
\newcommand{\pagenote}{ \newline \textit{Used on page:} }

\PassOptionsToPackage{style=long,section,numberedsection=autolabel,
translate}{glossaries}
\usepackage{glossaries}
\renewcommand{\glossarysection}[2][]{}

\makeglossaries

\newglossaryentry{full_load_hours}
{
	name={full load hours},
	description={Defines the number of hours that a power source runs on maximum power during a year. The full load hours $t_{flh}$ are obtained by dividing the energy generated in one year $E_{tot}$ with the peak power $P_{peak}$. ($t_{flh}=E_{tot}/P_{peak}$). For a nuclear power plant the full load hours can reach 7000 hours, whereas photovoltaics varies between 1000 and 2500 hours, depending on the location. \pagenote}
}

\newglossaryentry{price_experience_curve}
{
	name={price experience curve},
	description={The price experience curve describes how a price is lowered by the total amount of the produced quantity. The price experience factor hereby determines how much the price is lowered when the accumulated amount is doubled. For example: If the amount of produced PV modules is doubled the price is expected to go down by about 20\%. \pagenote}
}

\newglossaryentry{lcoe}
{
	name={levelized cost of electricity},
	text={levelized cost of electricity},
	description={The levelized cost of electricity (lcoe) defines the total cost to produce one kilowatthour of electricity including investment costs, interest, maintenance, insurance and taxes. \pagenote}
}

\newglossaryentry{monolithic_integration}
{
	name={monolithic integration},
	description={Monolithic integration is a method applied for thin-film solar cells to create modules from deposited cells. Since the current of one cell would be too high, the cell is divided into stripes and these sub-cells are connected in series. \pagenote}
}

\newglossaryentry{workfunction}
{
name={workfunction},
description={The workfunction of a metal is the energy required to remove one free electron into vacuum. If a semiconductor is brought in contact with a metal charge exchanges at the interface according to the energy level of the semiconductor and the workfunction of the metal. \pagenote}
}

\newglossaryentry{Fermi_level}
{
name={Fermi level},
description={Electrons are so called Fermions that distribute according to the Fermi statistics. The Fermi level is the median of the distribution. In semiconductor physics it is a measure of the energy of electrons that can be used. For example: The difference in Fermi levels of electrons and holes is equivalent to the open circuit voltage. \pagenote}
}

\newglossaryentry{RC_effects}
{
name={RC effects},
description={RC effects are parasitic currents caused by the interplay between the series-resistance (R) and the capacitance (C) of an electronic device. In transient electrical experiments the time constant $\tau = R \cdot C$ limits the dynamics that can be resolved by measurement. \pagenote}
}

\newglossaryentry{exciton}
{
name={exciton},
description={An exciton is a quasi-particle that consists of a bound electron and hole pair. Energy is required to dissociate it into free charge carriers, otherwise it is annihilated by geminate recombination. The energy for dissociation depends on its dielectric environment as 
$E = k \cdot 1/{\epsilon^2}$. \pagenote}
}

\newglossaryentry{dangling_bonds}
{
name={dangling bonds},
description={In a perfect crystal every atom has exact as many neighbour atoms as it needs to create covalent bounds. On the surface of the crystal, at a crystal boundary or at a defect not all atoms find a neighbour - this is called a dangling bond. In solar cells this is problematic as dangling bonds usually act as recombination center. \pagenote}
}

\newglossaryentry{phonon}
{
name={phonon},
description={A phonon is the quantum quasi-particle of thermal motion. Similar as the photon, which is the particle of light with energy depending on its wavelength, the phonon is the quasi-particle of heat, with energy depending on the wavelength of thermal motion. \pagenote}
}

\newglossaryentry{pn_junction}
{
name={pn-junction},
description={If a p-doped material (free holes) is put directly on a n-doped material (free electrons) a pn-junction is formed. The free electrons and the free holes diffuse towards each other and recombine, creating a space-charge zone. This junction type is the basis for most semiconductor devices. \pagenote}
}

\newglossaryentry{tandem}
{
name={tandem},
description={A so-called tandem structure solar cell is used to circumvent the general efficiency limitation of a single junction (Shockley-Queisser limit). Hereby two solar cells with different band-gaps are stacked on top of each other. \pagenote}
}

\begin{document}
\frenchspacing
\raggedbottom
%\selectlanguage{ngerman} % american ngerman
\selectlanguage{english}

%\DefineBibliographyStrings{ngerman}{andothers={et\ addabbrvspace al\adddot}}
%\renewcommand*{\bibname}{new name}
%\setbibpreamble{}
\renewcommand{\thempfootnote}{\arabic{mpfootnote}}
%\pagenumbering{roman}
\pagenumbering{arabic}

\pagestyle{plain}

% Title Page

\begin{titlepage}

\begin{addmargin}[-0.3cm]{-3cm} %printversion +1cm %screenversion -1cm
\begin{center}

% Logo University
%\begin{picture}(0,0)															
%\put(-270, -70){\scalebox{0.35}{
%	\includegraphics{images/logos/albert_Ludwigs_university_freiburg_logo.png}
%}}
%\end{picture}

% Logo Fraunhofer ISE
%\begin{picture}(0,0)
%\put(-130, -20){\scalebox{0.2}{
%	\includegraphics{images/logos/fraunhofer_ise_logo.png}
%}}
%\end{picture}

% Logo Fluxim
\begin{picture}(0,0)																			
\put(100, 0){ \scalebox{0.13}{
	\includegraphics{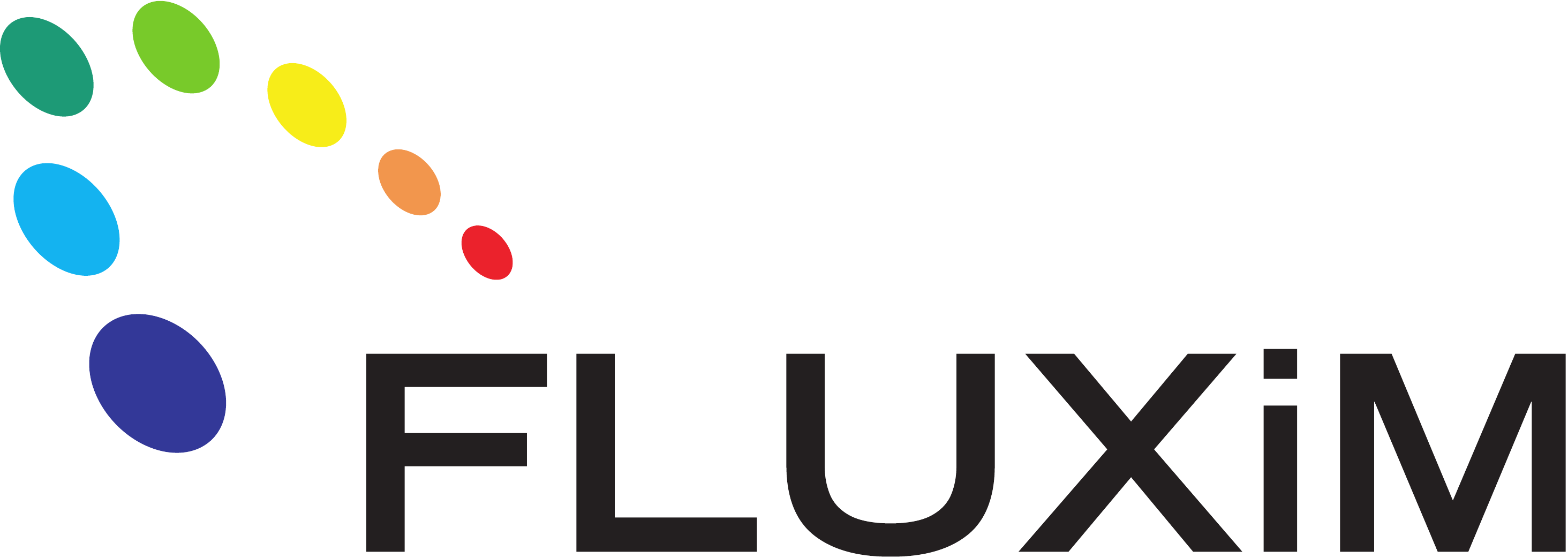}
}}
\end{picture}

\large

\hfill
\vfill

% Title Picture - Probe Tips
\scalebox{0.2}{
\includegraphics{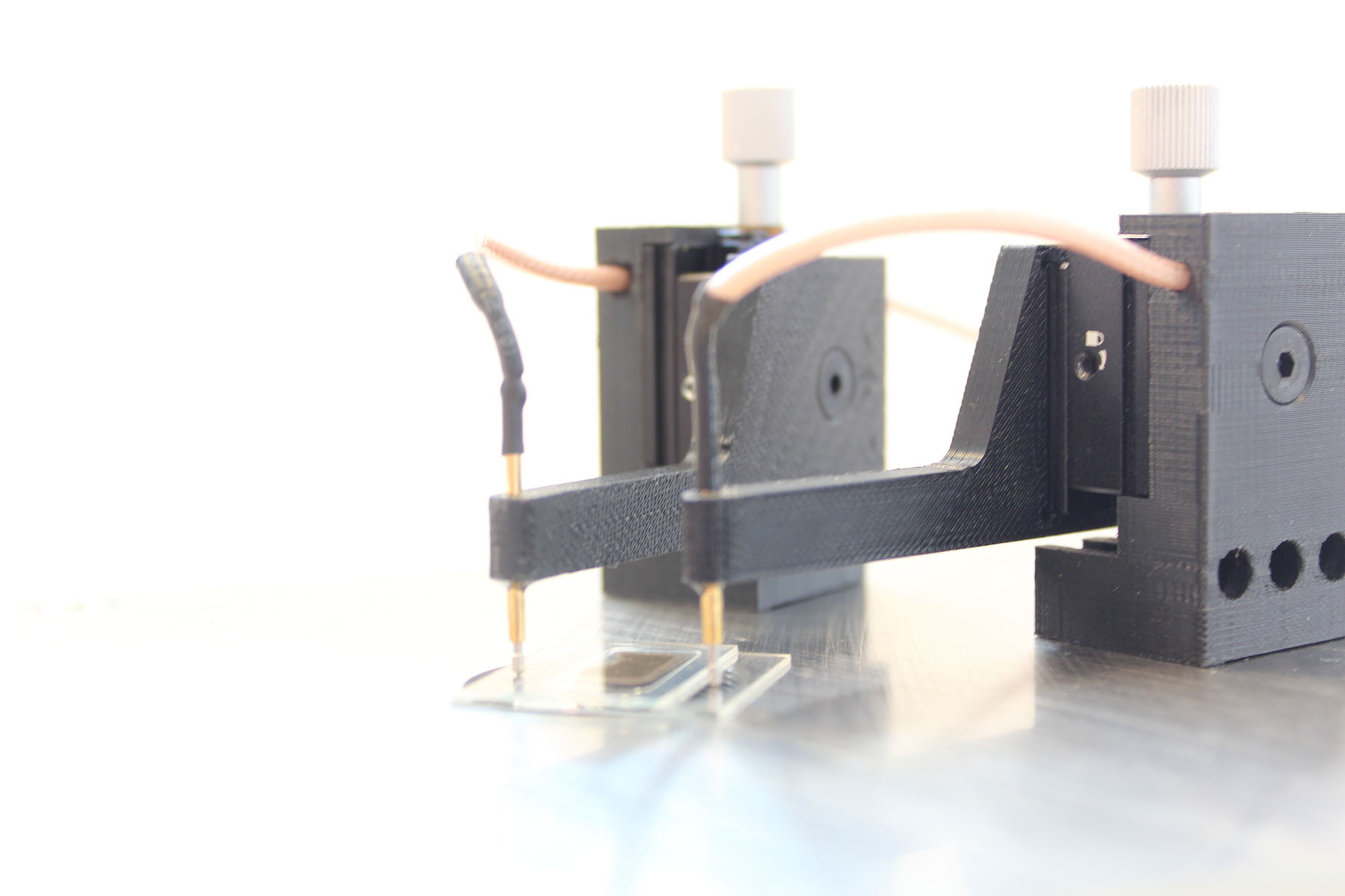}
}
\vspace{5mm}

\begingroup
\color{Maroon}\spacedallcaps{\myTitle} \\ \bigskip % Thesis title
\endgroup
\spacedlowsmallcaps{\mySubtitle}

\vfill

Master-Thesis\\
\myName \\
\vspace{5mm}
Albert-Ludwigs-University Freiburg \\
Master Online of Photovoltaics \\
\submissiondate

\end{center}

%\vfill

\end{addmargin}

\end{titlepage}

%% Backside Title  -----------------------------------------
\thispagestyle{empty}
\hfill
\vfill
\noindent\myName: \textit{\myTitle, \mySubtitle, \myYear}
%% Backside Title  -----------------------------------------

%\include{Context/titlePagePrint}

\cleardoublepage

\chapter{Abstract}\label{ch:abstract}

\section{English Abstract}

Transient opto-electrical measurements of \acf{MALI} \acfp{PSC} are performed and analyzed in order to elucidate the operating mechanisms. The current response to a light pulse or voltage pulse shows an extraordinarily broad dynamic range covering 9 orders of magnitude in time – from microseconds to minutes – until steady-state is reached. Evidence of a slowly changing charge density at the perovskite layer boundaries is found, which is most probably caused by mobile ions. \acfp{IV curve} are measured with very fast scan-rate after keeping the cell for several seconds at a constant voltage as proposed by Tress et al. Numerical drift-diffusion simulations reproduce the measured \acp{IV curve} using different distributions of ions in the model. Analysing the band diagram of the simulation result sheds light on the operating mechanism. To further investigate the effects at short time scales (below milliseconds) \acf{photo-CELIV} experiments are performed. We postulate that mobility imbalance in combination with deep hole trapping leads to dynamic doping causing effects from microseconds to milliseconds. Comprehensive transient drift-diffusion simulations of the \ac{photo-CELIV} experiments strengthen this hypothesis. This advanced characterization approach combining dynamic response measurements and numerical simulations represents a key step on the way to a comprehensive understanding of device working mechanisms in emerging perovskite solar cells.

\newpage

\section{German Abstract}

In der vorliegenden Arbeit werden Methyl-Ammonium Perovskite-Solarzellen analysiert mit transienten optisch-elektrischen Messungen, um die physikalischen Prozesse zu untersuchen.
Der transiente Photostrom, erzeugt durch einen Lichtpuls, zeigt eine ausserordentliche lange Dynamik über 9 Grössenordnungen in der Zeit - von Mikrosekunden bis Minuten - bis ein stationärer Zustand erreicht ist.

Es werden Belege präsentiert für langsam ändernde Ladungsträger-Dichten am Rand der Perovskit-Schicht, die sehr wahrscheinlich durch mobile Ionen innerhalb der aktiven Schicht verursacht werden. Es werden Strom-Spannungs-Kennlinien gemessen mit schnellen Spannungsrampen, nachdem die Zelle eine Weile auf bestimmter Spannung gehalten wurde, wie von Tress et al. vorgeschlagen. 
Die gemessenen IV-Kurven werden durch nummerische Drift-Diffusions-Rechnungen mit verschiedenen Ionenverteilungen reproduziert. Die Analyse der Banddiagramme der Simulationsresultate zeigen dabei die Funktionsweise der Solarzelle auf.

Um die Solarzelle im kurzen Zeitbereich (unterhalb von Millisekunden) zu untersuchen, werden \ac{photo-CELIV} Experimente gemessen. Dabei wird die Theorie aufgestellt, dass unausgewogene Ladungsträgermobilitäten in Kombination mit Loch-Traps zu dynamischer Dotierung des aktiven Materials führt, die Effekte im Zeitbereich zwischen Mikrosekunden und Millisekunden verursacht.
Diese Hypothese wird gestärkt durch umfassende transiente Drift-Diffusions Simulationen der \acs{photo-CELIV}-Ströme.

Diese umfassende Methode zur Charakterisierung, die dynamische Messungen mit nummerischer Simulation verknüpft, zeigt den Weg auf zu einem umfassenden Verständnis der physikalischen Prozesse innerhalb von Perovskite-Solarzellen.

%\include{Content/declaration}

%*******************************************************
% Table of Contents
%*******************************************************
%\phantomsection
\refstepcounter{dummy}
\pdfbookmark[1]{\contentsname}{tableofcontents}
\setcounter{tocdepth}{1} % <-- 2 includes up to subsections in the ToC
\setcounter{secnumdepth}{3} % <-- 3 numbers up to subsubsections
\manualmark
\markboth{\spacedlowsmallcaps{\contentsname}}{\spacedlowsmallcaps{\contentsname}}
\tableofcontents 
\automark[section]{chapter}
\renewcommand{\chaptermark}[1]{\markboth{\spacedlowsmallcaps{#1}}{\spacedlowsmallcaps{#1}}}
\renewcommand{\sectionmark}[1]{\markright{\thesection\enspace\spacedlowsmallcaps{#1}}}

\cleardoublepage

\pagestyle{scrheadings}
\acresetall

%\pagenumbering{arabic}
%\setcounter{page}{90}
% use \cleardoublepage here to avoid problems with pdfbookmark
\cleardoublepage

\part{Introduction}
\chapter{Economical and Political Introduction}\label{ch:introduction_pol}

\section{Climate Change}\label{ch:climate_change}

\marginpar{\newline From a scientific perspective climate change caused by greenhouse gas emission is evident. }

Stabilizing the global climate is one of the largest challenges of humanity in this century. The most recent report of the intergovernmental panel climate change IPCC \cite{stocker_climate_2013} shows that human influence on climate change is evident. 

\begin{quote}
\textit{"Human influence on the climate system is clear. This is evident from the increasing greenhouse gas concentrations in the atmosphere, positive radiative forcing, observed warming, and understanding of the climate system.", Working Group 1, Assessment Report 5, IPCC, 2013.}
\end{quote}

\begin{figure}[h]
\centering
\includegraphics[width=\textwidth]{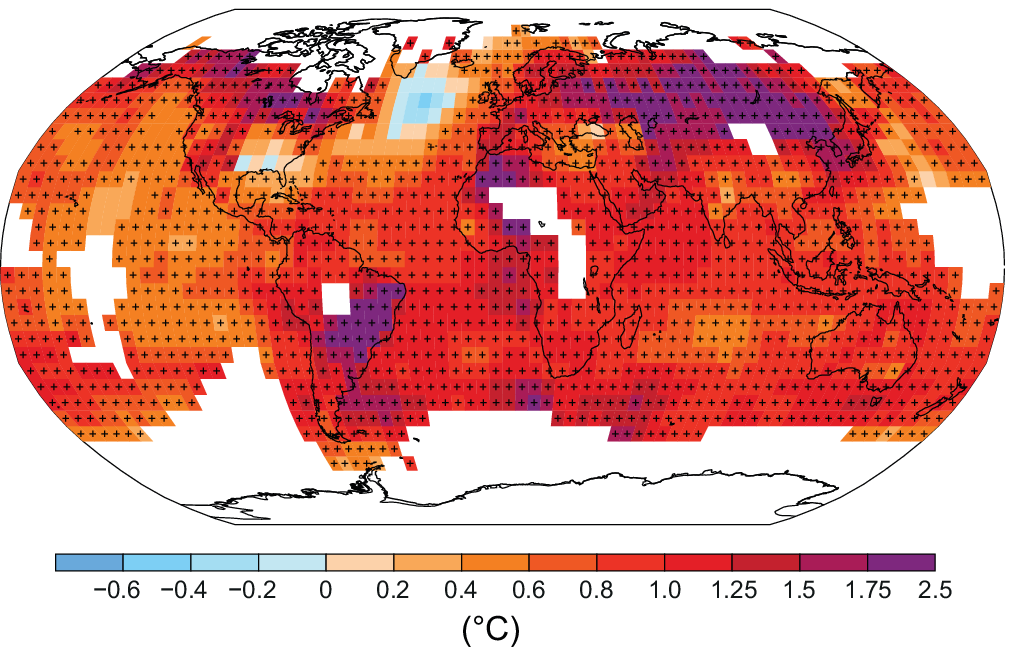}
\caption[Global surface temperature change]
{Map of the observed surface temperature change from 1901 to 2012.\\
\emph{Reprint from IPCC report WG1, AR5 \cite{stocker_climate_2013}.}}
\label{img:change_in_surface_temperature}
\end{figure}

\autoref{img:change_in_surface_temperature} shows the annual average temperature rise on the globe within the last century. Global temperature rose by 0.8 degree in average within this period. 

\marginpar{\newline Burning fossil fuels has lead to a strong increase in $CO_2$ concentration in the atmosphere.}

The combustion of fossil energy sources like oil, gas and coal forms carbon dioxide $CO_2$ that has accumulated in the atmosphere and in the ocean over the last century. This has lead to an increase in $CO_2$ concentration in the atmosphere from $280\,ppm$ ($0.028\%$) up to $400\,ppm$ ($0.040\%$). Greenhouse gases can be regarded as thermal isolation for the planet enabling a temperature range suited for life on this planet. The physical origin is the absorption and re-emission of thermal radiation.
Changing the concentration of greenhouse gases in the atmosphere leads to a change in isolation and therefore to a temperature increase.

\marginpar{$CO_2$ remains in the atmosphere for centuries. Climate change therefore persists over dozens of generations.}

As $CO_2$ is a highly non-reactive gas, it remains in the atmosphere for several hundred years. The human influence on the climate therefore persists over centuries and affects dozens of future generations.

\begin{quote}
\textit{"Warming of the climate system is unequivocal, and since the 1950s, many of the observed changes are unprecedented over decades to millennia. The atmosphere and ocean have warmed, the amounts of snow and ice have diminished, sea level has risen, and the concentrations of greenhouse gases have increased.“, Working Group 1, Assessment Report 5, \acs{IPCC}, 2013.}
\end{quote}

\begin{figure}[h]
\centering
\includegraphics[width=\textwidth]{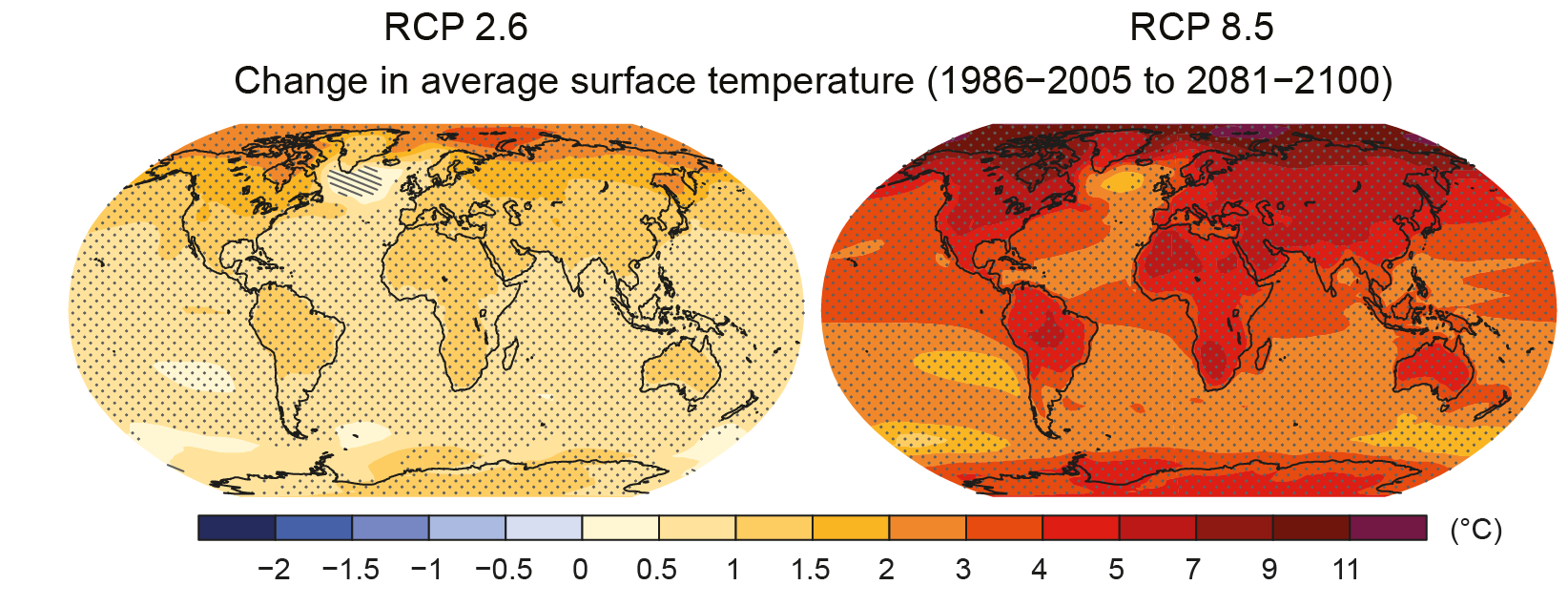}
\caption{Model result for average surface temperature change for scenario \acs{RCP} 2.6 and \acs{RCP} 8.5.\\
\emph{Reprint from \acs{IPCC} report WG1, AR5 \cite{stocker_climate_2013}.}}
\label{img:scenario_surface_temperature}
\end{figure}

The \acf{IPCC} has estimated the expected global temperature rise for different emission scenarios. The modelled mean temperature rise is shown in \autoref{img:scenario_surface_temperature} for two \acfp{RCP}. Please note that this temperature increase is in addition to the presently observed mean increase as shown in \autoref{img:change_in_surface_temperature}.

\acs{RCP} 2.6 corresponds to cumulative emissions between 2012 and 2100 of $1000\,Gt\,CO_2$, the scenario \acs{RCP} 8.5 to $6000\,Gt\,CO_2$, respectively.

\vspace{5mm}

\marginpar{An average temperature rise has major implications of life on earth, like water scarcity, more intense weather extremes and monsoons, global see level rise, droughts and major implications on food production.}

The $CO_2$ concentration in the ocean and the average temperature rise has major implications on life on earth. Since warmer air can take up more moisture, precipitation increases in some regions, whereas in others it decreases leading to droughts. Water as a resource will generally become more scarce and is expected to be a source of conflicts in the future. Water scarcity will furthermore lead to a reduction in agriculture.\\
Weather extremes occur more frequently and increase in intensity. Monsoons are expected to become more intense. The melt-down of glaciers, greenland and antarktis lead to a global sea level rise threatening whole island nations and civilizations on shallow land close to oceans.
Furthermore the take-up of $CO_2$ leads to an acidification of the oceanic water threatening corals and other species.

\marginpar{The issue of climate change is not longer a question of avoiding - it is a question of damage reduction. }

Most discussions about consequences of climate change are based on a 2 degree scenario, what corresponds to \acs{RCP} 2.6. Climate change is already in full progress and many of the mentioned consequences become more and more a reality. The issue of climate change is not longer a question of avoiding - it is a question of damage reduction. 
Scenarios like \acs{RCP} 8.5 would have enormous consequences and additional climate feed-backs would be triggered such as the release of methane from Siberian permafrost that is melting. The Arctic ice is completely molten by the end of this century in scenario \acs{RCP} 8.5.\\

\marginpar{Changing the societies from fossil-based to renewable is a major requirement to effectively protect the climate.}

To avoid the danger of climate change on cost of future generations it is of crucial importance to stop emitting greenhouse gases.
In the authors view it is the largest challenge of this centuries' societies, requiring change in social structure, change of political legislations as well as development and implementation of green technologies.

\section{Energy Markets and Photovoltaics}

\begin{figure}[h]
\centering
\includegraphics[width=\textwidth]{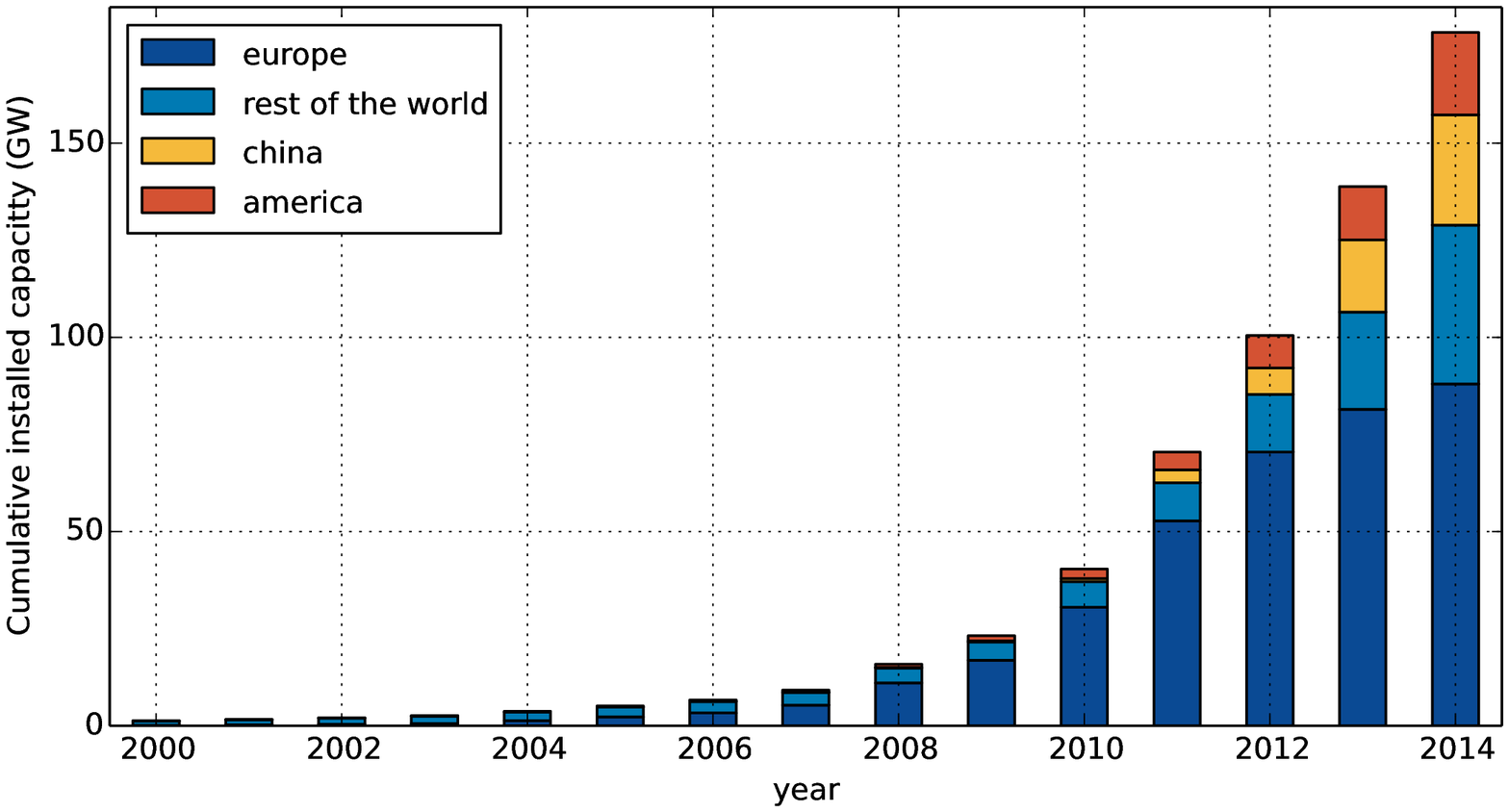}
\caption{Cumulative installed photovoltaic peak power worldwide.\\
\emph{Data source: SolarPowerEurope \cite{solar_power_europe_2015}.}}
\label{img:pv_market}
\end{figure}

The photovoltaic industry has grown tremendously in the past ten years. At the end of 2014 the total installed photovoltaic power reached $178\,GW_p$ as shown in \autoref{img:pv_market}. The industry association SolarPower Europe estimates that the cumulative installed power could reach half a terawatt in 2020 \cite{solar_power_europe_2015}. 

\marginpar{\newline At the end of 2014 the total installed PV power reached $178\,GW_p$}

Due to a strong reduction in support schemes for renewable energy the growth in Europe has slowed down. But Europe remains the region with the largest PV installation. In 2014 China was the biggest market - more than $10\,GW$ have been newly installed in 2014.\\

\begin{figure}[h]
\centering
\includegraphics[width=\textwidth]{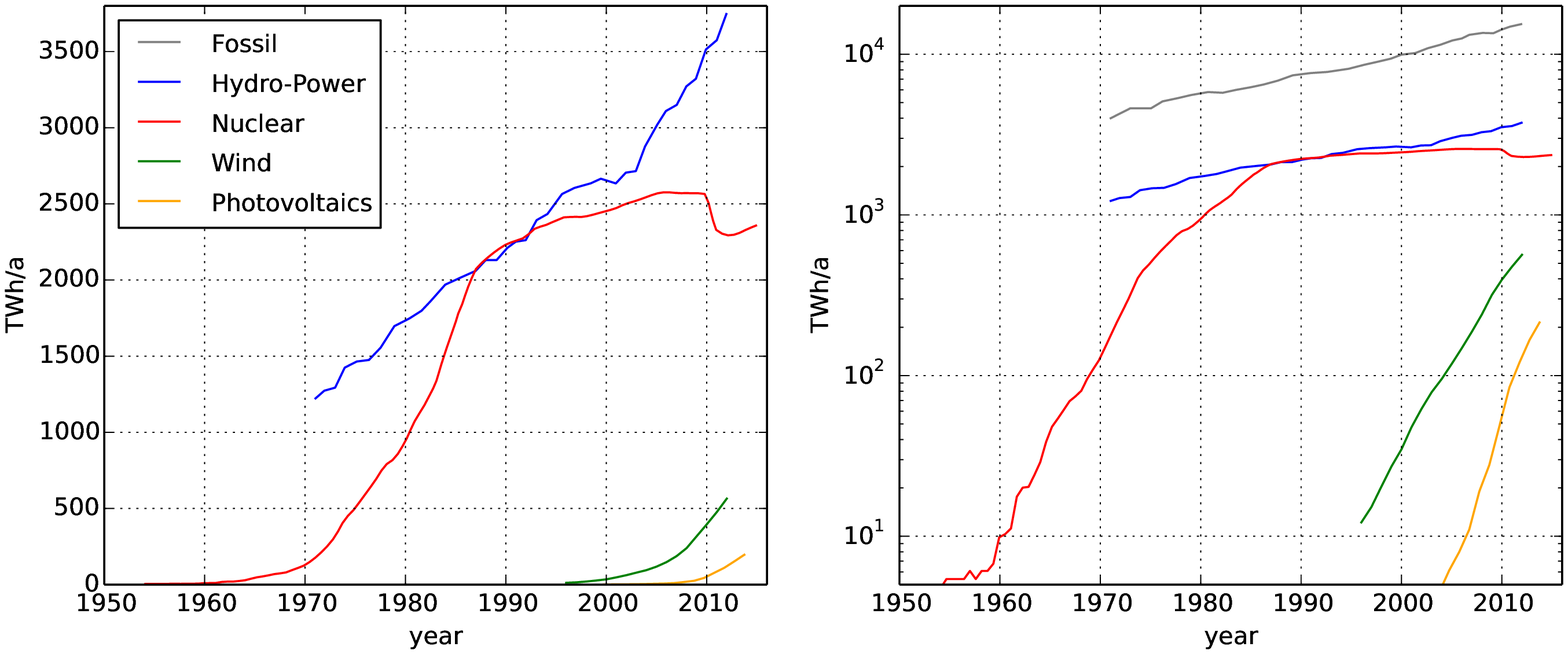}
\caption{Annual worldwide produced electricity by technology. Both graphs show the same information, on the left in linear y-scale, on the right in logarithmic y-scale.\\
\emph{Data Source: PV power from SolarPower Europe\cite{solar_power_europe_2015} with assumed \gls{full_load_hours} of $1100\,h$.
Wind power from Global Wind Statistics \cite{wind_statistics_2012} with assumed \gls{full_load_hours} of $2000\,h$.
Nuclear, hydro and fossil production from IEA Key World Energy Statistics \cite{iaea_statistics_2014}.
}}
%nuclear_instustry_2015
\label{img:energy_source_comparison}
\end{figure}

The amount of installed peak power of photovoltaics cannot be directly compared to other sources of energy since the \gls{full_load_hours} vary significantly.
Therefore the total amount of generated electricity per year is plotted  in \autoref{img:energy_source_comparison} to compare electricity generation from wind, PV, nuclear, hydro and fossil thermal electricity production. Please note that fossil fuels for heating or transportation is not considered in this graph.

\marginpar{\newline Different energy sources for electricity generation are compared by its global annual energy production. }

The nuclear industry experienced a boom starting in the 1970 increasing to an energy production of about $2500\,TW/a$. After the first nuclear catastrophe in Tschernobyl in 1986 its growth decreased whereas after the explosion of the nuclear power plant in Fukushima in 2011 the worldwide nuclear energy production even dropped.\\

\marginpar{\newline Nowadays wind and PV grow as fast as the nuclear electricity production during its boom phase in the 1980s. }

Today (2015) in Greece, Italy and Germany solar power produces more than $7\,\%$ of the total electricity demand \cite{solar_power_europe_2015}. On a global level the PV production is still on a low level compared to the annual hydro, nuclear and fossil electricity production. The total PV production reached about $10\,\%$ of nuclear and $1\,\%$ of fossil fuel production.
A look at the growth rates in \autoref{img:energy_source_comparison} shows that PV and wind are growing as fast as the nuclear energy did in the 1980. The trend is clear: In Europe in 2014 $19.9\,GW$ power of renewable energy sources was installed whereas nuclear power did not change and coal, oil and gas electricity production capacities have  been decommissioned\cite{solar_power_europe_2015}.

\section{Cost Development}\label{ch:cost-development}

\begin{figure}[h]
\centering
\includegraphics[width=\textwidth]{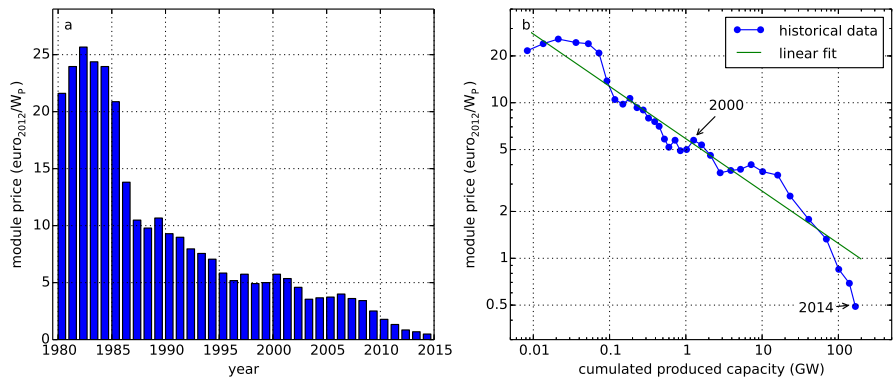}
\caption{a) Development of average module prices. b) \Gls{price_experience_curve} of PV modules with a price experience factor of $20\,\%$.\\
\emph{Data source: Agora Energiewende \cite{agora_pec_2015} and ISE PV report \cite{ISE_PV_report_2015}.}}
\label{img:price-development}
\end{figure}

The price of photovoltaic modules experienced an unforeseen decline while power conversion efficiencies went up. \autoref{img:price-development}a shows the price development from 25 Euros per Watt-Peak in 1980 down to below 50 Eurocent in 2014 \cite{agora_pec_2015}.

\marginpar{\newline The cost of PV modules decreased tremendously in the past 40 years. With every doubling of cumulated installed power prices decreased by $20\%$. }

The price experience factor for silicon PV modules was about $20\,\%$ in the last 35 years, meaning that for every doubling of the worldwide cumulated produced capacity the price was reduced about $20\,\%$. This effect is shown in the \gls{price_experience_curve} in \autoref{img:price-development}b.

\marginpar{\newline Many actors in the electricity production heavily underestimated the dynamics of cost development in the PV sector. Today PV electricity is less expensive than electricity from the grid (grid parity).}

\begin{figure}[h]
\centering
\includegraphics[width=\textwidth]{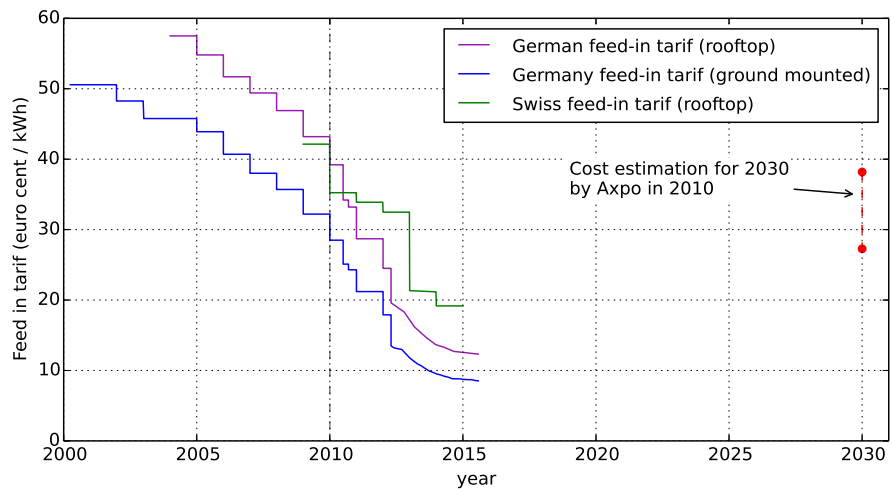}
\caption{Development of feed-in tariffs for photovoltaics in Germany and Switzerland\cite{wirth_fakten_2015, stickelberger_swisssolar_2013}. In red on the right the cost estimation for 2030 by Axpo, a major electricity supplier in Switzerland, is shown\cite{axpo_stromprespektiven_2010}.}
\label{img:feed-in-tarifs}
\end{figure}

With the prices of modules also the \gls{lcoe} dropped. When the \emph{Erneuerbare-Energien-Gesetz (EEG)} was introduced in 2004 in Germany the feed-in tariff for rooftop photovoltaics was close to 60 eurocent/kWh whereas in 2014 it was below 14 eurocent/kWh \cite{wirth_fakten_2015}. \autoref{img:feed-in-tarifs} shows the development of the feed-in tariffs for PV in Germany and Switzerland. This enormous decline of production cost was unforeseen or ignored by many conventional actors in the electricity production. The Swiss electricity supplier Axpo for example estimated in a report 2010 the cost of PV for the year 2030 to be between 30 and 42 Rp./kWh\cite{axpo_stromprespektiven_2010}. Already five years after the report the Swiss feed-in tariff was well below as shown in \autoref{img:feed-in-tarifs}.
This underestimation, which is by far not an exception, shows the huge dynamics of the PV industry in the past 10 years.\\

Meanwhile the cost of a PV module is less than half of the system costs. Planing, installation, cables and the inverter make up the other half. As the cost reduction potential of the installation is limited, the module efficiency gets more and more important for the total system costs. System costs are usually given in Franks per watt peak ($CHF/kW_p$) - Increasing the module power therefore decreases the total costs of the system.

\chapter{Technical Introduction}\label{ch:introduction_tech}

\section{Photovoltaic Technologies}\label{ch:technology}

In this section the main commercially relevant solar cell technologies are discussed and compared with the novel technologies \acfp{OSC}, \acfp{DSSC} and \acf{MALI} \acfp{PSC}.
To keep this section short other technologies and materials like organic hybrids, gallium-arsenide (GaAs), quantum dot or all types of concentrator cells are not treated.

\subsection{Crystalline Silicon Solar Cells}

\marginpar{\newline Silicon PV is currently the dominant technology in the market. Modules are produced from single wafers of crystalline silicon. Compared to other technologies it is rather thick ($150\,\mu m$).}

Crystalline silicon is the most widely used material to produce solar cells and covers more than 90 percent of the global market size. The highly purified silicon crystal is cut with a wire-saw in slices. These are then connected in series to create a module. The wafers are either made from a single crystal\footnote{Also called mono-crystalline.} (higher cost, higher efficiencies) or from a poly-crystalline\footnote{Also called multi-crystalline.} ingot (lower cost, lower efficiency).\\

Due to economy of scale effects this elaborate technique became extremely cheap, as shown in the previous section. The optical absorption of silicon solar cells is comparably weak, such that wafers need to be around $150\,\mu m$ thick to absorb most of the light. Due to the high material quality the charge carrier diffusion length is large enough for the carriers to leave the device via diffusion.

\subsection{CdTe and CIGS}

\marginpar{\newline \acs{CdTe} and \acs{CIGS} are are deposited on a substrate to create thin-film solar cells. A module is built by \gls{monolithic_integration}. }

The materials \ac{CdTe} and \ac{CIGS} are the two most prominent compound semiconductors used in photovoltaics. Both materials are strong absorbers such than they can be made thin - A \ac{CdTe} layer is about $8\,\mu m$ thick and \ac{CIGS} layer is about $2\,\mu m$ thick.
Both technologies are deposited on glass via co-evaporation, \ac{MOCVD} or other deposition techniques. In both cases \gls{monolithic_integration} is used to create a module.

Since cadmium is highly toxic the large scale application of \ac{CdTe} solar cells is controversial. The CdTe compound is however is non-toxic and strongly bound chemically, such that elevated temperatures above 1000 degree Celsius are required to dissolve the cadmium.

As indium and gallium are rare and expensive the use of combinations of more abundant materials is tested such as copper zinc tin sulfur (CZTS) reaching currently an record efficiency of $12.6\%$ \cite{green_solar_2015}.

\subsection{Amorphous Silicon}

\marginpar{\newline Amorphous silicon has a much stronger absorption than its crystalline form. It can therefore be used as thin film. Due to its low material quality efficiencies are much below classical wafer-based silicon. }

In amorphous silicon solar cells silicon is deposited on a substrate to form a thin amorphous film. Amorphous silicon is a direct semiconductor and has therefore a much higher absorption than its crystalline form. Due to a large amount of defects in the material the electrical quality is much lower than in crystalline silicon resulting in clearly lower efficiency.

Amorphous silicon can be used in a \gls{tandem} configuration with micro-crystalline silicon resulting in higher \acp{PCE}.
Since crystalline silicon has become very cheep in the recent years, amorphous silicon has lost of relevance for power production.

Other applications are possible since amorphous silicon can be deposited on flexible substrates.

\subsection{Novel technologies}

\marginpar{\newline Novel technologies offer cheaper production, light-weight or flexibility.}

A number of novel photovoltaic materials and concepts are investigated by researchers. The goal is to create solar cells that are either more efficient, less expensive or flexible, light-weight or transparent. Possible applications of such would be transparent solar cells\footnote{It may need to be mentioned, that transparent solar cells are by definition not very good absorbers. Efficient power conversion cannot be expected from this type of device.}, solar cells on mobile devices or the use as design element (for example Solarte by Belectric \cite{solarte_belectric}).\\

A \ac{DSSC} uses different materials for charge absorption and transport. A dye absorbs the photons - The electrons are rapidly transferred to an electron conductor (usually mesoporous $TiO_2$) and holes are transported by a liquid electrolyte.
\ac{MALI} perovskite solar cells were first fabricated as \ac{DSSC} \cite{park_perovskite_dssc}. Perovskite solar cells are introduced in the next chapter.\\

\marginpar{\newline Organic solar cells are very strong absorbers and can therefore be made very thin ($100\,nm$).}

An \acf{OSC} is made from organic molecules (a molecule that includes a carbon atom) that are either evaporated or solution-processed. The organic absorber materials have a very strong absorption such that it is sufficient to use an active layer thickness of $100\,nm$.
One main obstacle of \ac{OSC} is the high \gls{exciton} binding energy caused by the low dielectric constant. The thermal energy at room temperature is not sufficient to dissociate an \gls{exciton} into free charge carriers. An interface between two materials with different energy level is required. In state-of-the-art organic solar cells two materials are mixed in a so called bulk-heterojunction. Because some energy is lost at the dissociation of \glspl{exciton} the maximum achievable \ac{PCE} is lower than for other techniques \cite{gruber_thermodynamic_2012}.

\subsection{Technology Overview}

\marginpar{\newline The market share, record efficiency for modules and cells, the temperature coefficient and degradation rates of the mentioned technologies are compared.}

\begin{figure}[h]
\centering
\includegraphics[width=\textwidth]{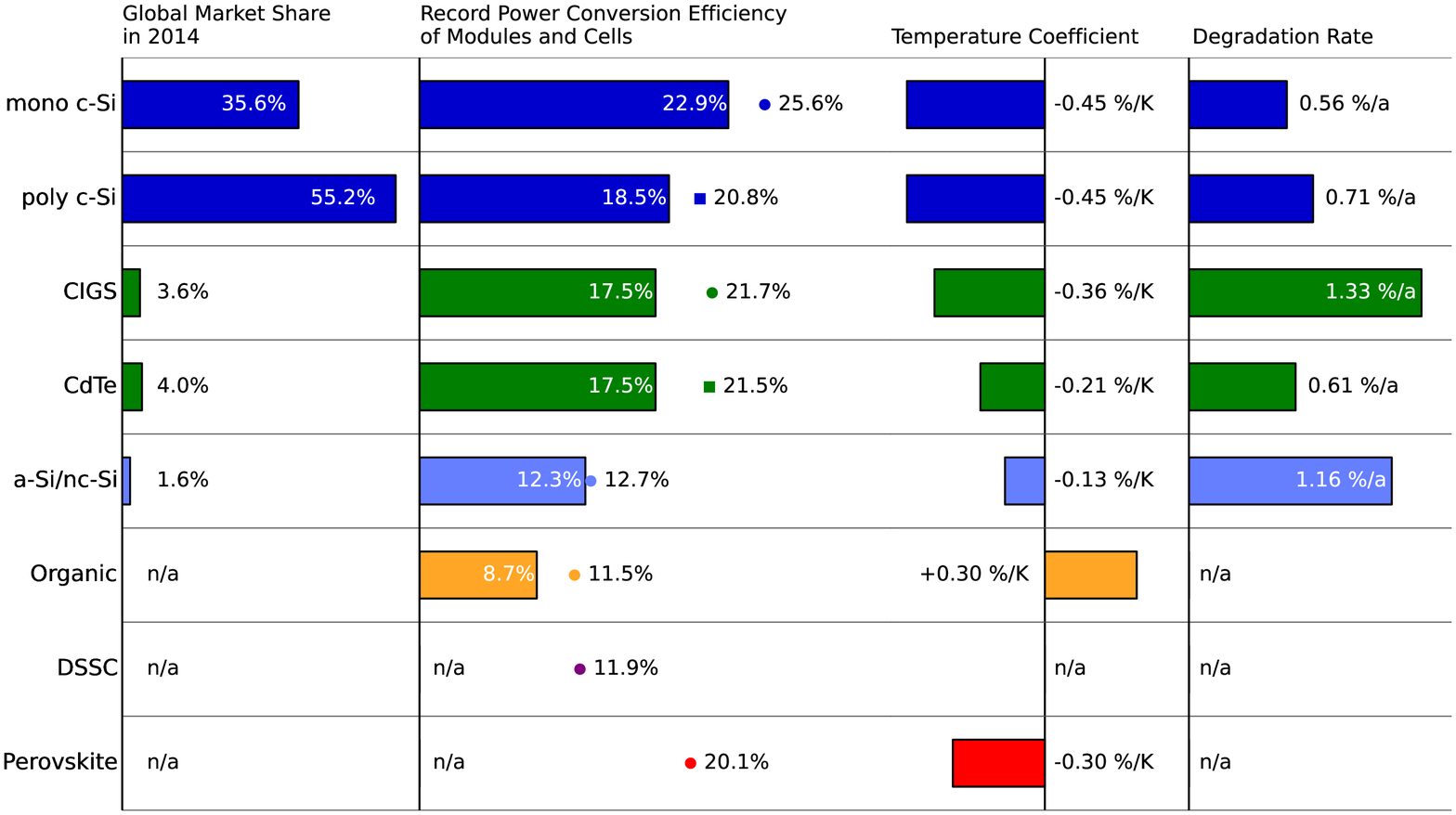}
\caption{Overview over the commercially relevant solar cell materials and the novel technologies. Dots indicate the record \ac{PCE}s for cells, the bars the module records. For the novel technologies (\ac{OSC}, \ac{DSSC} and perovskite) there is no reliable degradation data yet and they are not yet commercially available.\\
\emph{Data source: 
Market share from ISE PV report \cite{ISE_PV_report_2015}.
Record \acp{PCE} from solar cell efficiency tables version 46 \cite{green_solar_2015} and NREL \cite{nrel_pce}.
Temperature coefficients from Virtuani \cite{subsi_temperature_coeff_2010}, \ac{OSC} estimation from Katz and Riede \cite{katz_temperature_2001, riede_efficient_2011}, Perovskite from Jung \cite{jung_mat_to_devices_2015}.
Degradation data from Jordan \cite{jordan_photovoltaic_2013}.
}}
\label{img:technology_overview}
\end{figure}

\autoref{img:technology_overview} provides an overview over the various solar cell technologies. As mentioned above crystalline silicon dominates currently the market. With $22.0\,\%$ mono-crystalline modules have clearly the highest power conversion efficiency \cite{green_solar_2015}.\\

\marginpar{\newline Except for organic solar cells the power output of a module decreases with temperature. }

Most technologies have a negative temperature coefficient as shown in \autoref{img:technology_overview}, except \ac{OSC}. The charge transport of these systems is temperature activated leading to higher currents at elevated temperatures. This is an advantage over other technologies.\footnote{However, the crystalline silicon module with $22.9\,\%$ \acs{PCE} would need to reach a temperature 
of $163$ degree Celsius to have the same efficiency as the \ac{OSC} module at room temperature.}\\
The degradation rate stems from a literature survey about published degradation data \cite{jordan_photovoltaic_2013} and can vary significantly. With a degradation rate below $1\,\%$ it takes more than 22 years for the module to reach $80\,\%$ of its initial performance.

\section{The Rise of Perovskite Solar Cells}\label{ch:perovskite_intro}

\marginpar{\newline \acs{NREL} regularly publishes the evolution of record power conversion efficiencies of the major cell technologies.}

\begin{figure}[h]
\centering
\includegraphics[width=\textwidth]{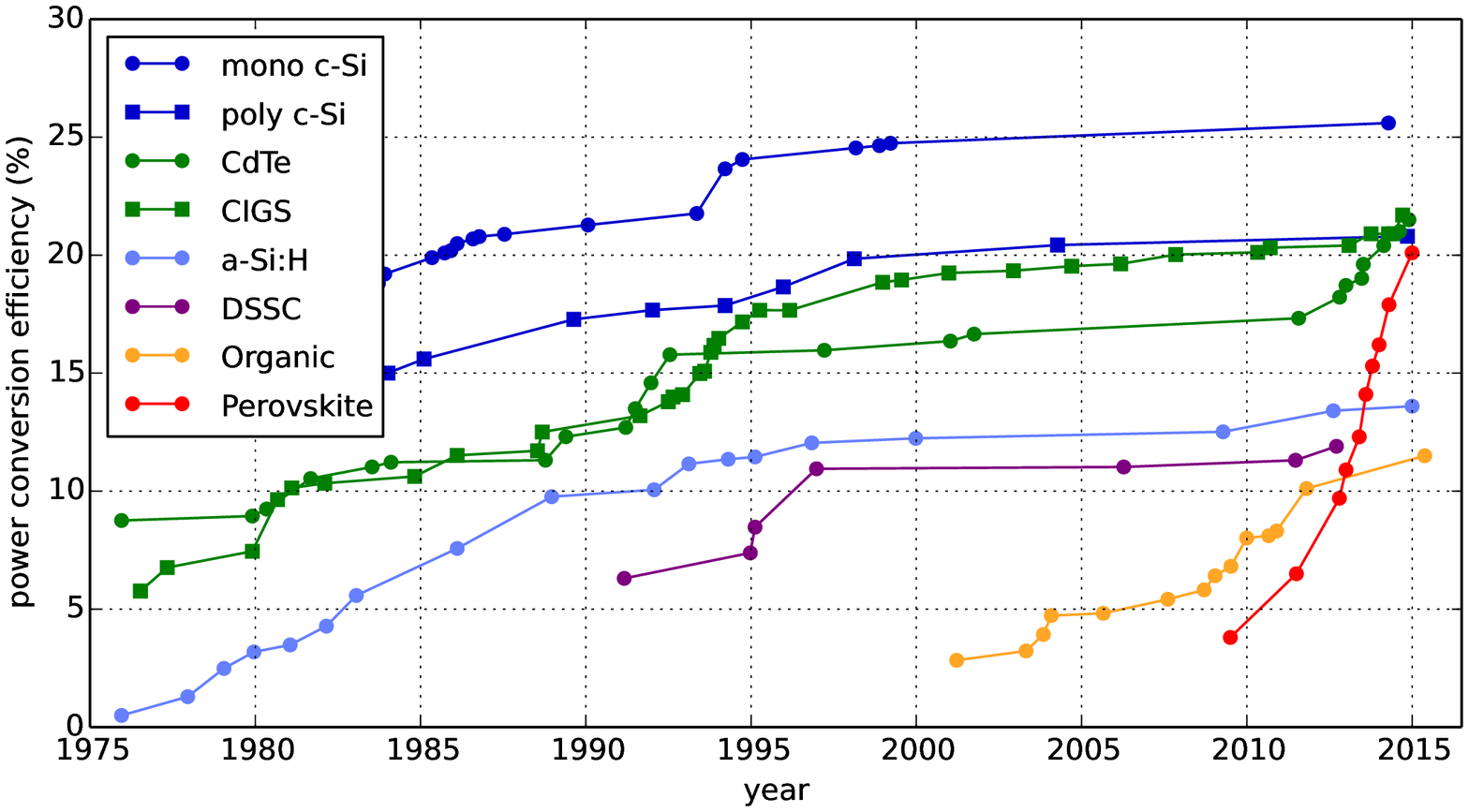}
\caption{Published record \aclp{PCE} for different cell technologies.\\
\emph{Data Source: Perovskite evolution from Park \cite{park_perovskite_2015} other data from NREL \cite{nrel_pce}.
}}
\label{img:record_pce}
\end{figure}

\autoref{img:record_pce} shows the evolution of published record \acfp{PCE} - a well known graph published regularly by the \acf{NREL}. It seems that for silicon solar cells the maximum achievable power conversion has been reached. For \ac{CIGS} and \ac{CdTe} there has been remarkable improvements in the last five years.\\

\marginpar{\newline Perovskite solar cells have experienced an unprecedented rise in efficiency up to $20.1\%$ and got a lot of attention in the research community.}

Looking at \autoref{img:record_pce} is it evident why \acf{MALI} \acfp{PSC} got a lot attention in the research community lately. The record power conversion efficiency went up to $20.1\%$ although this material was unknown in the PV community five years ago.\\
There are a couple of reasons why perovskite solar cells are highly efficient and of big interest for the community:
\begin{itemize}
	\item High optical absorption with a sharp absorption onset
	\item High charge carrier mobility
	\item Long charge carrier lifetimes
	\item Balanced charge transport
	\item High open-circuit voltages can be reached (up to $1.2\,V$)
	\item The band-gap can be adjusted (mixed halide peroskites)
	\item Low-cost fabrication
\end{itemize}

\marginpar{\newline The first perovskite solar cell was published in 2009.}

The material itself has been known in science for decades, but no attention was paid to its photovoltaic properties. In 2009 the \ac{MALI} perovskite was first used by Kojima in a \ac{DSSC} structure \cite{first_PSC_DSSC_2009}. This work was not particularly noticed since the efficiency was about $3\%$ and the device was unstable due to the liquid electrolyte. Perovskite solar cells got broad attention after Nam-Gyu Park's group published a stable device with $9.7\%$ efficiency \cite{kim_lead_2012}.

\subsection{Perovskite Material}

\begin{figure}[h]
\centering
\includegraphics[width=\textwidth]{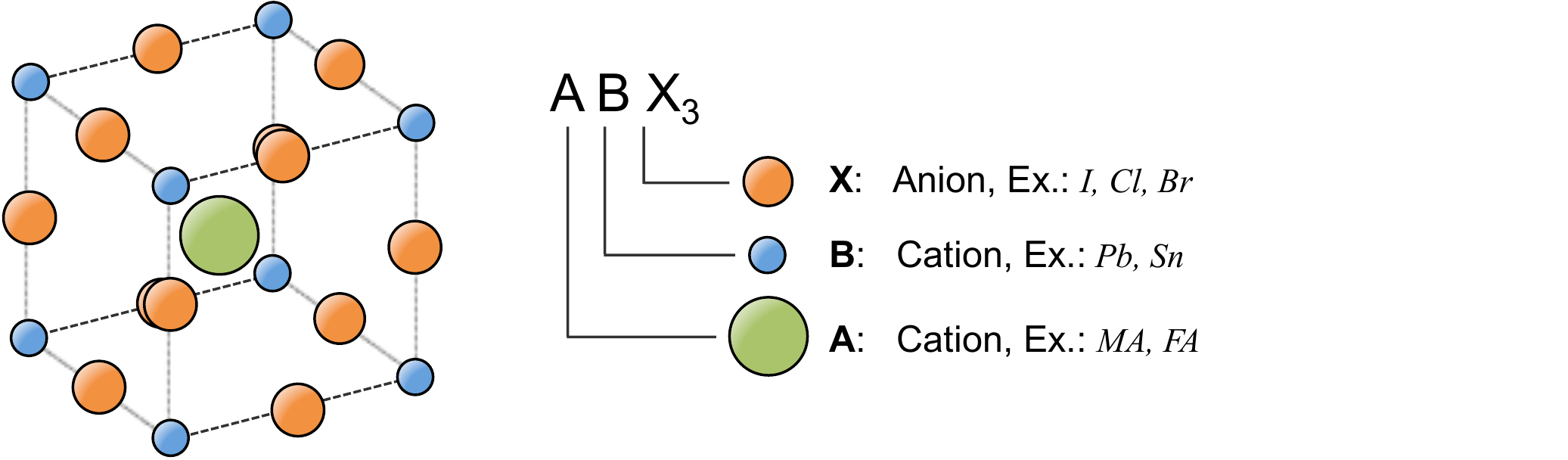}
\caption{Unit-cell of a perovskite structure.}
\label{img:psc_structure}
\end{figure}

\marginpar{\newline Perovskite is a crystal structure, not a material. The most common material combination for solar cells with perovskite structure is \acf{MALI}.}

Perovskite was initially the name for a mineral well-known as calcium titanium oxide ($CaTiO_3$) \cite{he_perovskite_2015}. Now the name perovskite is mainly used to describe a crystal structure. It has the general structure $ABX_3$ as shown in \autoref{img:psc_structure} where:
\begin{itemize}
  \item \textbf{A} is a cation. In most cases an organic molecule is used like methylammonium (MA) $CH_3NH_3^+$ or formamidinium (FA) $HC(NH_2)_2^+$.
  \item \textbf{B} is a cation that is normally either lead (Pb) or tin (Sn). To tune the material properties lead and tin can also be mixed.
  \item \textbf{X} is an anion where iodine (I), chloride (Cl) or Bromide (Br) is used. Often these atoms are mixed to tune the bandgap.
\end{itemize}

The standard material combination is \acf{MALI} $CH_3NH_3PbI_3$ but various other material combinations have been studied \cite{jung_mat_to_devices_2015} such as: 
$CH_3NH_3PbI_3$, 
$CH_3NH_3PbI_{3-x}Cl_x$, 
$CH_3NH_3PbBr_3$, 
$CH_3NH_3Pb(I_{1-x}Br_x)_3$,
$HC(NH_2)_2PbI_3$,\\
$HC(NH_2)_2Pb(I_{1-x}Br_x)_3$, 
$CH_3NH_3SnI_3$.

Recently a perovskite solar cell employing bismuth has been published \cite{park_bismuth_2015} reaching an efficiency below $1\%$.\\

\ac{MALI} perovskite can be produced with either by a solution process or by vacuum deposition \cite{gao_organohalide_2014}. The perovskite films deposited with these techniques are poly-crystalline with crystal sizes from several hundreds of nanometers to micrometers. Nie et al. published \ac{PSC} with grain sizes up to millimeters \cite{nie_high-efficiency_2015}.

\marginpar{Perovskite \acs{MALI} is poly-crystalline and deposited either by solution or with a vacuum process.}

Calculations by \ac{DFT} have shown that grain boundaries in perovskite show remarkably low mid-bandgap states \cite{yin_unusual_2014}. This is probably one of the origins of the charge carrier recombination in these materials.

\subsection{Perovskite Device Structure}

\begin{figure}[h]
\centering
\includegraphics[width=\textwidth]{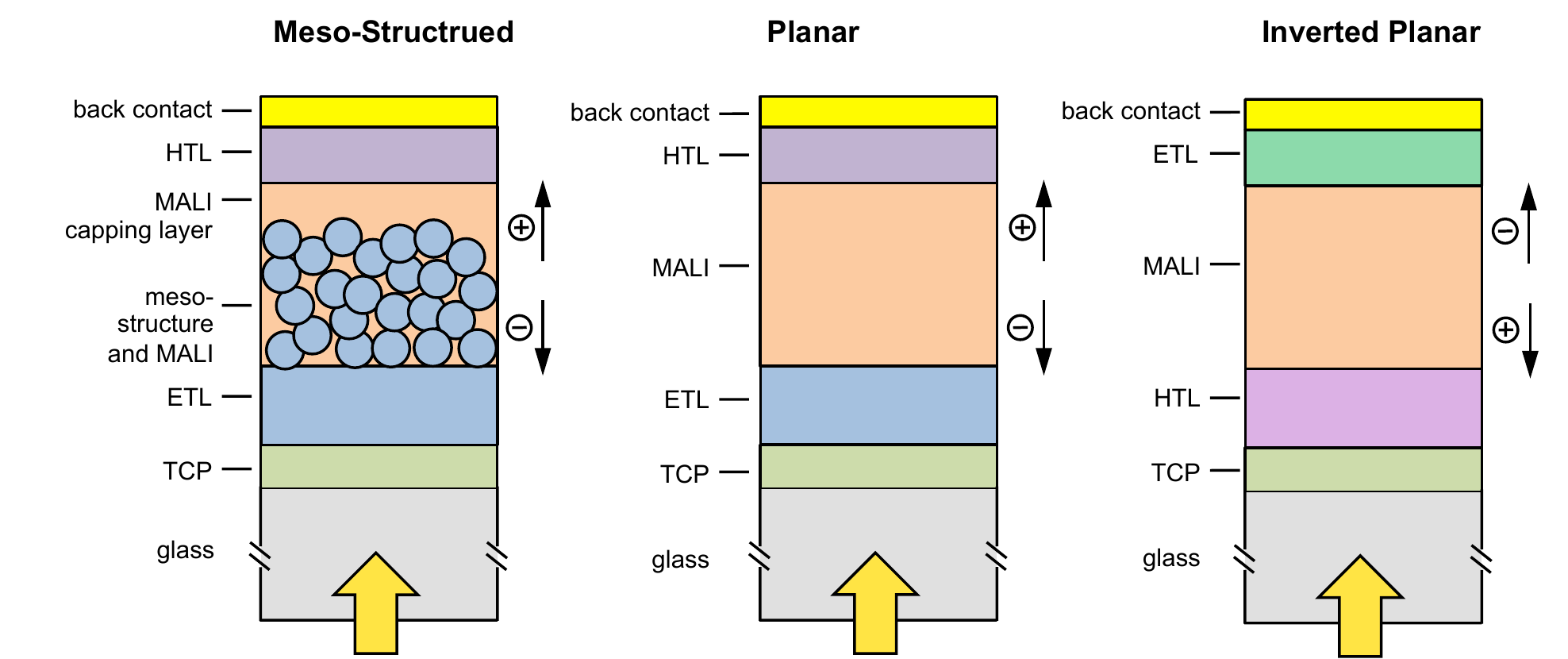}
\caption{Different solar cell architectures employing perovskite.}
\label{img:psc_device_structures}
\end{figure}

Perovskite can be applied in different solar cell architectures as illustrated in \autoref{img:psc_device_structures} using a variety of contact materials. In all three cases the layers are deposited on top of a glass with \ac{TCO}. The perovskite layer is then sandwiched between an \ac{ETL} and \ac{HTL} with a back contact of gold or aluminium.

\marginpar{\newline Perovskite solar cells are used either in a meso-porous scaffold with $TiO2$ or $Al_2O_3$ or in a planar structure.}

First published perovskite solar cells used porous $TiO_2$ structure where the perovskite was infiltrated into. In these structures electrons can be transported either in $TiO_2$ or in the perovskite whereas the holes are transported only in the perovskite. Lee et al. replaced the conducting $TiO_2$ with isolating $Al_2O_3$ resulting in a higher open-circuit voltage. \cite{lee_efficient_2012}. In these structures charge transport occurs solely in the perovskite layer.\\
These finding made clear that perovskite could also be used in a planar device structure omitting the mesoporous layer. The group of Henry Snaith published in 2013 the first planar structure reaching efficiencies of $15\%$ \cite{liu_efficient_2013}.\\

Following materials have been used as contact layers \cite{jung_mat_to_devices_2015}:\\
\ac{ETL}: $TiO_2$, PCBM, ZnO\\
\ac{HTL}: spiro-OMeTAD, P3HT, PEDOT:PSS, PCBTDPP, PTAA\\

Depending on the deposition process of the materials standard or inverted structures have been used. In the standard structure the electrons move to the front side (\ac{TCO}) and holes to the backside of the solar cell whereas in the inverted structure it is vice-versa as shown in \autoref{img:psc_device_structures}.

\subsection{Extraordinary Physical Effects}

\marginpar{\newline Many aspects of the physical working mechanism of \acsp{PSC} remains under debate.}

Despite this fast and unprecedented rise of \acp{PCE} many aspects of the device physics still remain under debate.\\
\acp{PSC} show extraordinary effects like current-voltage (IV) curve hysteresis \cite{snaith_anomalous_2014}, slow transient effects \cite{unger_hysteresis_2014, gottesman_extremely_2014, zhang_charge_2015, wu_charge_2015}, very high capacitance at low frequency \cite{juarez-perez_photoinduced_2014} and switchable photovoltaics \cite{park_perovskite_2014, xiao_giant_2014}. To explain the origin of these effects, different mechanisms have been proposed including interface traps and contact resistance between the $TiO_2$ and MALI layers \cite{snaith_anomalous_2014}, defects and traps at the $FTO-TiO_2$ layer interface \cite{jena_interface_2015}, ferroelectricity and ordering of dipole orientations

\marginpar{To explain the \ac{IV curve} hysteresis different physical effects like ferro-electricity and mobile ions have been proposed in the literature.}

\cite{chen_interface_2015, liu_hysteretic_2015, kim_ferroelectric_2015, frost_molecular_2014, wei_hysteresis_2014, chen_emergence_2015}, 
structural changes \cite{gottesman_extremely_2014}, bulk photovoltaic effect \cite{zheng_first-principles_2015} and ion migration. There is indeed increasing evidence that ion migration is taking place and has implications on cell physics and achievable performance 
\cite{unger_hysteresis_2014, zhang_charge_2015, xiao_giant_2014, tress_understanding_2015, eames_ionic_2015, haruyama_first-principles_2015}.
\\

Perovskite solar cells show further remarkable effects. Depending on growth conditions \ac{MALI} can be n-type or p-type as shown by first principle calculations \cite{yin_unique_2014}. Reversible photo-induced trap formation has been shown for mixed halide perovskites \cite{hoke_reversible_2015}. Like all ferroelectric materials MALI is piezoelectric and pyroelectric. MALI shows exceptionally low recombination \cite{wehrenfennig_high_2014} and very high PL efficiency. Furthermore optically pumped lasing has been demonstrated in $CH_3NH_3PbI_{3-x}Cl_x$ perovskites
\cite{deschler_high_2014}.\\

%\hl{Application in Tandem with CIGS or Si bottom cell?}
%Within the last years the published record power conversion efficiencies of \acf{MALI} \acfp{PSC} have gone through the roof up to $20.1\,\%$ \cite{ahn_highly_2015, zhou_interface_2014, nie_highefficiency_2015, im_growth_2014, jeon_solvent_2014, green_solar_2015, correabaena_highly_2015}.

\part{Methods}

\chapter{Device Fabrication}\label{ch:fabrication}

The perovskite solar cells characterized in this study were produced at the Institute of Microengineering (IMT) in Neuchatel according to the following procedure.\\

\marginpar{\newline The devices studied in this thesis are produced by solution employing a meso-porous $TiO_2$ layer as scaffold and electron contact and spiro-OMeTAD as hole contact.}

A solution-processed $TiO_2$ hole blocking layer was deposited on cleaned ITO substrates by spin coating twice a solution of $0.11\,ml$ diisopropoxy-titanium bis(acetylacetonate) in $1\,ml$ butanol and subsequent annealing at $500\, degree\,C$ in air. 
On top of the blocking layer, a 270 nm thick $TiO_2$ mesoporous scaffold was deposited by spin coating a $TiO_2$ nanoparticle paste and sintering at $500\, degree\,C$ in air. 
The sample was transferred to a nitrogen-filled glovebox, and the perovskite absorber layer was spin coated from a solution of $1.2\,M\,PbI_2$ and $CH_3NH_3I$ dissolved in a 7:3 v/v mixture of $\gamma -butyrolactone$ and dimethyl sulfoxide at $1000\,rpm$ for $15\,s$ and $5000\,rpm$ for $30\,s$. At 10 seconds before the end of the second spin coating step, chlorobenzene was dripped onto the rotating substrate.

After annealing at $100\,degree\,C$ for $10\,min$, 2,2',7,7'-Tetrakis-(N,N-di-4-methoxyphenylamino)-9,9'-spirobifluorene (spiro-OMeTAD) doped with Lithium was spin coated at $4000\,rpm$ for $30\,s$. Finally, a $70\,nm$ thick Au rear electrode was deposited by thermal evaporation through a shadow mask, defining a cell area of $0.25\,cm^2$.

\chapter{Experimental Setup}\label{ch:experimental}

\marginpar{\newline All experiments presented in this thesis are performs with the measurment system \paios.}

All experiments are performed using the all-in-one measurement system \paios \textit{3.0} \cite{paios}. \paios performs steady-state, transient and impedance measurements automated after each other. A function generator controls the light source - a white LED. A second function generator controls the applied voltage. The current and the voltage of the solar cell are measured with a digitizer. The current is measured via the voltage drop over a $20\,\Omega$ resistor or a transimpedance amplifier\footnote{With the transimpedance amplifier currents down to $1\,nA$ can be resolved.}, depending on the current amplitude.

%Between the measurements a waiting-time of 20 seconds is used to provide time for the device to relax. The required waiting time for complete relaxation would probably be half an hour. The waiting time of 20 seconds is chosen for practical reasons.
%\subsection{Measurement System Overview}

\subsection{Flexible Time-Resolution (Flex-Res)}

\begin{figure}[h]
\centering
\includegraphics[width=\textwidth]{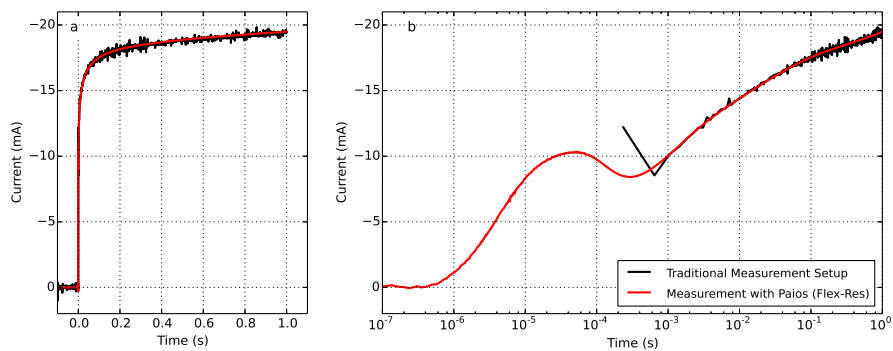}
\caption{a) \acf{TPC} response of a perovskite solar cell with linear time scale.
b) The same data but with logarithmic time scale. With linear time sampling 3 orders of magnitude can be resolved, whereas 7 orders of magnitude are possible with logarithmic time sampling (\textit{Flex-Res}). }
\label{img:flex_res}
\end{figure}

In this thesis transient current measurements are performed with very high dynamic range. Within the scope of this thesis the measurement system \paios was extended for this purpose.\\
\marginpar{To measure \ac{TPC} experiments over 7 orders of magnitude in time a special technique was developed named Flex-Res.}
Traditional oscilloscopes use fixed time steps. If such a measurement is plotted with logarithmic time $90\,\%$ of the measurement points are in the last time decade. This leads to a low dynamic range as shown in \autoref{img:flex_res} (black line). The total number of measurement points is $3000$. Between $100\,ms$ and $1\,s$ there are $2700$ points whereas between $1\,ms$ and $10\,ms$ only $27$ points are measured.

Therefore a procedure was developed where up to $30$ million points are measured and a time-logarithmic re-sampling is performed. The time-logarithmic re-sampling is necessary because $30$ million points take up too much memory and can hardly be processed.\footnote{If current, voltage and light intensity are measured with $30$ million points in double-precision (64 Bit) $720\,MB$ of memory is used. Efficient memory handling during the re-sampling is therefore crucial.}

With this technique, named \textit{Flex-Res}, transient currents can be resolved over 7 orders of magnitude. \autoref{img:flex_res} compares the traditional measurement with a measurement with \textit{Flex-Res}.
Apart from the higher time-resolution the second advantage is the increased current resolution at higher times. In the last decade of the \textit{Flex-Res} measurement in \autoref{img:flex_res}b, the current resolution is increased from $12\,Bit$ to $20\,Bit$ due to the averaging of $62'000$ points.

\subsection{Integration with Numerical Simulation}

\marginpar{\newline Numerical simulations are very useful to gain physical understanding.
When simulating several experiments of several devices, data handling and optimisation can get very complex.}

Numerical simulation of the absorption and charge transport is a very powerful tool to gain insight into physical processes. It is however also complex as many parameters play a role that are not precisely known. In this thesis more than 20 different experiments were performed on 10 different devices. Some experiments contain 20 measurement curves as experimental parameters are varied.

\begin{quotation}
\textit{With this amount of data manual data handling, comparison with simulation and plotting is not only tedious - it is close to impossible.}
\end{quotation}

This problem is solved by tight and seamless integration of measurement, processing, data storage and simulation. These improvements are made available in the measurement system \paios \citep{paios} of the company \textit{Fluxim AG}.\\

\marginpar{\newline The simulation software (\setfos) was seamlessly integrated into the measurement software (\paios) to keep the overview over devices, experiments and parameters. Measurement and simulation can hereby directly be compared.}

\begin{figure}[h]
\centering
\includegraphics[width=\textwidth]{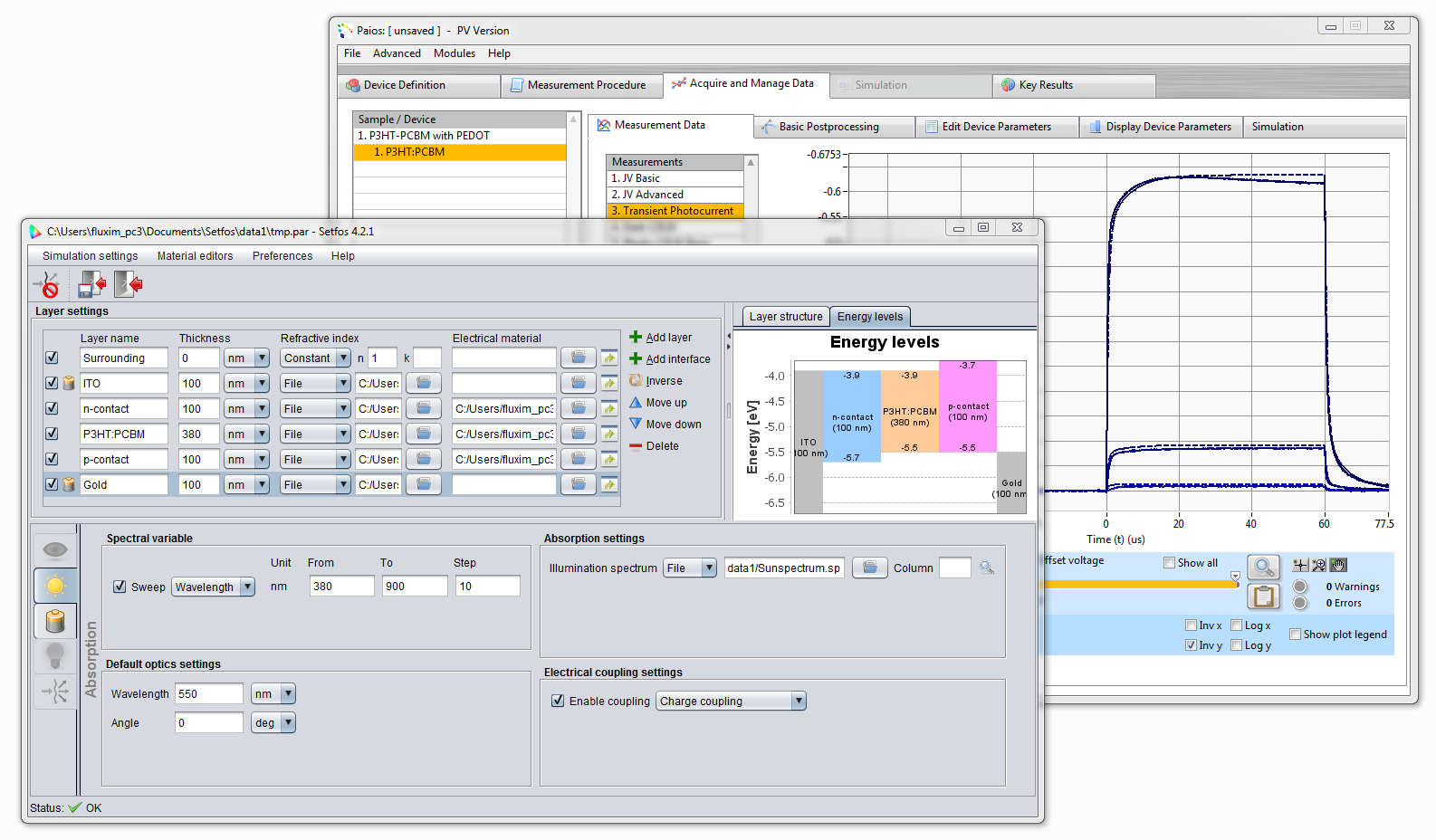}
\caption{Seamless integration of the simulation software \setfos \cite{setfos} and the measurement software \paios \cite{paios}. Measurement and simulation can directly be compared in the measurement software (simulation: dashed line, measurement: solid line).}
\label{img:spi}
\end{figure}

\autoref{img:spi} shows a screenshot of the \paios measurement software with the integrated simulation software \setfos \cite{setfos}. Hereby the \paios measurement software is the master, that stores and manages all measurement data, simulation parameter and simulation results.

\begin{figure}[h]
\centering
\includegraphics[width=0.7\textwidth]{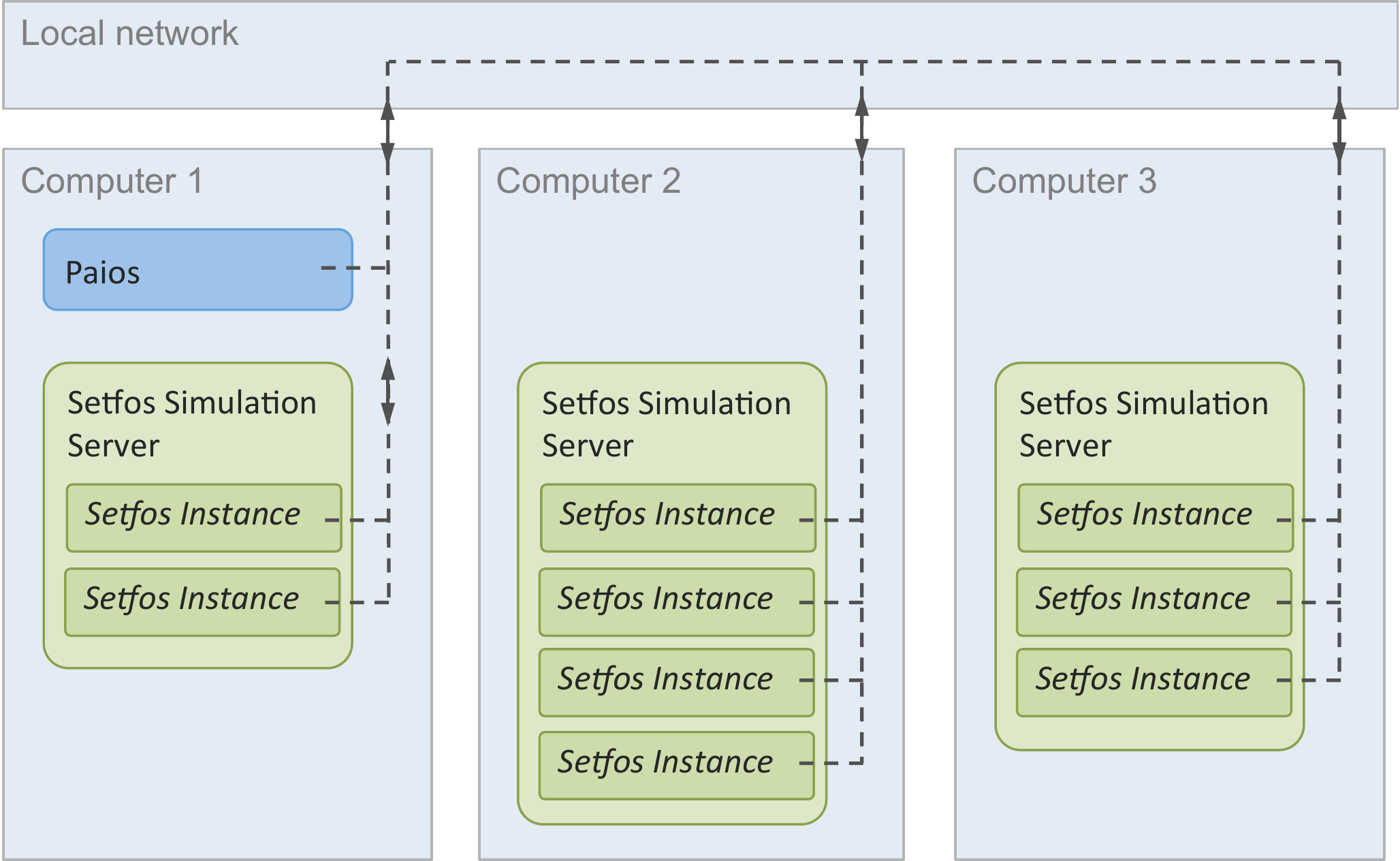}
\caption{Schematic illustration of the communication of \paios with \setfos over the local network. \paios can manage several Setfos-Servers on different computers that allows parallel distributed computing.}
\label{img:spi_network}
\end{figure}

The \paios software written in \textit{LabView} and the \setfos software written in \textit{Java} communicate over the local network. First \paios searches the local network for available Setfos-Servers. As schematically shown in \autoref{img:spi_network} several \setfos servers can run on the same computer. To perform simulations \paios can distribute simulation tasks among all \setfos instances enabling parallel computing.

The development of the \setfos - \paios integration (SPI) was a major step that facilitated the analysis of perovskite solar cells presented in this thesis.

\chapter{Simulation Model}\label{ch:simulation}

\marginpar{\newline First the absorption of photons is calculated. Then the drift and diffusion of charge carriers is computed. The models presented in this chapter are implemented in the software \setfos.}

The physical model used in this thesis is implemented in the numerical simulation software \setfos  \textit{4.2} \cite{setfos}. The absorption profile is calculated with a thin film optics algorithm \cite{lanz_extended_2011, hausermann_coupled_2009} considering the full solar cell stack from glass substrate to rear gold electrode. Complex refractive indices of perovskite are taken from Ref \cite{loper_complex_2015}.\\

The charge continuity-equation in the drift-diffusion model is solved for electrons and holes coupled by the Poisson-equation. Charge doping is simulated by adding fixed space charge in the Poisson-equation. Direct electron–hole recombination is considered in the model. For simplicity \ac{SRH} recombination is neglected. The model is solved in a one-dimensional domain.

\section{Model Equations}\label{ch:equations}

In this section the governing equations of the drift-diffusion model implemented in \setfos \cite{setfos} are listed and explained. The quantities are listed at the end of the section.\\

\marginpar{\newline The continuity equation describes how electron and hole concentration changes over time.}

The \textbf{continuity equation} for electrons and holes governs the change in charge carrier density due to current flow, recombination and generation.
\begin{equation}
\label{eq:contn}
\frac{\partial{n_e}}{\partial{t}}(x,t) = \frac{1}{q} \cdot \frac{\partial{j_e}}{\partial{x}}(x,t) - R(x,t) + G_{opt} \cdot g(x)
\end{equation}
\begin{equation}
\label{eq:contp}
\frac{\partial{n_h}}{\partial{t}}(x,t) = -\frac{1}{q} \cdot \frac{\partial{j_h}}{\partial{x}}(x,t) - R(x,t) + G_{opt} \cdot g(x)
\end{equation}
For the calculation of the charge generation $g(x)$ \setfos considers the measured illumination spectrum and refractive indices of each layer of the cell stack.\\

As \textbf{boundary conditions} the electron density at the anode and the hole density at the cathode are set to fixed values $n_{e0}$ and $n_{h0}$.
\begin{equation}\label{eq:nboundary}
n_e(0,t) = n_{e0}
\end{equation}
\begin{equation}\label{eq:pboundary}
n_h(d,t) = n_{h0}
\end{equation}

\marginpar{\newline Bimolecular recombination is used to annihilate electrons and holes.}
The radiative \textbf{recombination} is proportional to the charge carrier density of electrons and holes.
\begin{equation}
\label{eq:rec}
R(x,t) = B \cdot n_e(x,t) \cdot n_h(x,t)
\end{equation}

\marginpar{\newline The charge current consists of a drift- and a diffusion-part.}
The \textbf{currents of electron and holes} consist of drift in the electric field and diffusion due to the charge carrier density gradient.
\begin{equation}
\label{eq:ddn}
j_e(x,t) = n_e(x,t) \cdot q \cdot \mu_e \cdot E(x,t) + \mu_e \cdot k \cdot T \cdot \frac{\partial{n_e}}{\partial{x}}(x,t)
\end{equation}
\begin{equation}
\label{eq:ddp}
j_h(x,t) = -n_h(x,t) \cdot q \cdot \mu_h \cdot E(x,t) + \mu_h \cdot k \cdot T \cdot \frac{\partial{n_h}}{\partial{x}}(x,t)
\end{equation}

\marginpar{\newline The total current includes particle- and displacement-current.}
The \textbf{total current} is the sum of electron, hole current and the displacement current. This total current is constant in x.
\begin{equation}
\label{eq:jtot}
j(x,t) = j_e(x,t) + j_h(x,t) + \frac{\partial{E}}{\partial{t}}(x,t) \cdot \epsilon_r \cdot \epsilon_0
\end{equation}

\marginpar{\newline The Poisson equation is the first Maxwell equation, relating electric field and charge.}
The \textbf{Poisson equation} relates the electric field with the charges inside the layer. Charge doping is added in form of fixed charge densities $N_A$ and $N_D$.
\begin{equation}
\label{eq:poisson}
\frac{\partial{E}}{\partial{x}}(x,t) = -\frac{q}{\epsilon_0 \cdot \epsilon_r} \cdot (n_h(x,t) - n_e(x,t) + N_A - N_D)
\end{equation}

\marginpar{\newline The electric field is the gradient of the potential.}
The \textbf{cell voltage} is defined as the source voltage minus the built-in voltage and the voltage drop over the series resistance.
\begin{equation}
\label{eq:phi}
V_{Cell} = \int_{0}^{d}{E(x,t) \cdot dx} = V_{Source}(t) - j(t) \cdot S \cdot R_s - V_{bi}
\end{equation}

\marginpar{\newline The built-in field is the difference in \glspl{workfunction}. }
The \textbf{built-in voltage} is defined as the difference in \glspl{workfunction} of the electrodes. The workfunctions are calculated according to the boundary charge carrier density $n_{h0}$ and $n_{e0}$.
\begin{equation}
\label{eq:Vbi}
V_{bi} = \frac{\Phi_A - \Phi_C}{q}
\end{equation}
\begin{equation}\label{eq:workfunction_a}
\Phi_A = E_{LUMO} - ln \Big(\frac{n_{e0}}{N_0} \Big) \cdot k \cdot T
\end{equation}
\begin{equation}\label{eq:workfunction_c}
\Phi_C = E_{HOMO} + ln \Big(\frac{n_{h0}}{N_0} \Big) \cdot k \cdot T
\end{equation}

The equations above are solved on a one-dimensional grid either in steady-state, time-domain or in frequency-domain. The following quantities can be evaluated as post-processing.\\

The \textbf{potential} is evaluated according to:
\begin{equation}
\varphi(x_1,t)=\int_0^{x_1} E(x,t) \cdot dx
\end{equation}
\marginpar{Potential, effective bands and quasi \glspl{Fermi_level} are calculated as post-processing after the simulation.}

The \textbf{effective bands} are:
\begin{equation}
E_{CB}(x,t)=E_{LUMO} - q \cdot \phi (x,t)
\end{equation}
\begin{equation}
E_{VB}(x,t)=E_{HOMO} - q \cdot \phi (x,t)
\end{equation}

The \textbf{quasi \glspl{Fermi_level}} of electrons and holes are:
\begin{equation}
E_{fe}(x,t)=E_{CB}(x,t) + k \cdot T \cdot ln \Big( \frac{n_e(x,t)}{N_0} \Big)
\end{equation}
\begin{equation}
E_{fh}(x,t)=E_{VB}(x,t) - k \cdot T \cdot ln \Big( \frac{n_h(x,t)}{N_0} \Big)
\end{equation}

Further details of the simulation model can be found in our previous publications \cite{neukom_transient_2010, neukom_charge_2011, neukom_reliable_2012}.

\section{Model Parameters and Quantities}
\autoref{tab:sim_parameters} lists all parameters including the equation they influence and all other quantities occurring in the equations of the previous chapter.\\

\tablehead{\toprule Symbol & Parameter & Unit \\ \midrule}
\tabletail{\midrule\multicolumn{3}{c}{\emph{table continues on the next page}}\\}
\tablelasttail{\bottomrule}
\bottomcaption{Parameter and quantities used in equations in section \nameref{ch:equations}.}
\begin{center}
\begin{supertabular}[h]{l p{0.6\textwidth} l}\label{tab:sim_parameters}

$d$ & active layer thickness \newline
\small{\emph{equation for applied voltage (Eq. \ref{eq:phi}), \newline
boundary hole density (Eq. \ref{eq:pboundary})}} & $nm$ \\ 
\hline

$N_A$ & 
acceptor doping density (p-type) \newline
\small{\emph{Poisson-equation (Eq. \ref{eq:poisson})}} & $cm^{-3}$ \\
\hline 

$N_D$ & donor doping density (n-type)\newline
\small{\emph{Poisson-equation (Eq. \ref{eq:poisson})}} & $cm^{-3}$ \\ 
\hline

$\mu_e$ & electron mobility \newline
\small{\emph{electron drift-diffusion equation (Eq. \ref{eq:ddn})}} & $cm^2 \cdot V^{-1} \cdot s^{-1}$ \\ 
\hline

$\mu_h$ & hole mobility \newline
\small{\emph{hole drift-diffusion equation (Eq. \ref{eq:ddp})}} & $cm^2 \cdot V^{-1} \cdot s^{-1}$ \\ 
\hline

$R_S$ & cell series resistance \newline
\small{\emph{equation for applied voltage (Eq. \ref{eq:phi})}}  & $\Omega$ \\ 
\hline

$S$ & device surface \newline
\small{\emph{equation for applied voltage (Eq. \ref{eq:phi})}}  & $cm^2$ \\ 
\hline

$G_{opt}$ & photon to charge conversion efficiency \newline
\small{\emph{continuity-equations (Eq. \ref{eq:contn} and Eq. \ref{eq:contp})}} & $1$ \\ 
\hline

$B$ & radiative recombination coefficient \newline
\small{\emph{continuity-equations (Eq. \ref{eq:contn} and Eq. \ref{eq:contp}) via \newline
term for recombination (Eq. \ref{eq:rec})}} & $m^3 \cdot s^{-1}$ \\ 
\hline

$\epsilon_r$ & relative electric permittivity \newline
\small{\emph{Poisson-equation (Eq. \ref{eq:poisson}), \newline
equation for the total current (Eq. \ref{eq:jtot})}}& $1$ \\ 
\hline

$E_{LUMO}$ & energy of the \acf{LUMO} \newline
\small{\emph{equation for applied voltage (Eq. \ref{eq:phi}) via \newline
built-in voltage (Eq. \ref{eq:Vbi}, Eq. \ref{eq:workfunction_a})
}} & $eV$ \\ 
\hline

$E_{HOMO}$ & energy of the \acf{HOMO} \newline
\small{\emph{equation for applied voltage (Eq. \ref{eq:phi}) via \newline
built-in voltage (Eq. \ref{eq:Vbi}, Eq. \ref{eq:workfunction_c})
}}  & $eV$ \\ 
\hline 

$N_0$ & density of chargeable sites \newline
\small{\emph{equation for applied voltage (Eq. \ref{eq:phi}) via \newline
built-in voltage (Eq. \ref{eq:Vbi}, Eq. \ref{eq:workfunction_a}, Eq. \ref{eq:workfunction_c})
}}   & $cm^{-3}$ \\ 
\hline 

$T$ & device temperature \newline
\small{\emph{drift-diffusion equations (Eq. \ref{eq:ddp}, Eq. \ref{eq:ddn}), \newline
equation for applied voltage (Eq. \ref{eq:phi}) via \newline
built-in voltage (Eq. \ref{eq:Vbi}, Eq. \ref{eq:workfunction_a}, Eq. \ref{eq:workfunction_c})
}}   & $K$ \\ 
\hline 

$V_{source}$ & voltage of the voltage source that is connected to the device \newline
\small{\emph{equation for applied voltage (Eq. \ref{eq:phi})
}}    & $V$ \\ 
\hline

$n_{e0}$ & electron density at the left electrode \newline
\small{\emph{boundary condition for continuity-equation at x=0 (Eq. \ref{eq:contn}), \newline
equation for applied voltage (Eq. \ref{eq:phi}) via \newline
built-in voltage (Eq. \ref{eq:Vbi}, Eq. \ref{eq:workfunction_a})
}} & $cm^{-3}$ \\ 
\hline 

$n_{h0}$ & hole density at the right electrode\newline
\small{\emph{boundary condition for continuity-equation at x=d (Eq. \ref{eq:contp}), \newline
equation for applied voltage (Eq. \ref{eq:phi}) via \newline
built-in voltage (Eq. \ref{eq:Vbi}, Eq. \ref{eq:workfunction_c})
}} & $cm^{-3}$ \\ 
\hline

$\Phi_A$ & \gls{workfunction} anode & $eV$ \\ 
\hline 
$\Phi_C$ & \gls{workfunction} cathode & $eV$ \\ 
\hline 
$V_{bi}$ & built-in voltage & $V$ \\ 
\hline
$n_e$ & electron density & $cm^{-3}$ \\ 
\hline 
$n_h$ & hole density & $cm^{-3}$ \\ 
\hline 
$j_e$ & electron current & $mA \cdot cm^{-2}$ \\ 
\hline 
$j_h$ & hole current & $mA \cdot cm^{-2}$ \\ 
\hline 
$j$ & total current & $mA \cdot cm^{-2}$ \\ 
\hline 
$E$ & electric field & $V \cdot m^{-1}$ \\ 
\hline 
$\varphi$ & potential & $V$ \\ 
\hline 
$R$ & recombination & $cm^{-3} \cdot s^{-1}$ \\ 
\hline 
$g(x)$ & absorption profile & $cm^{-3} \cdot s^{-1}$ \\ 
\hline
$x$ & dimension in layer direction & $nm$ \\ 
\hline 
$t$ & time & $s$ \\ 
\hline 
$E_{CB}$ & energy of the conduction band – In comparison with $E_{LUMO}$ this band includes the band bending caused by the electric field. & $eV$ \\ 
\hline 
$E_{CB}$ & energy of the valence band – In comparison with $E_{VB}$ this band includes the band bending caused by the electric field. & $eV$ \\ 
\hline 
$E_{fe}$ & quasi \gls{Fermi_level} of electrons & $eV$ \\ 
\hline 
$E_{fh}$ & quasi \gls{Fermi_level} of holes & $eV$ \\ 
\hline 

% Calculation quantities

$\epsilon_0$ & vacuum permittivity & $F \cdot m^{-1}$ \\ 
\hline 
$q$ & unit charge & $C$ \\ 
\hline 
$k$ & Boltzmann constant & $J \cdot K^{-1}$ \\ 

% Constants
\end{supertabular}
\end{center}

\section{Perovskite Simulation Model}

\marginpar{\newline The solar cell structure is simplified for the numerical simulation. Drift-diffusion is calculated within one homogeneous layer neglecting the meso-porous scaffold and the interface layers.}

\begin{figure}[h]
\centering
\includegraphics[width=\textwidth]{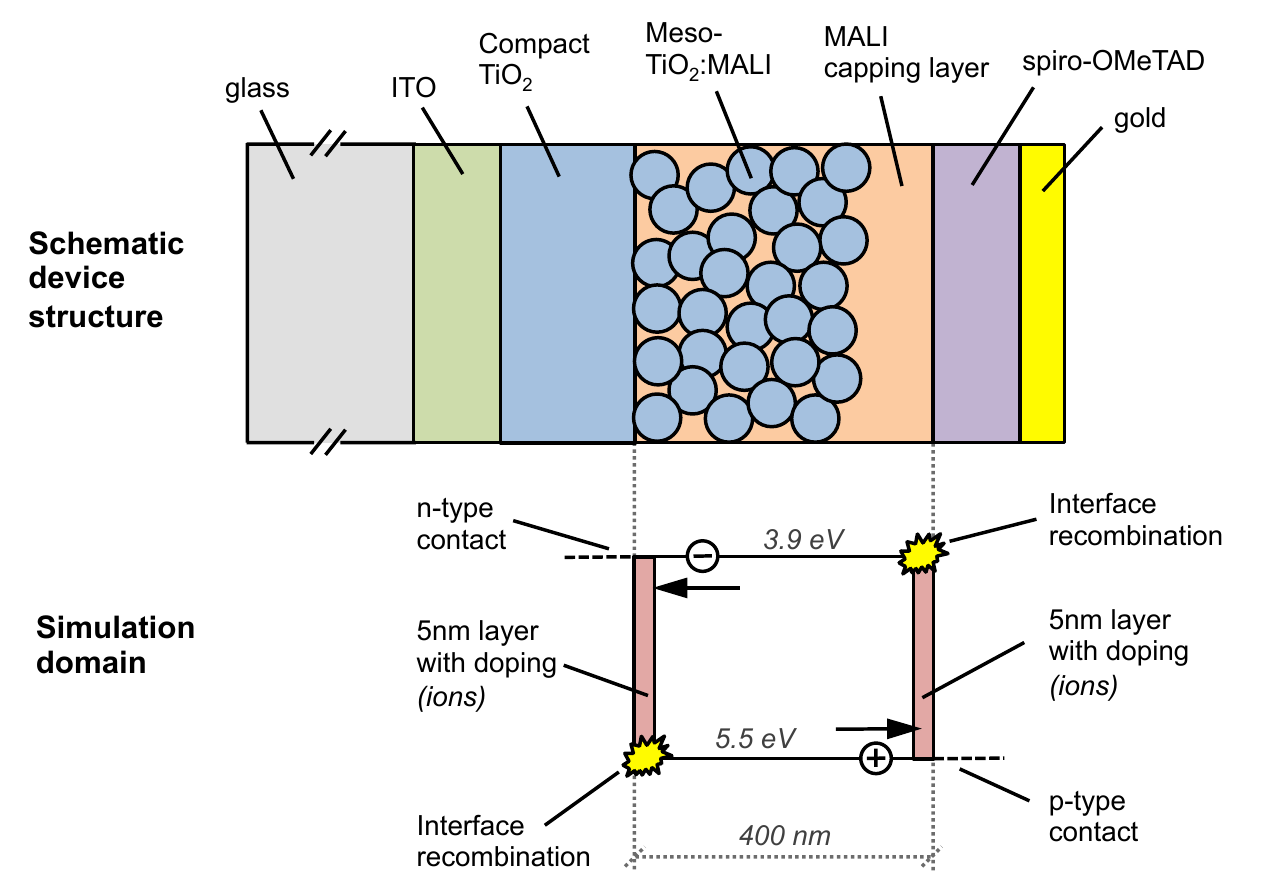}
\caption{Device structure and simulation domain. The perovskite layer \ac{MALI} and the mesoporous $TiO_2$ is simulated as one effective medium with one electron and one hole transport level.
}
\label{img:sim_scheme}
\end{figure}

The simulation scheme used for transient and steady-state simulation is shown in \autoref{img:sim_scheme}. The mesoporous $TiO_2$ structure where the \ac{MALI} is infiltrated into is neglected. We model the absorbing layer as one material with one transport level for holes and one for electrons. We assume fast electron transfer to the $TiO_2$ and electron transport to happen only within the $TiO_2$. We explain this assumption in detail in section \nameref{ch:slow_regime}.

In section \nameref{ch:slow_regime} the device is simulated with different distributions of ions. Therefore on each side of the MALI containing layer an additional virtual layer is added with $5\,nm$ thickness to model the ions close to the interface. This layer is also used to control the surface recombination of charge carriers at the electrodes. For simplicity charge transport in the compact $TiO_2$ layer and in the Spiro-OMeTAD layer is disregarded.
In further research the influence of the mesoporous layer on trap-density, mobility and doping needs to be investigated in more detail. Also the distribution of the ionic species within the perovskite layer is subject of further investigations.

\chapter{RC Correction}\label{ch:rc_correction}

In this thesis a method to correct \gls{RC_effects} is applied to correct \acs{photo-CELIV} experiment in section \nameref{ch:fast_regime}. In this section the extraction of the series resistance and the geometrical capacitance of a device is explained and the derivation of the correction formula (\autoref{eq:rc_compensation}) is shown.\\

\marginpar{\newline A device with a dielectric has a capacitance that forms an RC-circuit with all resistances in series, like contact resistance and measurement resistance.}
Each device with a dielectric between two electrodes has a capacitance. Changing the voltage V on a capacitance C requires a current i to flow to charge the electrodes.
\begin{equation}
i(t)=C \cdot \frac{dV}{dt}
\end{equation}

With a resistance $R_S$ in series an RC-circuit is formed with a time constant $\tau$.
\begin{equation}
\tau=R_S \cdot C
\end{equation}

Physical effects that happen on a time scale shorter than $\tau$ will therefore be hidden by \gls{RC_effects}.

\section{Determining series resistance and capacitance}\label{ch:measuring_rc}

\marginpar{\newline The geometrical capacitance and the series resistance are determined by applying a small voltage step in reverse direction.}

\begin{figure}[h]
\centering
\includegraphics[width=0.5\textwidth]{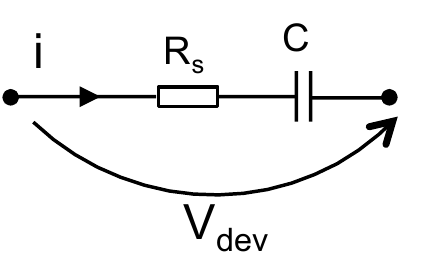}
\caption{RC-circuit used to fit the measured current.}
\label{img:rc_circuit_1}
\end{figure}

In order to determine the series resistance $R_S$ and the geometrical capacitance $C$ of the solar cell, a voltage step of -0.3 Volt is applied to the device. In reverse direction the device is blocking, so only an RC-current flows charging the capacitance. \autoref{img:rc_circuit_1} shows the simple electric circuit used to extract $R_S$ and $C_{geom}$. Compared to the scheme used to correct for \gls{RC_effects} (\autoref{img:rc_circuit_2_reprint} on page \pageref{img:rc_circuit_2_reprint}) the \emph{cell} is omitted as it is considered to be fully blocking in reverse direction.

\autoref{eq:rc_current} describes the relation of current and voltage of the circuit in \autoref{img:rc_circuit_1}. The measured transient voltage $V_{dev}(t)$ is used to calculate the RC-current.

\begin{equation}\label{eq:rc_current}
\frac{di}{dt} = \frac{1}{R_S} \cdot \frac{dV_{dev}}{dt} - \frac{1}{R_C \cdot C} \cdot i(t)
\end{equation}

\marginpar{\newline The current of an RC-circuit can be described analytically. Fitting the analytical solution to the measurement allows to extract the capacitance and the series resistance.}

\begin{figure}[h]
\centering
\includegraphics[width=\textwidth]{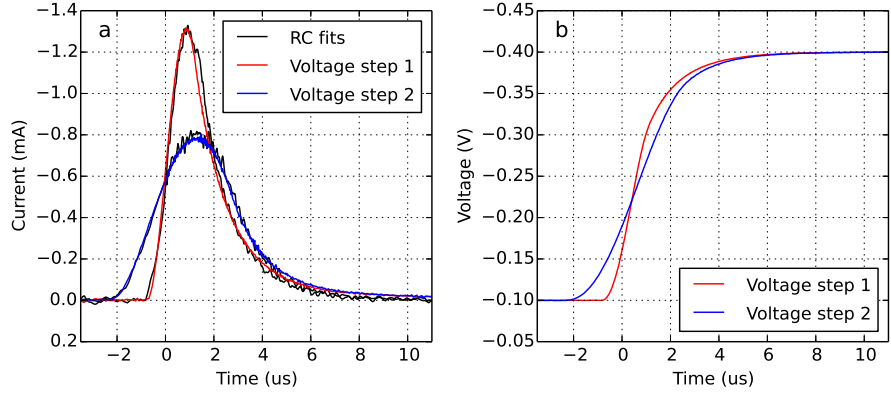}
\caption{a) Current response to a voltage step. b) applied voltage step with varied rise time.}
\label{img:rc_extraction}
\end{figure}

\autoref{img:rc_extraction}b shows two voltage steps with different rise-times that are applied to the \acf{PSC}. A rise-time is used to avoid high current peaks. In \autoref{img:rc_extraction}a the current response to the voltage steps is shown. The series resistance $R_S$ and the capacitance $C$ of \autoref{eq:rc_current} are now adjusted such to fit both currents. The result is shown in black using values of $R_S=59.9\,\Omega$ and $C=10.0\,nF$. 

\begin{quotation}
\textit{The RC-fits show more noise than the original current because the RC-fits are calculated using the measured voltage and its derivative.}
\end{quotation}

\section{Derivation of the RC-Current Correction}\label{ch:rc_correction_derivation}

\marginpar{\newline The current of a device with a capacitance and a series resistance can be corrected for the RC-current with the model presented in this section.}

\begin{figure}[h]
\centering
\includegraphics[width=0.5\textwidth]{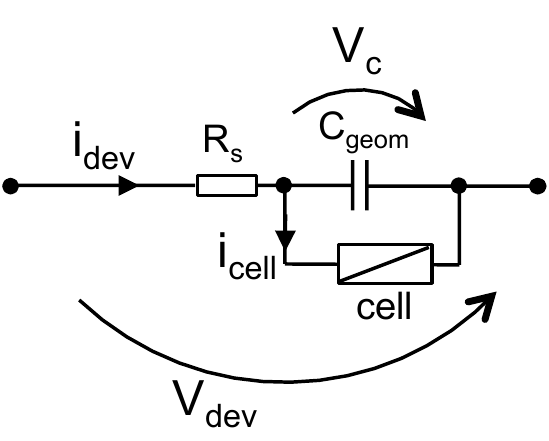}
\caption{Electrical circuit used to correct the cell current for the RC displacement current. This figure is a reprint of \autoref{img:rc_circuit_2}}
\label{img:rc_circuit_2_reprint}
\end{figure}

In the previous section the series resistance $R_S$ and the geometric capacitance $C$ were determined. These values are now used to calculate the RC-current and correct the device-current. 
\autoref{img:rc_circuit_2_reprint} shows the model for the RC-correction. Hereby the measured voltage of the device $V_{dev}(t)$ and its current $i_{dev}(t)$ are used to calculate the current that flows in the cell $i_{cell}(t)$.

The current that flows into the geometric capacitance $C_{geom}$ is
\begin{equation}\label{eq:calc_ic}
i_C(t) = C_{geom} \cdot \frac{dV_C}{dt}
\end{equation}

with $V_C(t)$ being
\begin{equation}\label{eq:calc_Vc}
V_C(t) = V_{dev}(t) - R_S \cdot i_{dev}(t)
\end{equation}

The current that flows into the capacitance $i_C$ is now subtracted from the device current to obtain the corrected current $i_{cell}$.

\marginpar{\newline Using the series resistance $R_S$, the geometrical capacitance $C_{geom}$, the measured current $i_{dev}(t)$ and the measured voltage $V_{dev}(t)$ the cell current $i_{cell}(t)$ can be calculated according to \autoref{eq:rc_compensation_methods}.}

\begin{equation}\label{eq:calc_icell}
i_{cell}(t) = i_{dev}(t) - i_C(t)
\end{equation}

Filling in \autoref{eq:calc_ic} and \autoref{eq:calc_Vc} in \autoref{eq:calc_icell} results in the final formula for the RC-current correction of a device with a geometrical capacitance $C_{geom}$, a series resistance $R_S$, a measured device voltage $V_{dev}(t)$ and a measured device current $i_{dev(t)}$.

\begin{equation}\label{eq:rc_compensation_methods}
i_{cell}(t) = i_{dev}(t) - C_{geom} \cdot \frac{dV_{dev}}{dt}(t) -
R_S \cdot C_{geom} \cdot \frac{di_{dev}}{dt}(t)
\end{equation}

\part{Solar Cell Physics}

\chapter{Solar Cell Physics}

\marginpar{\newline Solar cell physics with focus on charge transport is discussed in this chapter.}

In this section basic physical principles of solar cells are discussed. 
Not to overload this section the general introduction to semiconductors explaining holes, doping, bands and absorption as well as the general introduction to optics explaining the sun spectrum and Shockley-Queisser limit are omitted. The focus of this section lies in the charge transport.

\section{General Principle}

\marginpar{\newline The absorption of a photon leads to the excitation of an electron to a state with higher energy.}

In all solar cell types an incident photon that is absorbed, leads to an excitation of an electron from the valence band (or \acs{HOMO}-level) to the conduction band (or \acs{LUMO}-level). The electron can in this state fall back to the valence band with a probability depending on the possibilities to get rid of its energy via photon- or \gls{phonon}-emission.

%\marginpar{An exciton is a bound energy hole pair.}

The hole in the valence band and the electron in the conduction band attract each other due to coulomb interaction. This bound state of electron and hole can be described as quasi-particle that is called \gls{exciton}. Thermal energy, an electric field or a material interface is required to dissociate an \gls{exciton} into a free electron and a free hole.\\

\marginpar{\newline The main principle of a solar cell is, that an exited electron leaves the device at high potential doing work on an external load and recombining with the hole on low potential.}

\begin{figure}[h]
\centering
\includegraphics[width=\textwidth]{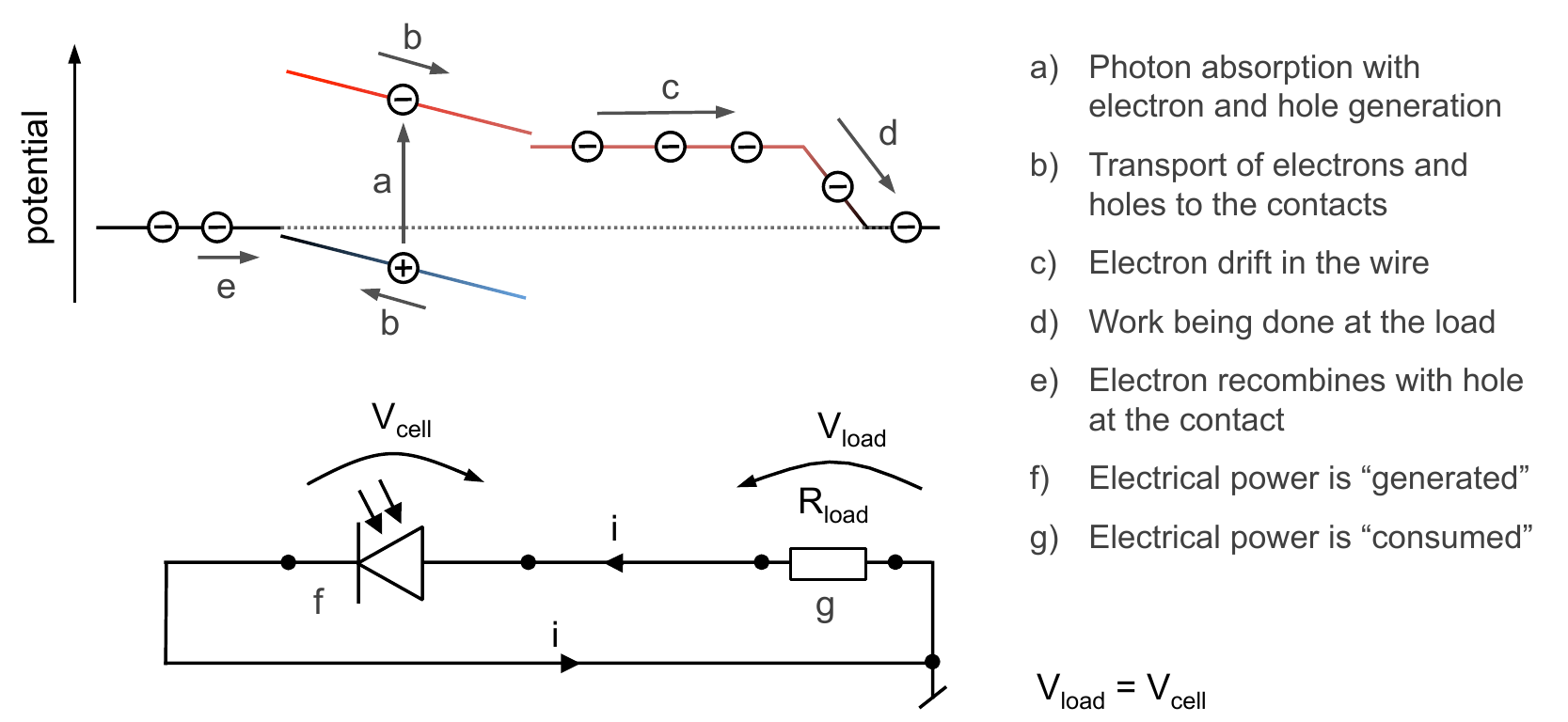}
\caption{General working principle of a solar cell illustrated with an extended band diagram (upper graph) and an electric circuit including a load (lower graph).
Please note that the current $i$ is defined positive, resulting in electrons moving in the opposite direction.}
\label{img:general-principle}
\end{figure}

\autoref{img:general-principle} shows the main operation principle of a solar cell. In step $a$ light is absorbed and the \gls{exciton} separated as described above. According to the driving forces explained in the next section the electrons and holes move to the contacts (step $b$). The wire connected to the solar cell is metallic and has therefore many free electrons transporting the charge (step $c$). In step $d$ the electron performs work on the load by going from the high potential to the low potential.
In this illustration the cell is at its \ac{MPP} somewhere between the short-circuit current and the open-circuit voltage. To reach this state the load $R_{load}$ must be chosen respectively.

In the electric circuit in \autoref{img:general-principle} the voltage of the solar cell $V_{cell}$ is in the opposite direction as the current $i$ (see $f$). From an electrical point of view in the solar cell power is generated whereas in the load resistor $R_{load}$ the power is dissipated (see $g$).

\section{Driving Forces and Band Diagrams}

\marginpar{\newline The gradient of the \gls{Fermi_level} is the driving force for electrons and holes and includes drift and diffusion.}

The driving force for electrons and holes is the gradient of the \gls{Fermi_level} \cite{wurfel_physics_2009}. The total particle currents $j_e$ and $j_h$ are described in \autoref{eq:df_je} and \ref{eq:df_jh}.

\begin{gather}\label{eq:df_je}
j_e = n_e \cdot \mu_e \cdot grad (E_{fe}) \\
j_h = -n_h \cdot \mu_h \cdot grad (E_{fh}) \label{eq:df_jh}
\end{gather}

where $n_e$ and $n_h$ are the electron and hole densities, $\mu_e$ and $\mu_h$ are the electron and hole mobilities and $E_{fe}$ and $E_{fh}$ are the \glspl{Fermi_level} for electrons and holes, respectively.\\

The \gls{Fermi_level} is equal to the electro-chemical potential\footnote{The electro-chemical potential is in contrast to its name not a potential but an energy.} $\eta$ that consists of the electrical potential $\varphi$ and the chemical potential $\gamma$ as shown in \autoref{eq:df_fermi_e} and \ref{eq:df_fermi_h}.

\begin{gather}\label{eq:df_fermi_e}
E_{fe} = \eta_e = \gamma_e + q \cdot \varphi \\
E_{fh} = -\eta_h = -\gamma_h - q \cdot \varphi
\label{eq:df_fermi_h}
\end{gather}

The chemical potential of electrons and holes ($\gamma_e$ and $\gamma_h$) is dependent on their charge carrier density ($n_e$ and $n_h$) according to \autoref{eq:df_chem_pot_e} and \ref{eq:df_chem_pot_h},

\marginpar{\newline The chemical potential depends logarithmically on the charge carrier density.}

\begin{gather}\label{eq:df_chem_pot_e}
\gamma_e = -\chi_e + k \cdot T \cdot ln \Big( \frac{n_e}{N_C} \Big) \\
\gamma_h = -\chi_h + k \cdot T \cdot ln \Big( \frac{n_h}{N_V} \Big)
\label{eq:df_chem_pot_h}
\end{gather}

where $\chi_e$ is the electron affinity, $\chi_h$ the ionization potential, $N_C$ and $N_V$ is the effective density of states, $k$ is the Boltzmann constant and $T$ is the temperature.

Combining \autoref{eq:df_je}, \ref{eq:df_fermi_e} and \ref{eq:df_chem_pot_e} results in \autoref{eq:df_n1} for electrons. Replacing the gradient with a one dimensional derivative finally results in \autoref{eq:df_n3} - the well-known drift-diffusion equation as described in chapter \nameref{ch:simulation} in \autoref{eq:ddn}.

\begin{gather}\label{eq:df_n1}
j_e = n_e \cdot \mu_e \cdot grad \big( -\chi_e + k \cdot T \cdot ln \Big( \frac{n_e}{N_C} \Big) + q \cdot \varphi \big) \\
j_e = n_e \cdot \mu_e \cdot k \cdot T \cdot \frac{\partial{}}{\partial{x}}
\Big( ln \Big( \frac{n_e}{N_C} \Big) \Big) + 
n_e \cdot \mu_e \cdot q \cdot \frac{\partial{\varphi}}{\partial{x}}\\
j_e = \mu_e \cdot k \cdot T \cdot \frac{\partial{n_e}}{\partial{x}} + 
n_e \cdot \mu_e \cdot q \cdot \frac{\partial{\varphi}}{\partial{x}}
\label{eq:df_n3}
\end{gather}

\marginpar{Inserting the equations of the \gls{Fermi_level} into the equation defining the current, results exactly in the drift-diffusion equation used in the numerical simulation.}

The result for the equation for holes is shown in \autoref{eq:df_h3}.
\begin{equation}\label{eq:df_h3}
j_h = \mu_h \cdot k \cdot T \cdot \frac{\partial{n_h}}{\partial{x}} - 
n_h \cdot \mu_h \cdot q \cdot \frac{\partial{\varphi}}{\partial{x}}
\end{equation}

To make the long story short: The gradient of the \gls{Fermi_level} is the driving force for the charge carriers combining forces of diffusion (chemical potential) and forces of drift (electrical potential).
In solar cell physics the \glspl{Fermi_level} and band structures are often illustrated to understand the device operating mechanisms. Band structures are explained in the next section.

\section{Band Diagrams and Basic Solar Cell Operation}

\begin{figure}[h]
\centering
\includegraphics[width=\textwidth]{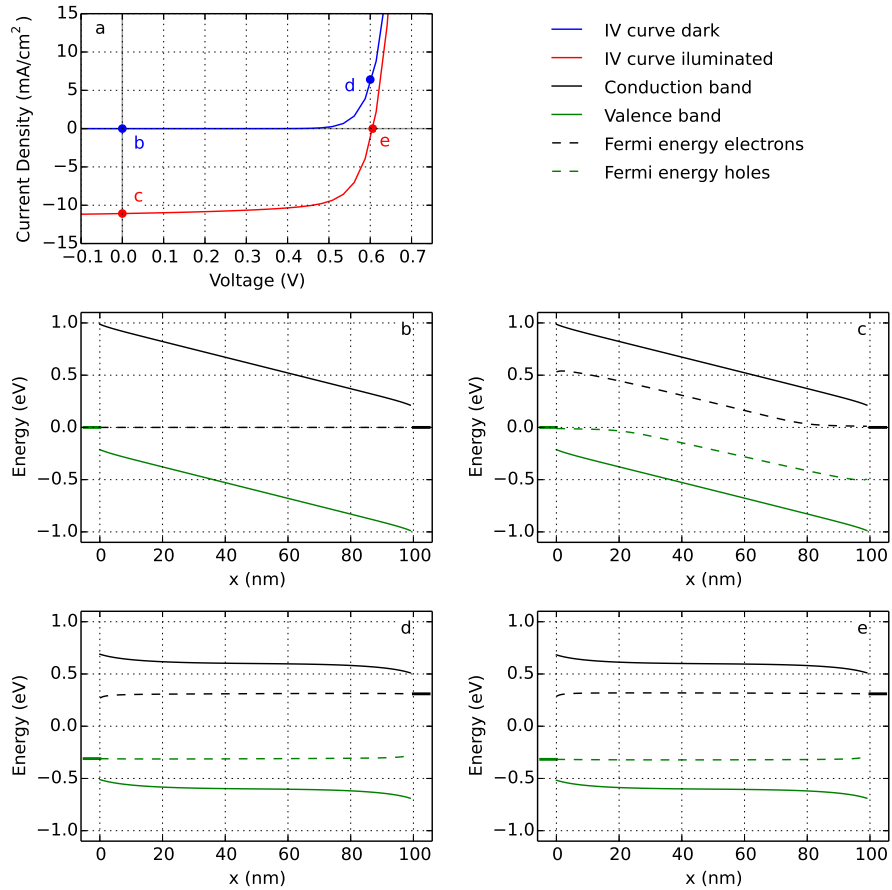}
\caption{a) Simulated \ac{IV curve} in the dark and illuminated. b) Band diagram at zero volt in the dark. c) Band diagram at zero volt illuminated. d) Band diagram with forward voltage in the dark. e) Band diagram at open-circuit voltage.}
\label{img:bands_principles}
\end{figure}

To understand band structures we look at a simple solar cell with good charge transport, low recombination and a built-in voltage that drives the charge carriers to the electrodes. The device is not doped and has no traps. The \acsp{IV curve} of such an idealised device is shown in \autoref{img:bands_principles}a.

\marginpar{\newline A band diagram is the illustration of the effective bands with the two quasi \glspl{Fermi_level} and is often used to explain solar cell physics.}
In a band diagram the energy is plotted versus the position in the layer. In our case the device is illuminated from the left. On the left at $x=0$ is the anode where the holes are extracted. On the right at $x=100\,nm$ is the cathode where the electrons are extracted.\\
In a general view one can say electrons tend to go up and holes go down.\\

\vspace{5mm}

At \textbf{short-circuit in the dark} in \autoref{img:bands_principles}b no current flows. The gradient of the \gls{Fermi_level} is zero. The Fermi-level for electrons and holes coincides. The bands are inclined meaning a constant electric field throughout the device.

\marginpar{\newline The illumination causes a \gls{Fermi_level} splitting.}

Under \textbf{illumination at short-circuit} in \autoref{img:bands_principles}c the absorbed photons lead to a Fermi-level splitting. There is no voltage drop on the cell. The two contacts (indicated with thick lines) have the same potential. As there is a gradient in both Fermi-levels an electron and a hole current flows.

At an \textbf{applied forward voltage in the dark} in \autoref{img:bands_principles}d the internal field is compensated and the bands become flat. The difference between the Fermi-level of holes on the left and the Fermi-level of electrons on the right is the applied voltage at the device.
A forward current flows. As the charge carrier density is very high, a very small gradient in the Fermi-levels is sufficient to drive the charges.

At \textbf{open-circuit} under illumination in \autoref{img:bands_principles}e no current flows - all charges recombine. The charge carrier density is high. This can be seen in the band diagram because the Fermi-levels are much closer to the bands as in case c.

\vspace{5mm}

Please note: The total current consisting of electron, hole and displacement current is always constant in x throughout the hole electric circuit (Kirchhoff's law).

\section{Majority versus Minority Carrier Devices}

\marginpar{\newline This section categorizes solar cells into majority- and minority-carrier devices.}

Solar cells can be categorised in many aspects. In this section the distinction is made between the main driving forces into \textit{minority carrier devices} governed by diffusion and \textit{majority carrier devices} governed by drift. This distinction may be uncommon to many solar cell specialists as they only deal with one or the other type. Nevertheless, this distinction is helpful to understand the physics of \acfp{PSC}.

\marginpar{\newline A majority carrier device is not doped and both charge carrier types have a similar density. In a minority carrier device one charge carrier type is dominant (ex: due to doping) and the other type is in the minority.}

\begin{figure}[h]
\centering
\includegraphics[width=\textwidth]{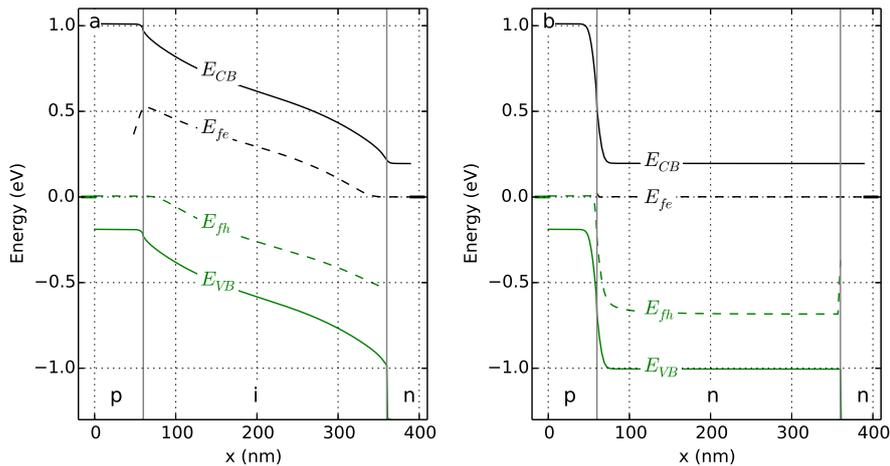}
\caption{a) Majority carrier device in pin-structure. Steep bands indicate a high electric field. b) Minority carrier device in pn-structure. Within the device the bands are flat.}
\label{img:minority_vs_majority}
\end{figure}

\autoref{img:minority_vs_majority} shows a comparison of the two device types. The majority carrier device is shown as \textit{pin}-structure\footnote{The pin stands for p-type doped, intrinsic (undoped) and n-type doped.}, the minority carrier device with the same structure but the intrinsic region $i$ replaced with n-type doping. Both band diagrams are shown at zero volt under illumination. Charge carriers are created homogeneously throughout the device.

\subsection{Minority Carrier Devices}
Minority carrier devices are doped\footnote{As described in this thesis also traps can lead to a high charge carrier imbalance. Furthermore also a high difference in electron and hole mobility can lead to a high imbalance of charge carriers resulting in a minority carrier device working mechanism.} what leads to an imbalance of charge carriers in a large part of the device. The transport is limited by the diffusion of the minority carriers.\\

\marginpar{\newline The classical pn-junction is a good example of a minority carrier device. The potential mainly drops at the junction, the rest of the device is field-free.}

An example of the band structure in a minority carrier device is shown in \autoref{img:minority_vs_majority}b. At the \gls{pn_junction} the electric field is very high, whereas in the n-type region it is screened (close to zero). In the n-type region the electron \gls{Fermi_level} ($E_{fe}$) is much closer to the conduction band ($E_{CB}$) than the hole Fermi-level ($E_{fh}$) to the valence band ($E_{VB}$). This shows that fewer holes are present than electrons.
In this example the minority carriers are the holes that need to diffuse from the n-type region to the p-contact. Transport by drift is negligible. The electron transport is not limiting due to the high density.\\

Most commonly known solar cell technologies are minority carrier devices using a \gls{pn_junction}: crystalline silicon, \ac{CdTe}, \ac{CIGS}.

\subsection{Majority Carrier Devices}

\marginpar{\newline Majority carrier devices are not doped and have an therefore an electric field that drives charges.}

Majority carrier devices are not doped and therefore the densities of electrons and holes are in a similar order of magnitude. In the majority carrier device shown in \autoref{img:minority_vs_majority}a an electric field is created in the intrinsic region due to the charges of the n-type and p-type region. Alternatively such a field can be created using metals with different \glspl{workfunction} as contacts.

\begin{quote}
\textit{Please note that also minority carrier devices have a built-in potential. But this potential drops at the \gls{pn_junction}, the rest of the device is field-free.}
\end{quote}

In majority carrier devices the diffusion length is often too short such that an electric field is required to transport charge carriers to the electrodes.
%A majority carrier device structure requires selective contacts what is provided by p- and n-doped regions or metals with different \glspl{workfunction}.
Common majority carrier devices are amorphous silicon and \acfp{OSC}.

\subsection{Lifetime}
In \textit{minority carrier devices} the charge carrier lifetime $\tau$ is defined as

\marginpar{The lifetime is inversely proportional to the recombination. In minority carrier devices the lifetime defines how long a minority carrier "lives" in average before it recombines.}

\begin{equation}
\tau_e = \frac{n_e}{R}
\end{equation}

where $R$ is the recombination and $n$ is the charge carrier density of the minority. For radiative recombination $R=\beta \cdot n_e \cdot n_h$ the lifetime is independent of the minority carrier density. Assuming a p-doped device ($n_h = N_A$) the lifetime results in
\begin{equation}
\tau_e = \frac{n_e}{\beta \cdot n_e \cdot n_h} = \frac{1}{\beta \cdot N_A}
\end{equation}
where $\beta$ is the recombination prefactor, $n_h$ is the hole density and $N_A$ is the doping density. Assuming $n_h\,>>\,n_e$ (which is the case in a doped device) the hole density is unaffected by the recombination and can be considered as constant. Therefore the charge carrier lifetime can be considered as a constant material parameter.

\marginpar{\newline The diffusion length defines how far a charge carrier can diffuse in average.}

Knowing the diffusion constant $D$ the diffusion length $L_D$ can be calculated according to
\begin{gather}
L_D = \sqrt{D_e \cdot \tau_e} = \sqrt{\mu_e \cdot k \cdot T \cdot \tau_e} = \sqrt{\frac{\mu_e \cdot k \cdot T}{\beta \cdot N_A}}
\label{eq:diff_length}
\end{gather}
Also the minority carrier diffusion length can be regarded as constant material parameter if the material is doped. For an efficient device the diffusion length needs to be significantly larger than the device thickness.

\vspace{5mm}

\marginpar{\newline The concept of lifetime is not physically meaningful for majority carrier devices. It is not constant in time and space due to similar charge carrier densities.}

In \textit{majority carrier devices} the concept of charge carrier lifetime cannot be applied directly. It cannot be assumed that a majority carrier density is constant and the lifetime of the minority carriers is limiting. Both the electron lifetime $\tau_e$ and the lifetime of holes $\tau_h$ depend on each other as shown in \autoref{eq:lifetime_majority}.

\begin{align}
\tau_e(n_h)& = \frac{1}{\beta \cdot n_h}&
\tau_h(n_e)& = \frac{1}{\beta \cdot n_e}
\label{eq:lifetime_majority}
\end{align}

Both the electron density $n_e$ and the hole density $n_h$ can vary over orders of magnitude depending on the position in the device and depending on time. Charge carrier lifetime can therefore not be regarded as constant material parameter in a majority carrier device like an \ac{OSC}.
Consequently the product of diffusion constant $D$ and lifetime $\tau$ as shown in \autoref{eq:diff_length} is not physically meaningful. A charge carrier travelling through the device will have different lifetimes depending on its position. Furthermore, in these types of devices charge carriers are mainly transported by drift.\\
Although the physical meaning is questionable the mobility-lifetime-product is sometimes used in publications about majority carrier devices like \acp{OSC} \cite{baumann_new_2012, dennler_charge_2006}.

\subsection{Traps and Doping}

The doping of a semiconductor can be intentional like in the case of a silicon solar cell or unintentional, as sometimes the case in organic photovoltaics. In the second case it is detrimental to device performance since the electric field is screened and charges cannot be transported to the electrodes as shown by Kirchartz et al.\cite{dibb_influence_2013}.\\

\marginpar{\newline The doping of a semiconductor results in additional free charge carriers and somehow fixed charges (ionic cores) of the other charge polarity.}

\begin{figure}[h]
\centering
\includegraphics[width=\textwidth]{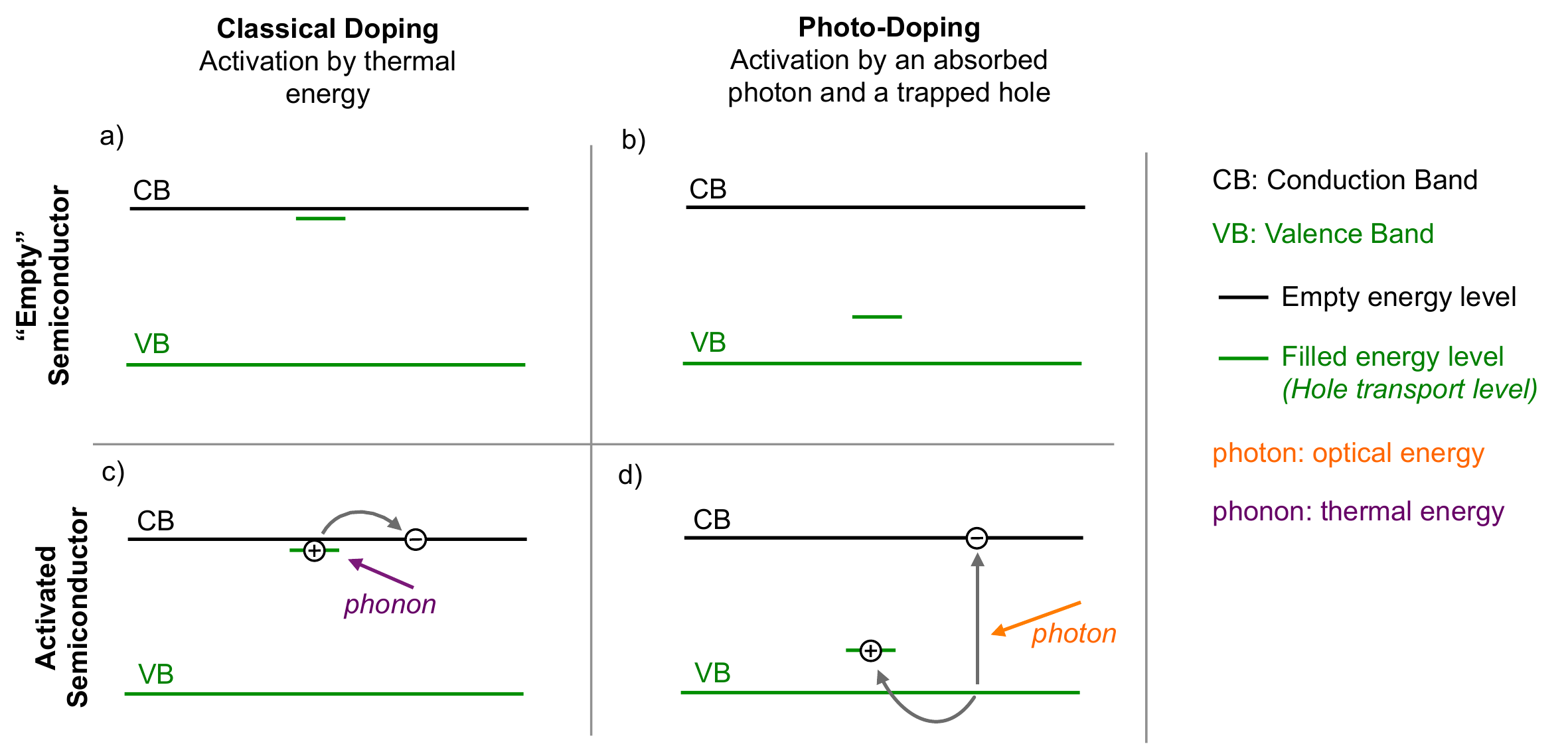}
\caption{
a) An "empty" semiconductor containing a filled energy level close to the conduction band.
b) An "empty" semiconductor containing a filled energy level somewhere in the band-gap (this is also called a hole-trap).
c) By thermal energy the electron is moved from its energy level to the conduction band (ionized dopant).
d) An electron and a hole are created by photon-absorption. After a while the hole gets trapped (photo-doping).}
\label{img:traps_vs_doping}
\end{figure}

Doping usually refers to the creation of free charge carriers leaving an ionized core of opposite charge polarity activated at room temperature. \autoref{img:traps_vs_doping}a and \ref{img:traps_vs_doping}c illustrate this process. An atom or molecule is placed in a semiconductor such that the atom's occupied energy level is close to the unoccupied conduction band of the semiconductor. Thermal energy at room temperature is sufficient to ionize this atom or molecule. As shown in \autoref{img:traps_vs_doping}c a free electron leaves behind an immobile positive charge (hole).\\

Charge trapping can however lead to the same effect what is sometimes referred to as photo-doping \cite{leijtens_electronic_2014}.
\autoref{img:traps_vs_doping}b shows a semiconductor with an additional energy level somewhere in the band-gap acting as hole-trap. Without any activation it is neutral as in the case of \autoref{img:traps_vs_doping}a.

If a photon is absorbed a free electron and a free hole are created. If the hole falls into the hole-trap it is immobile. The two situations (classical doping and photo-doping) lead to the same result.

\marginpar{If traps get filled they can have a very similar effect like classical doping - Often this is referred to as photo-doping.}

\vspace{5mm}

In numerical drift-diffusion modelling doping is added as fixed charge in the Poisson equation as shown in \autoref{eq:poisson} in Section \ref{ch:simulation}. The condition of charge balance automatically leads to additional free charge carriers of the same amount.

\section{Recombination and Open-Circuit Voltage}

\marginpar{\newline In radiative recombination a photon is emitted. It is usually the weakest form of recombination. It can be directly determined by measuring the \ac{EL}-signal at forward bias.}

\begin{figure}[h]
\centering
\includegraphics[width=\textwidth]{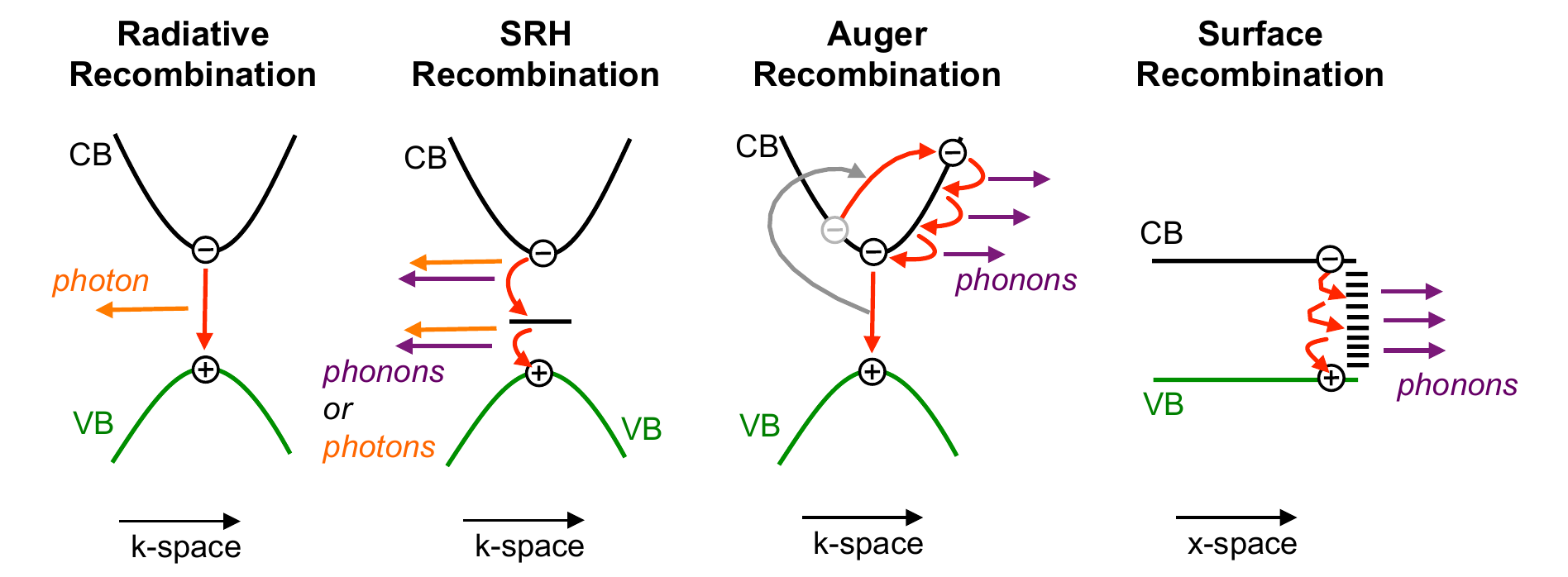}
\caption{Schematic illustration of recombination types in semiconductors.}
\label{img:recombination_types}
\end{figure}

Recombination is the annihilation of an electron and a hole. The potential energy of the electron in the conduction band thereby needs to be transferred elsewhere - in \gls{phonon} and/or photon emission. 
\autoref{img:recombination_types} shows the four recombination types that can be present in semiconductors.

\vspace{5mm}

\textbf{Radiative recombination} is physically inherent in absorbing materials, meaning it is the recombination type that cannot be avoided. In the dark at thermal equilibrium all rates of generation are equal to the rates of recombination - this is the principle of \textit{detailed balance} \cite{wurfel_physics_2009}.
The probability of an electron and a hole finding each other increases with both densities. Therefore the recombination rate is proportional to both charge carrier densities and consequently depends on the \gls{Fermi_level} splitting as shown in \autoref{eq:radiative_rec}.

\begin{equation}\label{eq:radiative_rec}
R_{rad} = \beta \cdot n_e \cdot n_h = 
\beta \cdot e^{\frac{E_{fe}-E_{fh}}{k \cdot T}}
\end{equation}

As indicated in \autoref{img:recombination_types} a photon is emitted in radiative recombination. The radiative recombination can therefore directly be determined by measuring the \ac{EL}-signal.\\
Generally one can say that a good solar cell with a high open-circuit voltage shows also a high \ac{EL}-signal. Radiative recombination is generally weaker than other forms of recombination. A high \ac{EL}-signal therefore indicates that all other recombination types are not dominant leading to a high charge carrier density and a high open-circuit voltage.

\vspace{5mm}

\marginpar{\newline Trap assisted recombination is a two-stage process requiring an energy level within the band-gap.}

\textbf{\acf{SRH} recombination} (also known as trap-assisted recombination) requires an intermediate energy level within the band-gap. The energy is either emitted by a \gls{phonon} or a photon as shown in \autoref{img:recombination_types}. Such mid-gap states occur by \gls{dangling_bonds} of a crystal or by impurities. In a perfect material \ac{SRH} recombination is absent.\\
Its mathematical description is more complex than for radiative recombination. For a cell in steady-state the \ac{SRH} recombination is described as

\begin{equation}\label{eq:srh_rec}
R_{SRH} = n_{imp} \cdot \frac{n_e \cdot n_h - n_i^2}
{\frac{n_e + N_C \cdot e^{-\frac{E_{CB} - E_{imp}}{k \cdot T}}}{ c_h } + 
\frac{n_h + N_V \cdot e^{-\frac{E_{imp} - E_{VC}}{k \cdot T}}}{ c_e }}
\end{equation}

where $n_{imp}$ is the density of states of the impurities, $n_i$ is the intrinsic density, $E_{imp}$ is the energy level of the impurity, $c_h$ and $c_e$ are the capture rates of electrons and holes and $N_C$ and $N_V$ are the effective density of states of the conduction band and valence bands, respectively.

\vspace{5mm}

\marginpar{\newline Auger recombination is only relevant for high charge carrier densities in minority carrier devices.}

\textbf{Auger recombination} can be regarded as the reverse of impact ionization. An electron transfers its energy to another electron that is moved up to a higher state in the conduction band. The second electron afterwards thermalises down to the conduction band edge by emitting \glspl{phonon} as shown in \autoref{img:recombination_types}.
\autoref{eq:auger_rec} describes the Auger recombination for electrons and holes

\begin{align}
R_{Aug,e}& = C_e \cdot n_e^2 \cdot n_h&
R_{Aug,h}& = C_h \cdot n_e \cdot n_h^2
\label{eq:auger_rec}
\end{align}

where $C_e$ and $C_h$ are Auger recombination constants. As Auger recombination scales with the cube of the charge carrier density, it is large for devices with high doping. It is practically irrelevant for majority carrier devices. In silicon solar cells it largely determines the efficiency limits of the record devices \cite{wurfel_physics_2009}.

\vspace{5mm}

\marginpar{\newline Surfaces contain many defects that create states within the band-gap, leading to higher recombination at the surface.}

\textbf{Surface recombination} happens via one or more states that are present at the surface of an interface to another material. In crystals surfaces always contain \gls{dangling_bonds} or defects that create states within the band-gap. A charge carrier reaching a surface hops from state to state loosing its energy by \gls{phonon} emission as shown in \autoref{img:recombination_types}. Surface recombination is described as

\begin{align}
R_{Sur,e}& = \nu_e \cdot n_e&
R_{Sur,h}& = \nu_h \cdot n_h
\label{eq:surface_rec}
\end{align}

where $\nu$ is the surface recombination velocity.

\begin{quote}
\textit{Please note: The expression surface recombination is used to describe the detrimental recombination of charge carriers at the opposite electrode - not the charge extraction. Holes recombine at the electron contact, electrons at the hole contact.}
\end{quote}

Metals are considered to have an infinite surface recombination velocity $\nu$, meaning that all charge carriers of the wrong type reaching the surface are 'immediately' lost.\\
The term surface passivation means applying measures to lower the surface recombination velocity. This can be done by coating additional buffer layers that are blocking one charge carrier type or by additional local doping at the contact\footnote{In silicon solar cells this concept is called back surface field BSF. The p-type wafer is doped with p+ at the contact to reduce the electron concentration and passivate the surface.}.

\vspace{5mm}

\marginpar{\newline Surface passivation is required for minority-carrier devices to reach charge selectivity.}

Minority carrier devices require passivated\footnote{A passivated surface has only few mid-gap levels leading to a low surface recombination.} surfaces as charges would run into the opposite electrode. We name this \textit{charge selectivity}. In majority carrier devices charge selectivity is ensured by the electric field. Surface passivation is therefore less detrimental in these devices.

\subsection{Recombination Order}

\marginpar{\newline The recombination order is often measured to analyse the dominant recombination type.}

The recombination order provides information about the dominant recombination type. The recombination order is a simplified view on the cell as the charge carrier concentration is zero-dimensional and similar electron and hole densities are assumed. The recombination order $k$ is defined as

\begin{equation}
R = n^k
\end{equation}

where $R$ is the recombination and $n$ is the charge carrier density.\\
Assuming the simplification $n = n_e = n_h$ the radiative recombination can be simplified to:

\marginpar{\newline The recombination order for radiative recombination or bimolecular recombination is $2$.}

\begin{equation}\label{eq:rec_order_rad}
R_{rad} = \beta \cdot n_e \cdot n_h = 
\beta \cdot n^2
\end{equation}

The recombination order for radiative recombination is therefore 2. For \ac{SRH} recombination the recombination order results in 1 as shown in \autoref{eq:rec_order_srh}. For the Auger recombination the recombination order is 3 as shown in \autoref{eq:rec_order_auger}.

\marginpar{\newline The recombination order for \ac{SRH} recombination is $1$.}

\begin{equation}\label{eq:rec_order_srh}
R_{SRH} = n_{imp} \cdot \frac{n_e \cdot n_h}
{C_1 \cdot n_e + C_2 \cdot n_h + C_3} =
C_4 \cdot n
\end{equation}

\marginpar{\newline The recombination order for Auger recombination is $3$.}

\begin{equation}\label{eq:rec_order_auger}
R_{Aug,e} = C_e \cdot n_e^2 \cdot n_h =
C_e \cdot n^3
\end{equation}

\subsection{Ideality Factor}

\marginpar{\newline The ideality factor is directly related with the recombination order.}

The ideality factor\footnote{There are two kinds of ideality factors that can differ from each other. The dark ideality factor is measured in the IV-curve forward direction in the dark. The light ideality factor is determined at the slope of the open-circuit voltage versus the light intensity. In this thesis only the second type is treated.} has the same purpose as the recombination order, that is to distinguish recombination types. In this section the relation between ideality factor and recombination order is shown.\\

The ideality factor $n_{id}$ is used in the equation to describe an \ac{IV curve} analytically

\begin{equation}\label{eq:analytical_iv}
j(V) = j_S \cdot \big( exp \Big( \frac{ V \cdot q}{k_B \cdot T \cdot n_{id}} \Big) - 1\big) - j_{ill}
\end{equation}

\marginpar{\newline The \ac{IV curve} of a solar cell can be described analytically using the ideality factor.}

where $j_S$ is the reverse saturation current and $j_{ill}$ is the current due to illumination. If the ideality factor $n_{id}$ is one, the cell is \textit{ideal} and at its optimum. The 'ideal' ideality factor is always larger than one\footnote{This however is the theory. Further below we will see that this is not always the case.}.

\vspace{5mm}

%\begin{quotation}
%\textit{"Experimentally, nongeminate recombination is often studied at the open %circuit where, due to the lack of an external current flow, effects of series %resistances and charge carrier transport are minimized."\\
%Thomas Kirchartz, \cite{kirchartz_meaning_2012}}
%\end{quotation}

\marginpar{\newline The slope of $V_{OC}$ is measured to determine the ideality factor.}

The ideality factor can be determined by analysing the slope of the open circuit voltage versus the logarithmic light intensity.
The open-circuit voltage is the difference of the \gls{Fermi_level} at the contacts. Using \autoref{eq:df_fermi_e} and \ref{eq:df_fermi_h} this results in \autoref{eq:voc} for the open-circuit voltage.

\begin{equation}\label{eq:voc}
V_{OC} = \frac{E_{BG}}{q} + \frac{k_B \cdot T}{q}
ln \Big( \frac{n_e \cdot n_h}{N_C \cdot N_V} \Big)
\end{equation}

Applying the same simplifications as used in the previous section $n = n_e = n_h$ and substituting all constants with $C$ results in \autoref{eq:voc2}.

\begin{equation}\label{eq:voc2}
V_{OC} = C + \frac{k_B \cdot T}{q}
ln \big( n^2 \big)
\end{equation}

At open-circuit the recombination $R$ of charge carriers equals the generation $G$ since no current can flow.

\begin{equation}\label{eq:gen_rec}
G = R = n^k
\end{equation}

Putting \autoref{eq:gen_rec} into \autoref{eq:voc2} results in

\marginpar{\newline \newline $V_{OC}$ depends logarithmically on the light intenstiy.}

\begin{equation}
V_{OC} = C + \frac{k_B \cdot T}{q}
ln \Big( G^{\frac{2}{k}} \Big) = 
C + \frac{k_B \cdot T}{q} \cdot \frac{2}{k} \cdot ln(G)
\end{equation}

This result shows that the open-circuit voltage depends logarithmically on the charge carrier generation and therefore the light intensity.\\

Solving the equation that describes the \ac{IV curve} (\autoref{eq:analytical_iv}) for $j(V)=0$ results in \autoref{eq:nid_voc}.

\begin{equation}\label{eq:nid_voc}
V_{OC} = n_{id} \cdot \frac{k_B \cdot T}{q} \cdot
ln \big(1 + \frac{j_{ill}}{j_S}\big)
\end{equation}

As the short circuit current is much larger than the saturation current ($j_{ill} >> j_S$) the expression $1 + j_{ill}/j_S$ can be regarded as proportional to the generation $G$. This leads to the relation 

\begin{equation}\label{eq:nid}
n_{id} = \frac{2}{k} = 
\frac{dV_{OC}}{d (ln(G))} \cdot 
\frac{q}{k_B \cdot T}
\end{equation}
\marginpar{Relation between ideality factor and recombination order.}

With pure radiative recombination the recombination order is $2$ and the ideality factor is $1$. If \ac{SRH} recombination is dominant the recombination order is $1$ and the ideality factor is $2$.

\vspace{5mm}

\marginpar{\newline If radiative recombination is dominant the $V_{OC}$ versus light intensity is exactly logarithmic in the simulation.}

\begin{figure}[h]
\centering
\includegraphics[width=\textwidth]{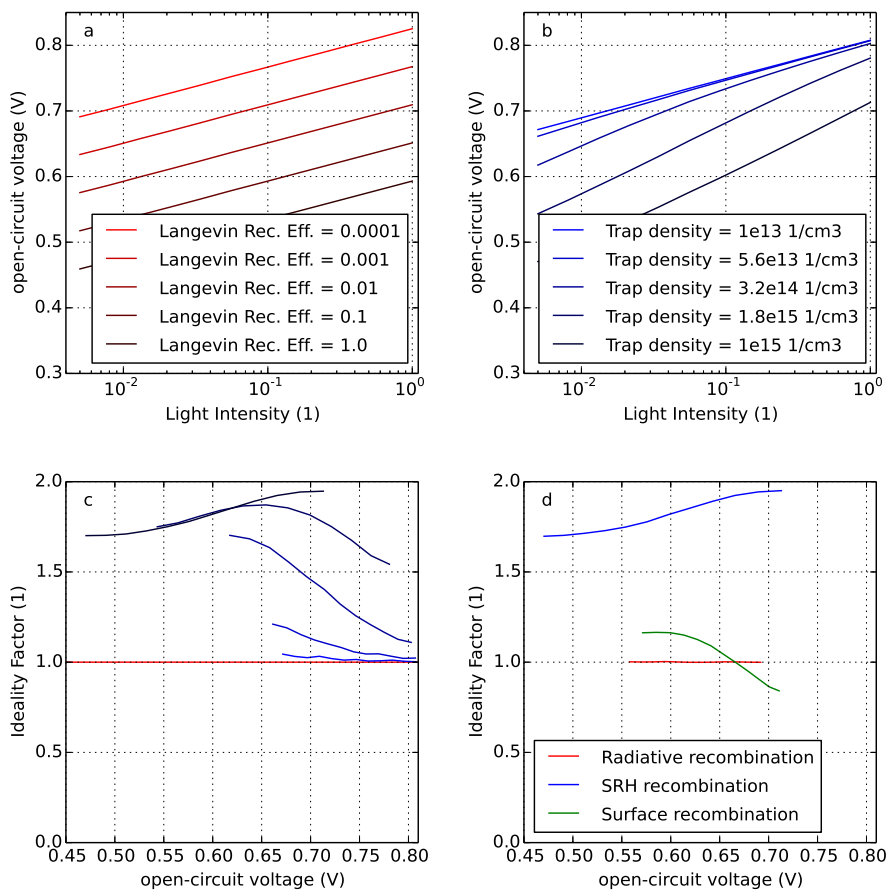}
\caption{a) Simulated open-circuit voltage with varied radiative recombination efficiencies.
b) Simulated open-circuit voltage with \ac{SRH} recombination and varied trap-density.
c) Simulated ideality factors for the radiative recombinations and the \ac{SRH} recombinations of a and b.
d) Simulated ideality factors for different recombination types.}
\label{img:voc_and_nid}
\end{figure}

To understand the applicability and limitations of the ideality factor numerical simulations are performed.
\autoref{img:voc_and_nid}a shows simulated open-circuit voltage versus light intensity with varied radiative recombination efficiency\footnote{In this model Langevin recombination is used, that is proportional to radiative recombination efficiency.}. A device without doping, traps and with well passivated contacts is simulated such that radiative recombination is the only relevant recombination.
The open-circuit voltage is therefore exactly proportional to the logarithm of the light intensity. The ideality factor calculated from \autoref{eq:nid} is exactly $1$ as shown in \autoref{img:voc_and_nid}c.\\

\ac{SRH} recombination in comparison has a different slope (open-circuit voltage versus logarithmic light intensity) with varying trap density (\autoref{img:voc_and_nid}b). The ideality factor changes between one and two depending on the light intensity as shown in \autoref{img:voc_and_nid}c.\\

\marginpar{\newline With \ac{SRH}-recombination the simulated ideality factor varied between $1$ and $2$ depending on the light intensity. }

\autoref{img:voc_and_nid}d compares the ideality factors of radiative recombination, \ac{SRH} recombination and surface recombination. In this example the ideality factor of the simulation with surface recombination gets even below one. 

%This should not be possible according to the theory. In other examples the ideality factor can reach values above $3$ \cite{kirchartz_meaning_2012}.\\

The concept of recombination order and ideality factor is strongly simplified. Due to spatial variation of charge carrier density, variation in energy levels or variation in charge carrier mobilities the ideality factor can reach values below one or above three \cite{kirchartz_meaning_2012}.
Conclusions from ideality factors should therefore be made carefully.

\part{Results And Discussion}
\chapter{Results}

In this thesis \acf{MALI} \acf{PSC} with mesoporous $TiO_2$ scaffold and Spiro-OMeTAD are investigated. More details about the device can be found in chapter \nameref{ch:fabrication}.

\section{IV Curve Hysteresis}

\marginpar{\newline Perovskite solar cells very often show a hysteresis when the \ac{IV curve} is measured upward or downwards.}

The shape of an \ac{IV curve} of a perovskite solar cell can vary depending on voltage ramp speed and direction \cite{snaith_anomalous_2014, unger_hysteresis_2014, tress_understanding_2015}. Though hysteresis effects have been observed and discussed at length, their origin is still subject of intense debate. 
Unger et al. showed that hysteresis effects are related to slow processes that occur on the same time scale as the voltage ramps used \cite{unger_hysteresis_2014} . We shine light on this puzzle by systematic IV curve acquisitions with voltage pulses and with pre-bias voltage.\\

\vspace{5mm}

\marginpar{\newline The investigated cell has an efficiency of $12\%$.}

The solar cells we investigate have generally a small hysteresis. For this study about charge transport dynamics and slow effects we chose a batch with a stronger hysteresis. The cell investigated has a \ac{PCE} of about $12\%$.\\

\marginpar{\newline The shape of the \ac{IV curve} depends on the slope speed and direction. Going from high to low voltage results in a higher fill factor.}

\begin{figure}[h]
\centering
\includegraphics[width=\textwidth]{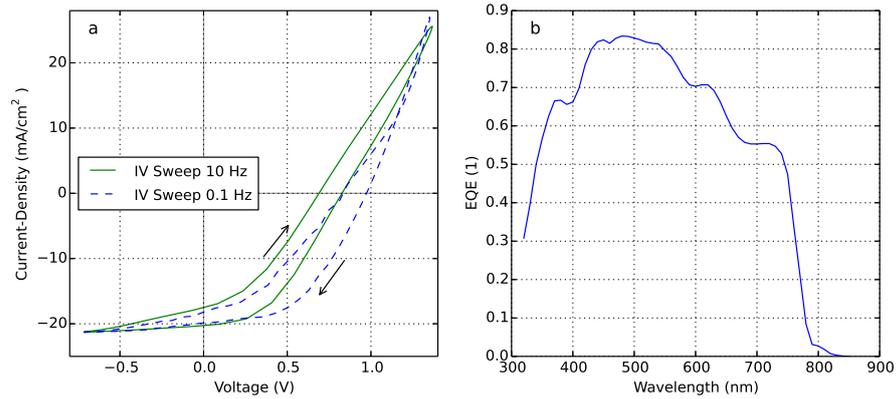}
\caption{a) Current-voltage characteristics measured with a sinusoidal voltage with frequency $10\,Hz$ and $0.1\,Hz$ under illumination. b) \ac{EQE} measurement of the \ac{PSC} investigated.}
\label{img:iv_and_eqe}
\end{figure}

\autoref{img:iv_and_eqe}a shows the \ac{IV curve} hysteresis measured with a slow sinusoidal voltage with $10\,Hz$ and $0.1\,Hz$. This results in a measurement time of 100 milliseconds and 10 seconds, respectively. \emph{Please note that this hysteresis measurement was not performed under AM1.5 but with an illumination from a high-power white LED.} The cycle going from high voltage to low voltages results in a higher open-circuit voltage and higher fill factor. The physical origin of this effect will be addressed further below in this thesis.\\
\autoref{img:iv_and_eqe}b shows the \ac{EQE} of the investigated \ac{PSC}.

\section{Transient current response from microseconds to minutes}
\label{ch:TPC_and_VP}

\marginpar{\newline Transient measurements provide important information about the dynamics of physical processes.}

To gain knowledge about the origins of the hysteresis it is promising to perform transient experiments as they provide valuable information about charge transport and related processes such as recombination and trapping. Thus, on \ac{PSC} \acf{TEL} \cite{tress_understanding_2015},
\acf{TPV} \cite{baumann_persistent_2014, stranks_recombination_2014, tress_understanding_2015}, 
\acf{photo-CELIV} \cite{ahn_highly_2015, im_growth_2014}, 
\acf{TPC} \cite{tress_understanding_2015} and 
transient photo-conductivity \cite{gottesman_extremely_2014} 
have been measured. Except for the \ac{TPV} measurements of Baumann et al. \cite{baumann_persistent_2014} so far all transient experiments were reported with only two or three orders of magnitude in time – either in the regime of milliseconds or seconds.

\vspace{5mm}

\marginpar{\newline The \acf{TPC} of the perovskite solar cell increases over $8$ orders of magnitude in time and reaches steady-state only after $5$ minutes.}

\begin{figure}[h]
\centering
\includegraphics[width=\textwidth]{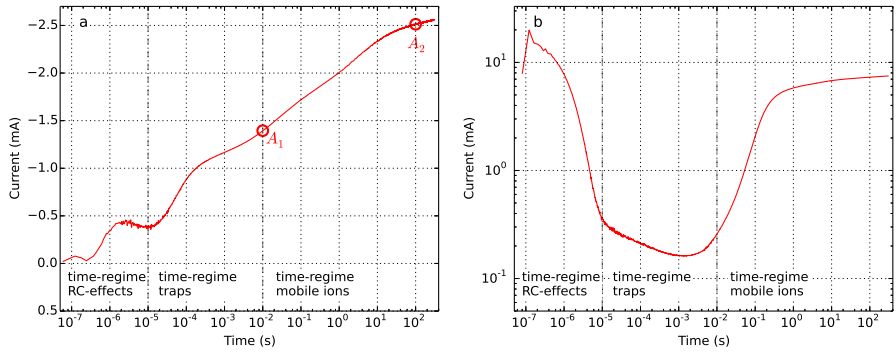}
\caption{a) \acf{TPC} measurement of a perovskite solar cell. The LED light is turned on at $t=0$ with an intensity of approximately one sun. The applied voltage is $V(t)=0\,V$. $A_1$ and $A_2$ mark two states of the solar cell that are referred to later in \autoref{img:pulsed_iv}. b) Current response to a voltage step in the dark with $V=2\,V$. The first regime is governed by \gls{RC_effects}, the second by space charge effects due to imbalanced charge mobilities and traps. The third regime is governed by mobile ions.}
\label{img:tpc_and_vp}
\end{figure}

We perform transient measurements with a broad dynamic range. \autoref{img:tpc_and_vp}a shows the transient photocurrent of a perovskite solar cell as a response to a light step (light is turned on at $t=0$, $V=0$). At 4 microseconds the current overshoots and is followed by a steady rise over 8 orders of magnitude in time lasting up to 5 minutes. This long rise is extraordinary. We have previously investigated micro-crystalline silicon solar cells, \ac{CdTe}, \ac{CIGS}, various \ac{OSC}, hybrid and \ac{DSSC}. In all these devices the current rise takes place on very different time scales but always ends within two or three orders of magnitude in time (for example $0.1\,\mu s$ to $10\,\mu s$, or $1\,\mu s$ to $100\,\mu s$).

\marginpar{\newline Extraordinary slow effects occur under illumination and in the dark.}
In \autoref{img:tpc_and_vp}b the current response to a voltage step in the dark (voltage is turned on at $t=0$) is shown. The initial current peak (up to $10\,\mu s$) is a RC displacement current charging the geometrical capacitance as we have confirmed by independent determination of the series resistance and capacitance (see Section \nameref{ch:measuring_rc}). After the RC peak the current is slightly decreasing and starts to rise again only after 10 milliseconds. From \autoref{img:tpc_and_vp} we conclude that slow effects on different time scales occur under illumination and in the dark.\\

\begin{figure}[h]
\centering
\includegraphics[width=\textwidth]{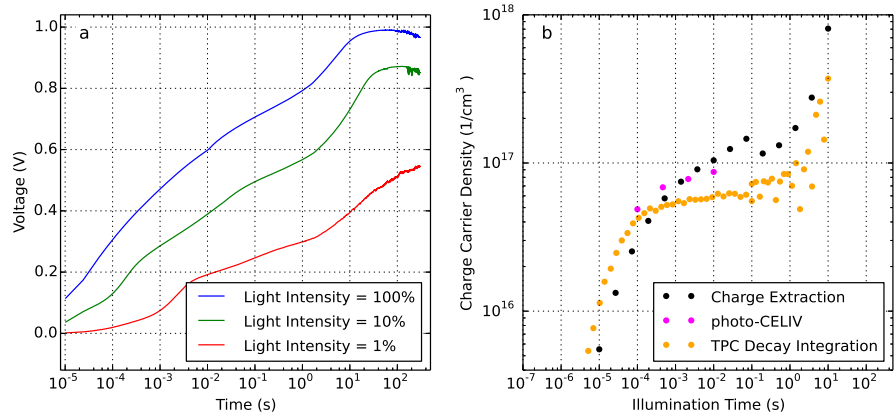}
\caption{a) \acf{TPV} rise for three different light intensities. b) Three different methods to extract the charge carrier density dependent on the illumination duration.}
\label{img:tpv_and_ce}
\end{figure}

\marginpar{\newline Slow transient effects are also present during the \ac{TPV} rise.}
\autoref{img:tpv_and_ce}a shows transient photovoltage for light intensities of $1\%$, $10\%$ and $100\%$. Also here a dynamic range covering 8 orders of magnitude in time is observed.\\

\autoref{img:tpv_and_ce}b compares different methods for charge carrier density determination plotted versus the illumination time. In the charge extraction technique the device is illuminated and kept at open-circuit. After light turn-off the voltage is switched to zero. To get the number of extracted charges the current is integrated. The \acs{photo-CELIV} experiment is explained in section \nameref{ch:fast_regime}. Here the \acs{CELIV} current is RC-corrected and integrated to obtain the charge carrier density. These two techniques keep the device at open-circuit before charge extraction. The extracted values are well comparable.\\
\marginpar{\newline The charge carrier density is measured with three techniques depending on the illumination duration. After $1\,s$ illumination the charge density gets unrealistically high - mobile ions may be the origin of this effect.}
In the technique \acs{TPC} decay integration a transient photocurrent is measured at short-circuit. After light turn-off the current is integrated to obtain the charge carrier density at short-circuit.
For all three charge extraction techniques a steady rise is observed for short illumination durations. This can be explained by the filling of traps. At one millisecond there is a certain saturation until 100 milliseconds. Then the extracted charge of \emph{charge extraction} and \emph{TPC decay integration} increase drastically up to almost $10^{18}\,cm^{-3}$. It has been speculated that a change in the ion distribution inside the cell can explain this extraordinarily high charge density of the charge extraction \cite{oregan_optoelectronic_2015}. The origin for this slow rise is subject to further investigations.

In this thesis three different mechanisms are postulated to be present. In the first time regime \gls{RC_effects} dominate, in the second imbalanced mobilities and charge trapping. In the third regime mobile ions changing the electric field influence the total charge current. We put our focus on the second and third regime. In section \nameref{ch:slow_regime} we show \acp{IV curve} with 10 millisecond and 100 second pulses. In section \nameref{ch:fast_regime} we discuss \ac{photo-CELIV} experiments (up to $10\,ms$).

\section{Slow Regime (Milliseconds to Minutes)}\label{ch:slow_regime}

\marginpar{\newline \acp{IV curve} are measured with voltage and light pulses - once with long and once with short pulse length.}

\begin{figure}[h]
\centering
\includegraphics[width=\textwidth]{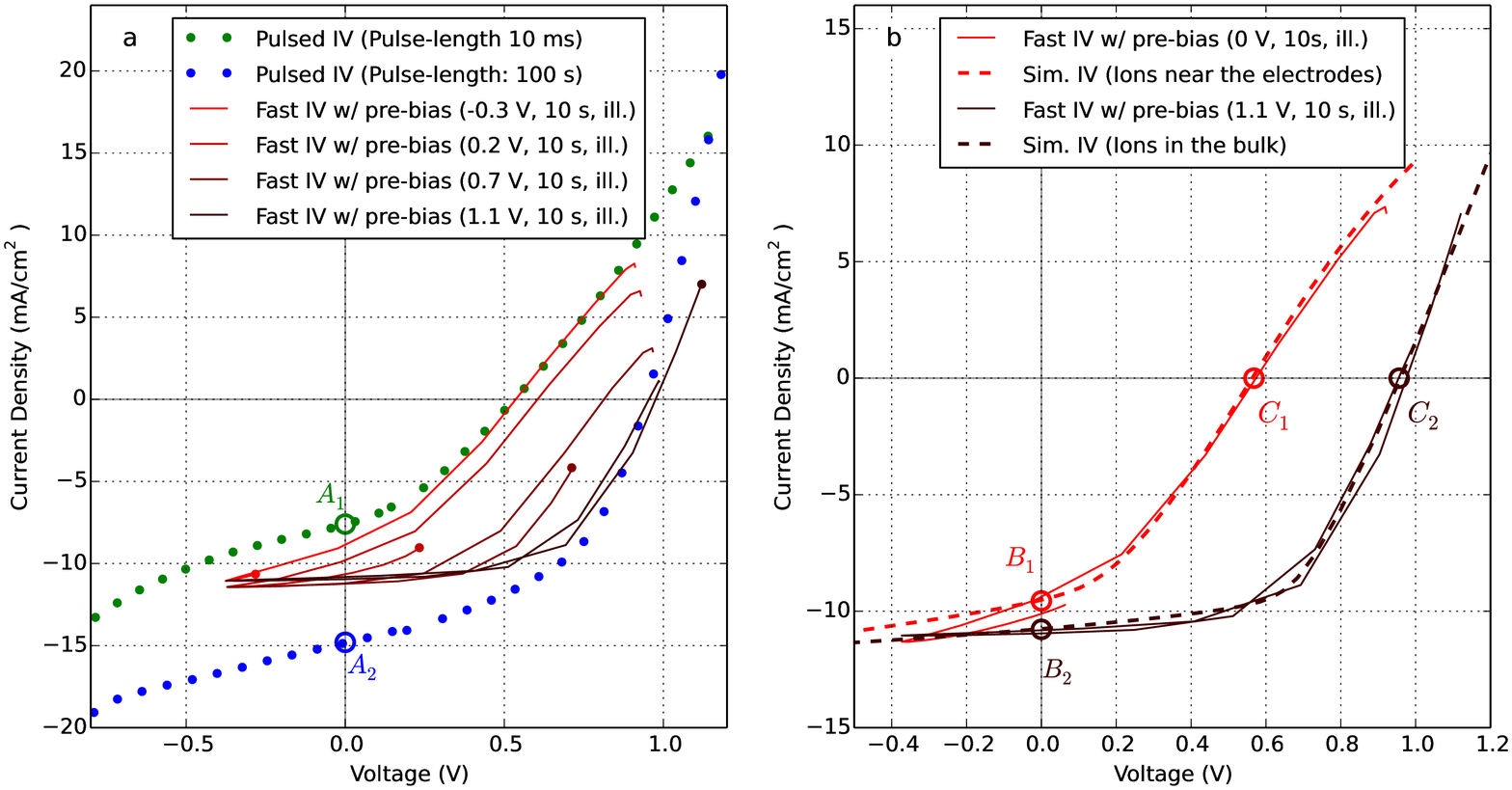}
\caption{a) Measured \acp{IV curve} with 10 millisecond and 100 second voltage and light pulses (points). The solid lines represent ramped IV measurements (70 V/s) with pre-bias under illumination for 10 seconds. States $A_1$ and $A_2$ are marked also in \autoref{img:tpc_and_vp}a for clarity. b) Numerical drift-diffusion simulations of the \acp{IV curve} with different ion distribution in comparison with the measured \acp{IV curve} with pre-bias (same curves as in a). The simulation with ions close to the electrodes reproduces the measured \ac{IV curve} with pre-bias at 0 Volt under illumination. Simulation with distributed ions reproduces the measured \ac{IV curve} with pre-bias at 1.1 Volt and illumination. The states $B_1$, $B_2$, $C_1$ and $C_2$ are explained in detail in \autoref{img:bands}.}
\label{img:pulsed_iv}
\end{figure}

The \emph{slow} effects (ranging from $10\,ms$ to $100\,s$) are investigated by measuring \ac{IV curve} with rectangular voltage pulses where light and voltage are switched on simultaneously. The current value is read out at the end of the pulse. Measuring \acp{IV curve} with pulsed voltages has the advantage that every voltage point is acquired with exactly the same time after light and voltage turn-on. Between the pulses we wait 20 seconds to give the device sufficient time to recover. Effects on different times scales can be isolated with this approach by choosing different pulse durations.
Pulsed \acp{IV curve} are shown in \autoref{img:pulsed_iv}a in dots with 10 millisecond and 100 second pulse-length. Each point represents a measurement with one pulse. The IV curve acquired with pulse-lengths of 10 milliseconds differs substantially from the one with pulses of 100 seconds duration. Tress et al. \cite{tress_understanding_2015} postulated that the charge transport properties of these perovksite solar cells depend on the \emph{state} in which the device is. This \emph{state} changes with the applied voltage. In the pulsed IV curve with 100 seconds pulse duration the device has time to change to the respective equilibrium \emph{state}. For clarity the short-circuit current after $10\,ms$ and $100\,s$ are marked as state $A_1$ and $A_2$ in \autoref{img:pulsed_iv}a and \autoref{img:tpc_and_vp}a.

\marginpar{\newline The cell current depends on an internal "state", that changes with time and applied voltage. This causes the hysteresis in the IV curve.}

Independent of the physical origin of these \emph{states} it is now clear that they change slowly and differ with voltage. During the acquisition of a classical IV curve the cell changes from \emph{state} to \emph{state}, giving rise to the \ac{IV curve} hysteresis. The \ac{IV curve} hysteresis is therefore not a hysteresis in the classical sense originating from a material property in steady-state, but rather a measurement artifact occurring in non-steady-state characterization. It has been shown that the hysteresis is small at either very fast or very slow IV ramps \cite{unger_hysteresis_2014, tress_understanding_2015}. This is explained if the ramp-rate is either significantly faster or slower than the time constant of these changing states.

\marginpar{\newline Mobile ions inside the perovskite material may be responsible for the internal "state".}
Tress et al. \cite{tress_understanding_2015} postulated mobile ions inside the perovskite layer being responsible for the internal \emph{state}. Indeed Eames et al. \cite{eames_ionic_2015} and Haruyama et al. \cite{haruyama_first-principles_2015} showed with \acf{DFT} calculations that iodide vacancies can move through the perovskite layer with an activation energy of $0.45\,eV$ to $0.6\,eV$.

\marginpar{\newline IV curves are measured with a fast ramp after keeping the cell at a pre-bias voltage. Hereby ions are first positioned. The following IV curve is fast enough not to influence the position of the ions. }
Tress et al. used a voltage pre-bias to condition the device followed by a fast IV ramp. Since the ramp-time is faster than the time-constant of the ionic movement, the full \ac{IV curve} is acquired without affecting the distribution of the ions. Using a pre-bias voltage allows to change the position of the ions prior to the measurement. During fast acquisition of the \ac{IV curve} the ion distribution can be considered to be “frozen” due to the small ion mobility. This is in contrast to the pulsed IV where for each voltage step the ions have time to migrate.

Fast \ac{IV curve} measurements with 10 seconds voltage and light pre-bias are shown in solid lines in \autoref{img:pulsed_iv}a. The \ac{IV curve} with -0.3 Volt pre-bias is close to the pulsed IV with 10 milliseconds duration. In both cases ions are close to their equilibrium distribution in the dark. Keeping the device even longer under illumination increases the current further to state $A_2$. This indicates that light has also a direct or indirect influence on the ion position. An indirect influence could be the creation of charge carriers that change the electric field inside the device. The curves measured with different pre-bias voltage are shifted in voltage and have a different slope at forward current.

\subsection{IV Curve Simulation}

Physical processes are often too complex to understand with simple and qualitative explanations. Numerical simulations provide more insight and enable a broader understanding of underlying physics by putting hypotheses under test. Steady-state simulations of \acp{PSC} have recently been reported by Nie and co-workers \cite{nie_high-efficiency_2015}. Sun et al. \cite{sun_physics-based_2015} presented an elaborate analytical model to describe \ac{PSC} \ac{IV curve}s.\\

\marginpar{\newline The \acp{IV curve} of the perovskite solar cell are simulated assuming ions at the interface or in the bulk.}

\begin{figure}[!]
\centering
\includegraphics[width=\textwidth]{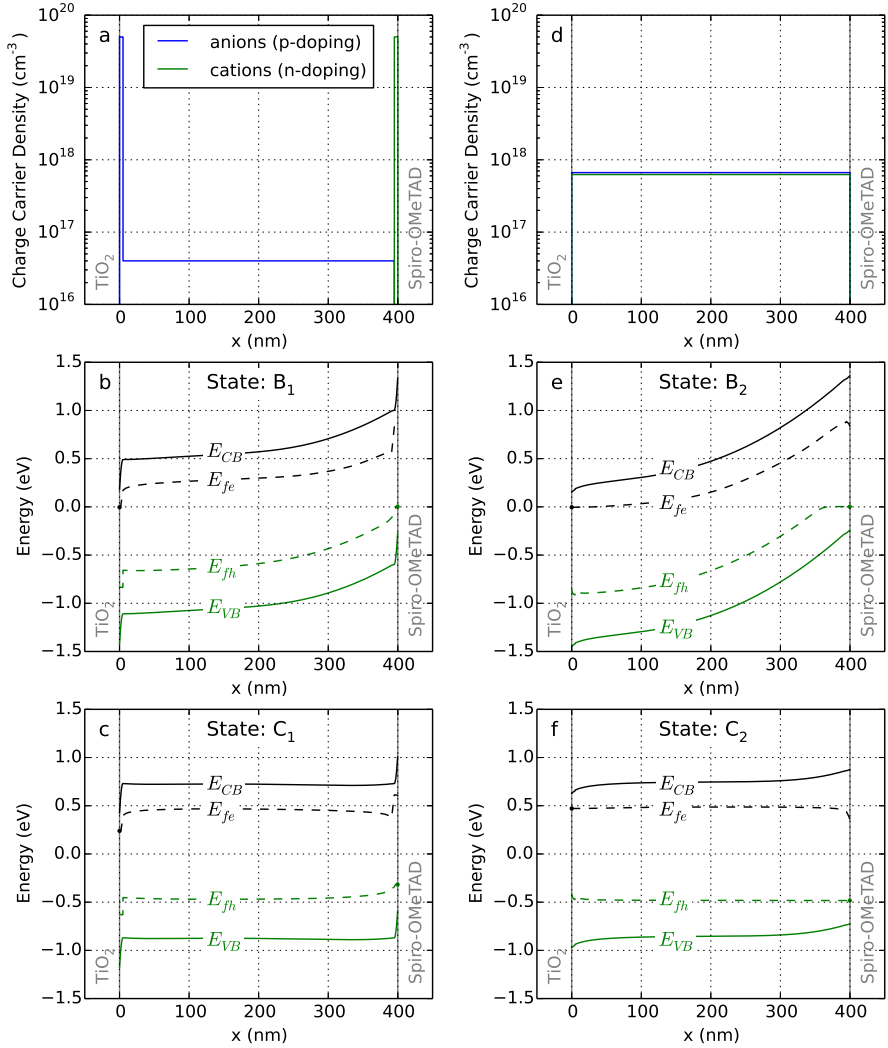}
\caption{a) Assumed ion distribution inside the perovskite layer used for state $B_1$ and $C_1$. Anions (-) near the $TiO_2$ layer and cations (+) near the spiroOMeTAD layer decrease the effective built-in field. Energy bands ($E_{CB}$ and $E_{VB}$) and \gls{Fermi_level}s ($E_{fe}$, $E_{fh}$) with ion distribution of (a) at short-circuit under illumination, state $B_1$ (b) and at open-circuit, state $C_1$ (c), respectively.\\
(d) Assumed ion distribution to simulate state $B_2$ and $C_2$ as shown in (e) and (f). States $B_1$, $B_2$, $C_1$, $C_2$ are marked in \autoref{img:pulsed_iv}b.}
\label{img:bands}
\end{figure}

To test the feasibility of the explanation with mobile ions we use numerical drift-diffusion to simulate \acp{IV curve} with different ion distributions. Two states are simulated, one state with ions near the electrodes and one state with ions distributed homogeneously in the bulk. The ion distribution used in the simulation is displayed in \autoref{img:bands}a (near the electrodes) and in \autoref{img:bands}d (distributed in the bulk). The total number of ions is identical in both cases. The ions are modelled by two interface layers of 5 nanometer thickness that are doped. Drift and diffusion of electrons and holes is calculated, using bulk- and surface recombination. More details about the model are presented in the Section \nameref{ch:simulation}.

\marginpar{\newline The \ac{IV curve} simulations with different ion position reproduce the measured curves using different pre-bias voltages.}

\autoref{img:pulsed_iv}b shows two measured \acp{IV curve} (solid lines) once with pre-bias at 1.1 Volt and light, once with 0 Volt and light. Four points are marked ($B_1$, $B_2$, $C_1$, $C_2$), which will be explained in \autoref{img:bands}. The simulated \acp{IV curve} are fitted to the fast \acp{IV curve} with pre-bias voltage. The simulation is in good agreement with the measurement data as shown in \autoref{img:pulsed_iv}b. All parameters for the two \acp{IV curve} are identical, except for the ion distribution, according to \autoref{img:bands}a and \autoref{img:bands}d. The simulation results show that the ions near the electrodes reduce the effective built-in field and therefore reduce the open-circuit voltage from $C_2$ to $C_1$. The open-circuit voltage is reduced by $40\%$ while the short-circuit current is reduced by $10\%$.\\

\marginpar{\newline Both IV curves are simulated with the same parameters except the ion distribution.}

Simulation parameters for both states are summarized in \autoref{tab:sim_parameters_iv}. The \emph{hole barrier at \ac{ETL}} is the difference between the \ac{HOMO} level of perovskite and the $TiO_2$ (\ac{ETL}). It is used to control the surface recombination at this contact. The same is the case for electrons at the spiro-OMeTAD (\ac{HTL}). A negative energy barrier as used in this table indicates a high surface recombination.

\begin{table}[h]
\centering
  \begin{tabularx}{\textwidth}{ X l }
\toprule
Parameter & Value \\
	\midrule
Hole mobility $\mu_h$			& $0.15\,cm^2/Vs$ \\ \hline
Electron mobility $\mu_e$		& $2.4 \cdot 10^{-4}\,cm^2/Vs$ \\ \hline
Recombination coefficient $B$	& $7.3 \cdot 10^{-8}\,m^3/s$ \\ \hline
n-doping	 $N_D$					& $4 \cdot 10^{16}\,cm^{-3}$ \\ \hline
Ion charge at \ac{ETL} for states ($B_1$, $C_1$) & $2.5 \cdot 10^{13} cm^{-2}$ \\ \hline
Ion charge at \ac{HTL} for states ($B_1$, $C_1$) & $2.5 \cdot 10^{13} cm^{-2}$ \\ \hline
Ion charge at \ac{ETL} for states ($B_2$, $C_2$) & $0$ \\ \hline
Ion charge at \ac{HTL} for states ($B_2$, $C_2$) & $0$ \\ \hline

Hole barrier at \ac{ETL}				& $-0.18\,eV$ \\ \hline
Electron barrier at \ac{HTL}	& $-0.28\,eV$ \\ \hline
Electron density at \ac{ETL}	 $n_{e0}$ & $2.3 \cdot 10^{18}\,cm^{-3}$ \\ \hline
Hole density at \ac{HTL} $n_{h0}$ & $8 \cdot 10^{16}\,cm^{-3}$ \\ \hline
Relative electrical permittivity
%\footnote{This is the total permittivity including the $TiO_2$ mesoporous layer.}
$\epsilon_r$ & $35$ \\ \hline
Bandgap $E_{BG}$					& $1.6\,eV$ \\ \hline
Series resistance $R_S$			& $59.9\,\Omega$ \\ \hline
Thickness $d$					& $400\,nm$ \\
	\bottomrule
\end{tabularx}
\caption{Parameters used for simulation of \acp{IV curve} in \autoref{img:pulsed_iv} and \autoref{img:bands}. }
\label{tab:sim_parameters_iv}
\end{table}

\autoref{img:bands}b shows the simulated energy bands at short-circuit (state $B_1$ of \autoref{img:pulsed_iv}b) with ions near the electrodes. \autoref{img:bands}c shows bands for the same ion distribution at open-circuit (State $C_1$ of \autoref{img:pulsed_iv}b). \autoref{img:bands}e and \autoref{img:bands}f show short-circuit and open-circuit (state $B_2$ and $C_2$ of \autoref{img:pulsed_iv}b) with ions distributed in the bulk. The \gls{Fermi_level}s in state $B_2$ are steep throughout the whole layer whereas in state $B_1$ \gls{Fermi_level}s are flat near the $TiO_2$ interface.

\marginpar{Ions at the interface reduce the effective built-in voltage leading to smaller \gls{Fermi_level} gradient.}

Therefore charge extraction is less efficient and the current is lower. The spikes in the bands close to the electrodes are caused by the charge of the ions. The ions close to the electrodes (state $C_1$) decrease the selectivity of both contacts, thus leading to an enhanced surface recombination and significantly lower open-circuit voltage \cite{wurfel_charge_2015, reinhardt_identifying_2014}.\\

\marginpar{\newline The internal field leads to a drift of mobile ions to the interface, partially compensating the internal field.
\newline \newline Under illumination the ions diffuse from the interface back to the bulk.}

\begin{figure}[h]
\centering
\includegraphics[width=\textwidth]{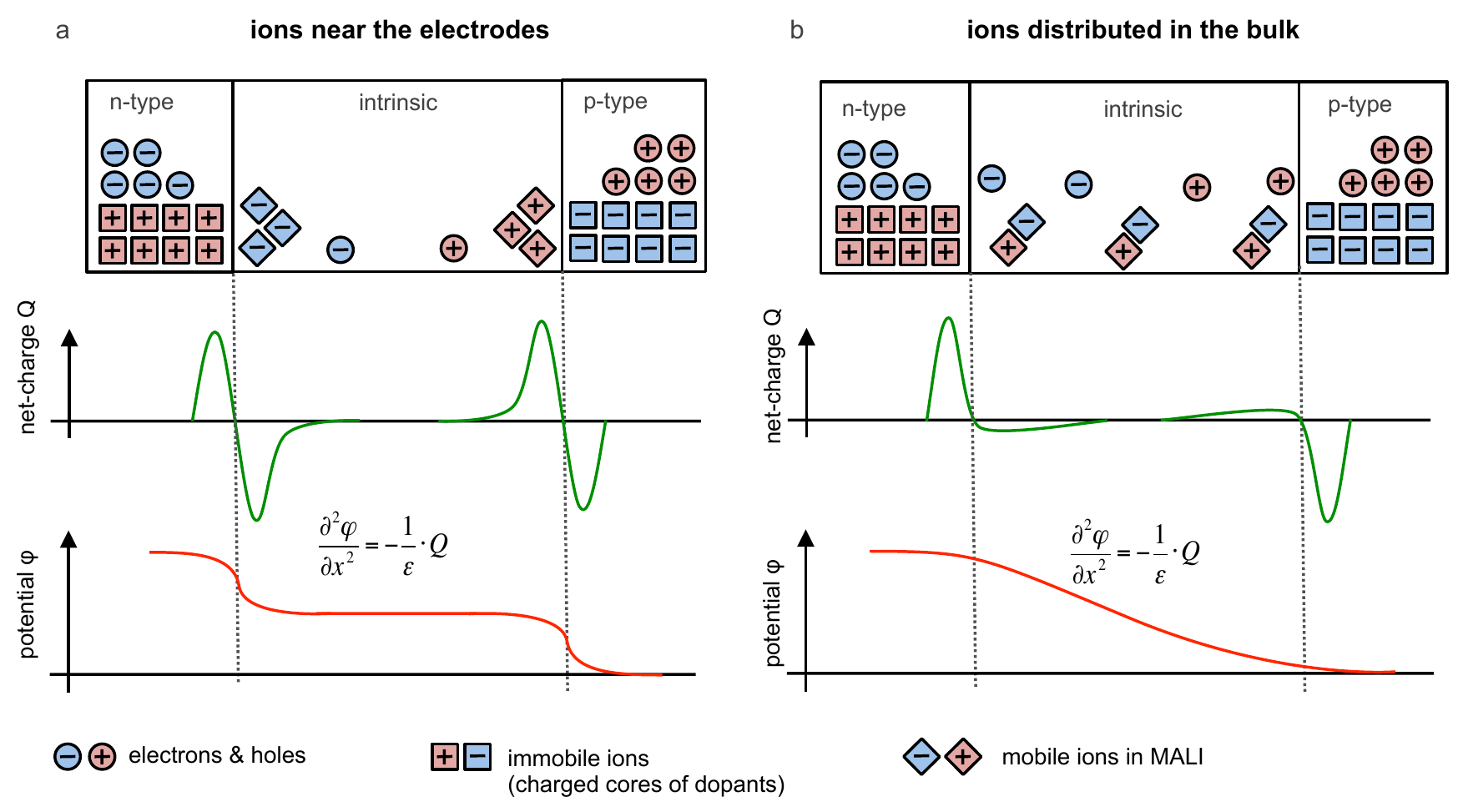}
\caption{Schematic illustration of the effect of mobile ions on the potential. a) Ions are close to the interfaces and screen the electric field inside the bulk. The band is therefore flat. This is the state in the dark. b) Ions are distributed in the bulk and compensate each other. The potential drops over the whole intrinsic region leading to efficient charge extraction.}
\label{img:ion_scheme}
\end{figure}

The reduced built-in voltage can be explained as follows: A p-type material has mobile holes and negatively charged ionic cores. In a p-i-n structure the negative ions of the p-type in combination with the positive ions of the n-type lead to an electric field in the intrinsic region. This field is called built-in field. With $TiO_2$ (n-type) and Spiro-OMeTAD (p-type) transport layers perovskite solar cells exhibit a built-in field. At short-circuit in the dark this built-in field leads to a drift of mobile cations (+) to the perovskite-spiro-OMeTAD interface and anions (-) to the perovskite-$TiO_2$ interface. The mobile cations (+) in the perovskite hereby partially compensate the fixed negative charge in the Spiro-OMeTAD. Therefore in the dark at short-circuit the built-in field is reduced. Under illumination the photo-generated charge disturbs the electric field. In this state the ions can diffuse away from the interface to the bulk. Therefore the built-in field increases over illumination time. We illustrate this effect schematically in \autoref{img:ion_scheme}.\\

\marginpar{\newline In our case ions lead to a reduction of the built-in voltage, which is in contrast with other publications investigating mobile ions. The difference lies in the contacts.}

In contrast to Zhang et al. \cite{zhang_charge_2015} that explained mobile ions schematically, the ions reduce the electric field in our case and the performance increases with time when ions relax. The difference lies in the cell architecture investigated. In our publication devices with selective contacts are used whereas Zhang et al. created a device with symmetric gold contacts. Zhang and co-workers used a bias voltage in forward direction to create an electric field and certain charge carrier selectivity. Therefore mobile ions enhance the performance and go to the opposite direction than in our case. The same effect is observed as switching PV \cite{park_perovskite_2014, xiao_giant_2014} and further confirms the mobile ion hypothesis.

\marginpar{\newline Mobile ions are a plausible explanation for the hysteresis in perovskite solar cells.}

With our proof-of-principle simulations we show that mobile ions are a plausible explanation for the observed \ac{IV curve} hysteresis and the slow effects observed. According to these findings it seems best to track the maximum power point until stabilization in order to determine the power conversion efficiency of a perovskite solar cell.

\section{Fast Regime (Microseconds to Milliseconds)}\label{ch:fast_regime}

\marginpar{\newline In the fast regime mobile ions can be considered as constant as they are too slow.}

In the previous section we have investigated the \emph{slow} effects of the \ac{TPC} measurement in \autoref{img:tpc_and_vp}a. In this section the \emph{fast} effects from 1 microsecond to 10 milliseconds are investigated. In this time regime mobile ions do not play a role, as they are too slow. We use the technique \acf{photo-CELIV} to analyse the charge carrier dynamics. \ac{photo-CELIV} is a technique that is frequently used to measure the charge carrier mobility, recombination or doping in organic solar cells \cite{dennler_charge_2006, juska_extraction_2000, bange_charge_2010, lorrmann_charge_2010, juska_extraction_2011, sandberg_direct_2014}.
In a previous publication we have shown that mobility extraction from \acs{CELIV} data is error-prone \cite{neukom_charge_2011, hanfland_physical_2013}. Nevertheless \acs{CELIV} experiments give valuable insight into underlying physics of transport and recombination of photo-generated charges.

\marginpar{\newline Charge carrier dynamics are investigated using the technique \acs{photo-CELIV}.}

In the \ac{photo-CELIV} experiment charge carriers are created by an LED pulse while the device is kept at open-circuit. Since the illumination time is short, a readjustment of the mobile ions is not expected. After turning off the light a voltage ramp in negative direction extracts all charge carriers. Usually one current peak is observed that is related with the faster charge carrier being extracted. The shorter the time of the current peak the higher the mobility of the charge carriers.\\

\marginpar{\newline The \acs{CELIV}-current of the \acs{PSC} shows two peaks indicating different electron and hole mobility.}

\begin{figure}[t!]
\centering
\includegraphics[width=\textwidth]{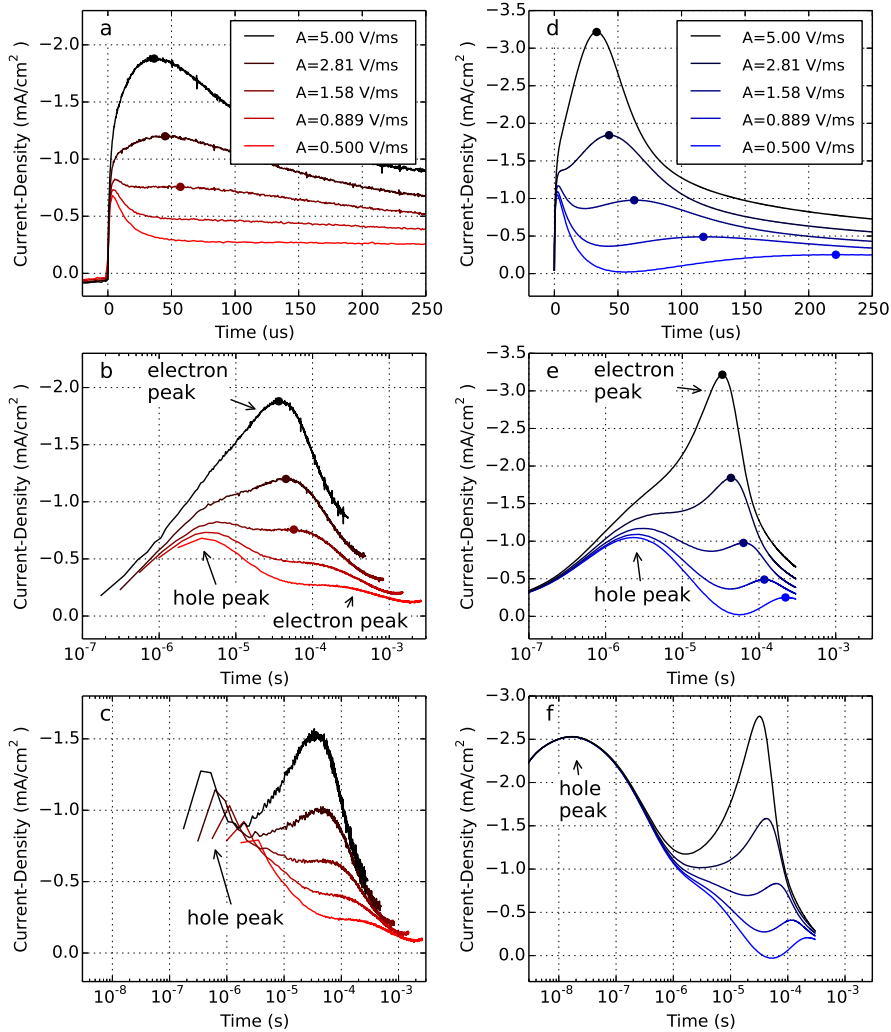}
\caption{Measured \ac{photo-CELIV} currents with 5 different ramp rates A in linear (a), in logarithmic time scale (b) and with correction of \gls{RC_effects} (c). The circle marks the second current peak. \setfos drift-diffusion simulations in (d), (e) and (f) respectively.}
\label{img:celiv_A}
\end{figure}

\autoref{img:celiv_A}a shows such \ac{photo-CELIV} currents of the \ac{PSC} with different voltage ramp rates A. The cell is illuminated for 1 millisecond and kept at open-circuit. At $t=0$ the light is turned off and the voltage ramp $V(t)=V_{OC} - A \cdot t$ starts. For certain ramp rates two subsequent current peaks are observed. For the ramp rate ($A=1.58\,V/ms$) a peak at about $4\,\mu s$ appears followed by a second peak at $60\,\mu s$. The second peak is visible in \autoref{img:celiv_A}b where the same data is plotted versus logarithmic time. In the measurement with the highest ramp rate ($A=5\,V/ms$) only one current-peak is visible. The \ac{photo-CELIV} results suggest that two charge carrier types are extracted causing the two current peaks. The amount of extracted charge varies significantly between the two carriers as indicated by the different peak height. Furthermore one carrier seems to be significantly faster than the other. From this we conclude a difference in the mobility.\\

\marginpar{\newline Different values for the charge carrier mobility of perovskite have been published ranging from $20\,cm^2/Vs$ to $160\,cm^2/Vs$. \newline \newline The cell investigated uses a $TiO_2$ scaffold. Electrons are most probably transported in the $TiO_2$ that has a mobility of around $10^{-2}\,cm^2/Vs$. This difference in mobility could explain the two \acs{CELIV}-peaks.}

Dong et al. \cite{dong_electron-hole_2015} measured the charge carrier mobility in a perovskite single crystal by \ac{SCLC} on mono-polar devices using the Mott-Gurney Law. They found a mobility of $160\,cm^2/Vs$ for holes and $20\,cm^2/Vs$ for electrons. Leijtens et al. \cite{leijtens_electronic_2014} however published balanced mobilities in planar devices of $20\,cm^2/Vs$ measured by photo-conductivity measurements. Balanced mobilities could also be expected from the similar effective mass of electron and holes \cite{giorgi_small_2013}. Ponseca et al. \cite{ponseca_organometal_2014} published time-resolved terahertz and microwave conductivity measurements and found balanced mobilities of $25\,cm^2/Vs$ in pure perovskite. They suggested however imbalanced mobilities in mesoporous scaffolds with $TiO_2$. It is energetically favourable for electrons to transfer to $TiO_2$. This process has been estimated to happen within picoseconds \cite{marchioro_unravelling_2014, giorgi_small_2013, wehrenfennig_high_2014}. Electrons are in this case transported in the $TiO_2$ scaffold and not in perovskite. Since the mobility in $TiO_2$ has been published to be around $10^{-2}\,cm^2/Vs$ \cite{hendry_local_2006} electrons are expected to be much slower in $TiO_2$ than in \ac{MALI}. We therefore assume that the holes are the faster carriers creating the first peak in the \ac{photo-CELIV} experiment whereas the electrons cause the second peak. Electron and hole peaks are marked with arrows in \autoref{img:celiv_A}.
An imbalance in mobility can also cause an imbalance in charge carrier density. The slower carrier is thereby limited by space charge effects and accumulates whereas the faster carrier leaves the device faster and accumulates on the electrodes outside of the bulk. This could in our case explain why more electrons than holes are extracted.

\marginpar{\newline \gls{RC_effects} can significantly perturb \acs{CELIV} measurements.}

The question remains why there appears to be only one peak for the highest ramp rates ($A=5\,V/ms$). In a previous publication we have shown that \gls{RC_effects} can significantly perturb \acs{CELIV} measurements \cite{neukom_charge_2011}. We therefore correct the \gls{RC_effects} by subtracting the displacement current of the measured current as proposed by Kettlitz and co-workers \cite{kettlitz_eliminating_2013}.
We assume that the device has a series resistance caused by the sheet resistance of the \ac{TCO} and the contact resistance that we summarize as $R_S$. The geometrical capacitance $C_{geom}$ of the device is determined by the dielectric constants of the layers. We determine $R_S$ to be $59.9\,\Omega$ and $C_{geom}$ to be $10.9\,nF$. Details on RC extraction can be found in Section \nameref{ch:measuring_rc}.\\

\marginpar{\newline The \acs{CELIV} current can be corrected for \gls{RC_effects} using a simple electric circuit.}

\begin{figure}[h]
\centering
\includegraphics[width=0.5\textwidth]{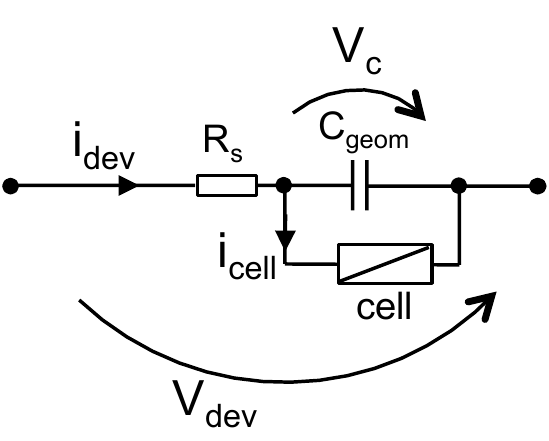}
\caption{Electric circuit used to correct \gls{RC_effects} of \acs{CELIV} measurements.}
\label{img:rc_circuit_2}
\end{figure}

We propose a simplified scheme of the solar cell to compensate \gls{RC_effects} as shown in \autoref{img:rc_circuit_2}. From $i_{dev}$ (the measured current) and $V_{dev}$ (the measured voltage) we calculate the corrected current $i_{cell}$ according to \autoref{eq:rc_compensation}. The derivation of \autoref{eq:rc_compensation} is found in Section \nameref{ch:rc_correction_derivation}.

\begin{equation}\label{eq:rc_compensation}
i_{cell}(t) = i_{dev}(t) - C_{geom} \cdot \frac{dV_{dev}}{dt} -
R_S \cdot C_{geom} \cdot \frac{di_{dev}}{dt}
\end{equation}

\marginpar{\newline With the RC-correction the hole current peak shifts to the left (shorter times).}

The RC correction is applied to the \ac{photo-CELIV} data in \autoref{img:celiv_A}a and \autoref{img:celiv_A}b. The RC corrected current is shown in \autoref{img:celiv_A}c. Since the displacement current is corrected the total current is lower. Comparing the current with the lowest ramp rate ($A=0.5\,V/ms$) in \autoref{img:celiv_A}c with \autoref{img:celiv_A}b it seems that the fast hole peak is amplified and at shorter time scales. \gls{RC_effects} have lowered and shifted the hole peak to longer times. With the RC correction this hole peak is recovered and visible in the currents of all ramp rates. The sub-microsecond hole peak is at the limit of the time resolution of our measurement, but can nevertheless be identified clearly. We show further below with numerical simulation that this peak shall not be taken as an artefact.\\

\marginpar{\newline The holes are extracted first, as they have the higher mobility, leading to the first current peak. Electrons cause the second peak.}

We conclude that before their extraction electrons and holes accumulate in the bulk. The amount of electrons in the bulk exceeds the amount of holes in the bulk. The rest of the holes has left the device and accumulates on the electrode. As the voltage ramp starts, holes move out of the device quickly. In the case of the high ramp rate ($A=5\,V/ms$) holes are not detectable in the outer device current. The holes only charge the capacitance of the cell. We will strengthen this hypothesis further below in the text when we discuss numerical simulations in \autoref{img:celiv_A}d, \autoref{img:celiv_A}e and \autoref{img:celiv_A}f.

\marginpar{\newline Why are there much more electrons than holes extracted? \newline \newline The charge imbalance increases with time as shown in the \acs{photo-CELIV} measurement with varied illumination duration.}

\begin{figure}[h]
\centering
\includegraphics[width=\textwidth]{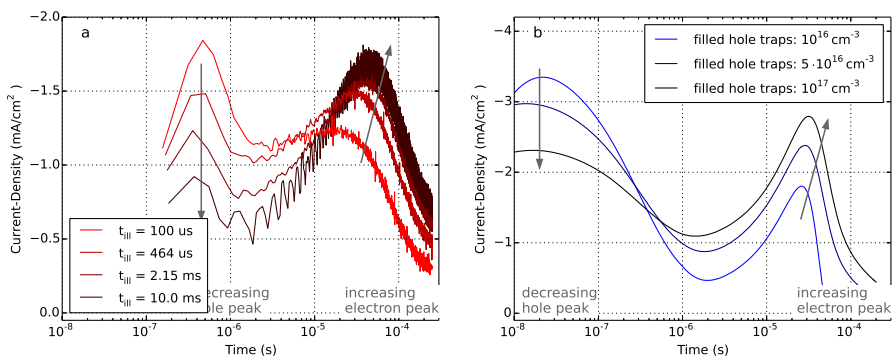}
\caption{a) Measured \ac{photo-CELIV} currents using constant ramp rate ($A=5\,mV/s$) and varied illumination duration prior to the ramp. b) Simulated \ac{photo-CELIV} currents using different densities of filled traps. Curves in a) and b) are RC corrected with \autoref{eq:rc_compensation}.}
\label{img:celiv_duration}
\end{figure}

We now discuss the origin of the imbalanced charge carrier density before the voltage ramp. Imbalanced charge in the device can be caused by an imbalance in charge carrier mobility (as mentioned above), asymmetric extraction barriers, charge doping or trapping. As shown above, there is evidence for imbalanced mobilities. The question therefore is, if other effects further enhance the imbalance.
To further investigate the imbalance of charge carrier density, the illumination duration prior to the \ac{CELIV} voltage ramp is varied. \autoref{img:celiv_duration}a shows RC-corrected \ac{photo-CELIV} currents with varied illumination duration prior to the voltage ramp. In the case of short illumination duration ($t_{ill}=100\,\mu s$) the hole peak is high and the subsequent electron peak is low. With longer illumination time the hole peak decreases whereas the electron peak increases. This result suggests that the imbalance of charge carrier density is increasing with illumination duration. Assuming the mobility to be constant in time the mobility imbalance cannot be the origin of this increase.

\marginpar{\newline With longer illumination duration holes get trapped leading to an increased ratio of free electrons to free holes.}

As commonly known charge doping leads to an imbalanced charge carrier density. In most solar cells, like crystalline silicon or \ac{CIGS}, this is the case. But classical doping is constant in time and would not lead to the effects observed in \autoref{img:celiv_duration}a. Captured charges in deep traps however behave like ionized dopants, too. Occupied traps have the same effect as the same amount of dopants. Leijtens and Stranks et al. \cite{leijtens_electronic_2014, stranks_recombination_2014} have observed a doping effect of traps in perovskite cells. They also showed evidence for trap-assisted recombination. Baumann et al. \cite{baumann_identification_2015} showed the presence of deep trap states, estimated at a depth of $0.5\,eV$, with \ac{TSC}. The question about the nature of the traps was addressed in the literature. \ac{DFT} simulations show that anion-cation substitutions (iodide replacing lead or MA) can act as deep hole traps \cite{yin_unusual_2014, yin_unique_2014}.

\marginpar{\newline Trapping can cause an imbalance in charge carrier density very similar to classical doping.}
Our qualitative understanding is thus as follows: If a hole is trapped it is immobile like an ionized dopant and leads indirectly to an additional free electron. This may not be intuitive: 

\begin{quotation}
\emph{Imagine an empty semiconductor with deep hole traps. One photon is absorbed - a free electron and a free hole are created. After the hole has been trapped, there is a free electron left. There is no difference compared to an empty semiconductor with one electron dopant.}
\end{quotation}

\marginpar{\newline The more holes get trapped the higher the ratio of free electrons to free holes. With illumination duration the charge imbalance increases and more electrons are extracted.}
With longer illumination duration traps become more and more filled. Therefore the amount of free electrons increases. The amount of free holes decreases since they encounter more recombination partners. Therefore deep hole trapping is assumed to lead to an additional imbalance of charge carrier density.
Further evidence for charge trapping is the long and steady current-decay in the voltage-pulse experiment shown in \autoref{img:tpc_and_vp}b. Knapp et al. \cite{knapp_role_2012} presented numerical simulations of \ac{DIT} (voltage pulses) with fast and slow traps leading to a similar current decay as observed in this thesis.

\subsection{Photo-CELIV Simulations}

\marginpar{\newline Transient simulations are used to reproduce the measured \acs{photo-CELIV} currents.}

Above \ac{photo-CELIV} measurements with qualitative explanations were presented. To confirm these findings we perform numerical drift-diffusion simulations of the \ac{photo-CELIV} experiments. Details on the simulation model can be found in section \nameref{ch:simulation}. 

In this state the mobile ions are close to the electrodes (see \autoref{img:bands}a). Therefore we use a small charge carrier density as boundary condition. For simplicity we use n-type doping to mimic filled hole traps. The series-resistance $R_S$ is considered in the simulation because \gls{RC_effects} play an important role in the transient measurements \cite{neukom_charge_2011, kettlitz_eliminating_2013}.

\marginpar{\newline The simulated \acs{photo-CELIV} currents reproduce the effects of the measurement qualitatively for the different ramp rates.}

\autoref{img:celiv_A}d and \autoref{img:celiv_A}e show the simulation results of the \ac{CELIV}-currents using the same experimental conditions as in the measurement. A good qualitative agreement is achieved for this variation of ramp rates and wide range of time scales. The simulation here is not meant to quantitatively extract material parameters but rather to derive an understanding of the operating mechanisms in a broad time range. The main dynamics are reproduced: For small ramp rates the hole current peak and the subsequent electron current peak are visible. For the largest ramp rate the hole peak vanishes and only the electron peak is observed.
In the simulation we consider the series resistance $R_S$ to cause \gls{RC_effects} and then apply the same RC correction to the simulated data, like in the case of the measured data (see \autoref{img:celiv_A}f). Thus \autoref{img:celiv_A}c and \autoref{img:celiv_A}f can directly be compared. This consistency validates our approach of the RC correction. The initial peak is similar as in the measurement.\\

\marginpar{\newline The simulation also reproduces the \acs{CELIV} with varied illumination duration. }
We use the same simulation parameters to explain the \ac{photo-CELIV} experiment with varied illumination duration (see \autoref{img:celiv_duration}b). Here the doping density is varied between $10^{16}\,1/cm^3$ and $10^{17}\,1/cm^3$ to control the amount of trapped holes. The effect of the imbalanced charge carrier density is reproduced. The first peak decreases as fewer holes are around and the second peak increases due to the accumulation of electrons in the bulk. 

\marginpar{\newline The simulation parameters for the IV curve simulation and the CELIV simulation are not identical. The simulations shall be seen as "proof-of-principle", showing the plausibility of the explanation of the physical effects discussed in this thesis.}

\autoref{tab:sim_parameters_celiv} summarizes the simulation parameters used for the simulation of the \ac{CELIV} currents. The parameters used here are not identical as the one used for the \ac{IV curve} simulations in \autoref{tab:sim_parameters_iv}. To optimize simulation convergence and speed the $5\,nm$ layers containing the ionic charge (ions at the interface) were omitted and the charge densities at the interface ($n_{e0}$ and $n_{h0}$) were reduced to $10^{12}\,cm^{-3}$. This results in a similar built-in voltage as in the case with the ion layers.
The charge carrier mobilities used in the transient simulations are significantly higher compared to the \ac{IV curve} simulations. Also the surface recombination is very low compared to \autoref{tab:sim_parameters_iv}.

\begin{table}[h]
\centering
  \begin{tabularx}{\textwidth}{ X l }
\toprule
Parameter & Value \\
	\midrule
Hole mobility $\mu_h$			& $1\,cm^2/Vs$ \\ \hline
Electron mobility $\mu_e$		& $0.2\,cm^2/Vs$ \\ \hline
Recombination coefficient $B$	& $5.2 \cdot 10^{-9}\,m^3/s$ \\ \hline
n-doping	 $N_D$ (varied in \autoref{img:celiv_duration})	& 
$1 \cdot 10^{16}$ to $1 \cdot 10^{17}\,cm^{-3}$ \\ \hline

Hole barrier at \ac{ETL}				& $0.3\,eV$ \\ \hline
Electron barrier at \ac{HTL}	& $0.3\,eV$ \\ \hline

Electron density at \ac{ETL}	 $n_{e0}$ & $1 \cdot 10^{12}\,cm^{-3}$ \\ \hline
Hole density at \ac{HTL} $n_{h0}$ & $1 \cdot 10^{12}\,cm^{-3}$ \\ \hline
Relative electrical permittivity
$\epsilon_r$ & $35$ \\ \hline
Bandgap $E_{BG}$					& $1.6\,eV$ \\ \hline
Series resistance $R_S$			& $59.9\,\Omega$ \\ \hline
Thickness $d$					& $400\,nm$ \\
	\bottomrule
\end{tabularx}
\caption{Parameters used for simulation of the \ac{photo-CELIV} experiments in \autoref{img:celiv_A} and \autoref{img:celiv_duration}. }
\label{tab:sim_parameters_celiv}
\end{table}

The hypothesis with imbalanced mobilities and deep trapping seems to be a plausible explanation for the \ac{photo-CELIV} currents presented in this thesis.

\section{Limitation of the approach}

\marginpar{\newline The presented model is used for qualitative understanding, not for parameter extraction. \newline \newline A fully consistent model has not been shown in the literature so far. The presented work shows that the current model is nevertheless able to describe the effects in the measurements.}

We would like to emphasize that we do not present a fully consistent model for transient and steady-state characterization as we published for an organic solar cell in Ref \cite{neukom_reliable_2012}. The model parameter set for the \acp{IV curve} and for the \ac{CELIV} currents is not identical. The focus of the model is on qualitative agreement and understanding. The dynamics of ionic motion is not modelled. The dynamics of trapping and detrapping are not modelled. Therefore the curves in \autoref{img:tpc_and_vp}a and \autoref{img:tpc_and_vp}b going from microsecond to minutes cannot yet be reproduced with one model. Future model refinements will address these issues. The presented work illustrates that the currently limited model is nevertheless able to describe the effects in the measurements.

\chapter{Discussion}
In summary we postulate imbalanced mobilities in combination with traps to be responsible for the space charge effects on short time scales and mobile ions for the change of field distribution on longer time scales. With our numerical model we show that this explanation is plausible and the measurement results can be reproduced. In this section we discuss additional physical effects that could play a role.

\section{Time Dependent Charge Generation}

\marginpar{\newline Time dependent charge generation is improbable because the \gls{exciton} binding energy is very low. }

Analysing the current rise in \autoref{img:tpc_and_vp}a one could speculate on time-dependent charge carrier generation. If \gls{exciton} dissociation was dependent on the electric field this could be the case. There is an on-going scientific debate about the \gls{exciton} binding energy in \ac{MALI} perovskite. Theoretical calculations by Frost et al. \cite{frost_atomistic_2014} predict $0.7\,meV$. Different measurements were published: $37\,meV$ by Hirasawa et al. \cite{hirasawa_magnetoabsorption_1994} already in 1994, $19\,meV$ at low temperature by Sun et al. \cite{sun_origin_2014}, $16\,meV$ at low temperature by Miyata and co-workers \cite{miyata_direct_2015}. \Gls{exciton} dissociation is expected to take place even without an electric field at room temperature ($kT = 26\,meV$).
Furthermore O’Regan et al. \cite{oregan_optoelectronic_2015} suggest no change in separation efficiency based on calculations of recombination fluxes. Deschler et al. \cite{deschler_high_2014} showed free charge carrier generation after $1\,ps$. We therefore consider direct charge generation from photons and neglect the intermediate \gls{exciton} state in our model.

\section{Dipoles, Interface-Traps and Ferroelectricity}

\marginpar{\newline Instead of mobile ions also dipoles or interface traps could cause the change in built-in voltage. With our approach this cannot be excluded.}

Our results suggest a slow change in the electric field caused by a change of interfacial charge. Ferroelectricity is expected to change the width of the space-charge region \cite{wei_hysteresis_2014} due to a change in electrical permittivity. But changing the relative electric permittivity in the model has only very little influence on the shape of the \ac{IV curve}. By rotation of dipoles in large ferroelectric domains however the same effect of a change in surface charge would be observed. With our model we can therefore not exclude ferroelectricity to play a role.
Electron traps at the $TiO_2$ interface could also lead to the same change in charge close to the interface. Questionable here is why trapped electrons at the $TiO_2$ interface would depopulate under illumination.
By looking at the dynamics of \autoref{img:tpc_and_vp}a, mobile ions seem the most plausible explanation. Interface trapping and ferroelectricity can however not be fully excluded.

\section{Excluding specific fabrication issues}

\marginpar{\newline Different device architectures were measured to exclude issues in fabrication.}

In order to exclude the possibility that specific issues in the device fabrication cause these effects we have measured \acp{PSC} from three different groups and with different structures ($TiO_2$ scaffold, $Al_2O_3$ scaffold, planar). All cells show a current rise with a dynamic range over at least 6 orders of magnitude in time – Similar as in \autoref{img:tpc_and_vp}. We conclude that traps and mobile ions are present in all these structures although the charge carrier dynamics varies between the device types. We measured 9 devices in total but show here only results of one device for consistency and simplicity.

\section{Open Questions}

\marginpar{\newline How do the bulk properties change when ions move? Why does the hysteresis depend on the contact material? What is the effect of surface passivation on the hysteresis? }

An open question remains on how ion migration affects the bulk properties. Is the trap and doping density inside the device dependent on the position of the mobile ions? Or to put it differently: How is the bulk material quality affected when interstitials or vacancies move to the electrodes due to an electric field?
It has been shown that devices with P3HT and PCBM as contact material show much less \ac{IV curve} hysteresis \cite{nie_high-efficiency_2015, zhang_charge_2015}. If the \ac{IV curve} hysteresis is caused by ion migration, why does it depend on the contact material? One could speculate that ion migration takes place independent of the contact material, but is more detrimental to cell performance if the surface recombination is high. With passivated surfaces the reduced built-in field might be irrelevant since the charge carriers can diffuse out of the device in any case – with or without an electric field.

\part{Summary and Outlook}

\chapter{Summary}\label{ch:summary}

Voltage pulse and light pulse measurements of \acf{MALI} perovskite solar cells are presented with a time range spanning 9 orders of magnitude. A “fast regime” from microseconds to milliseconds and a “slow regime” from milliseconds to minutes are identified.\\
Pulsed \acp{IV curve} measured with different pulse-lengths and fast ramped IV curves with voltage pre-bias explain the “slow regime”. Drift-diffusion simulations with different ion positions explain the experimental results and strengthen the hypothesis of mobile ions as explanation for the IV curve hysteresis in perovskite solar cells.
\ac{photo-CELIV} experiments with different ramp rates and different illumination times are used to shed light on and explain the “fast regime”. Transient drift-diffusion calculation is applied to understand the observed experimental results. The hypothesis of imbalanced charge carrier mobilities and deep hole traps is very plausible as the model-based analysis shows.\\

In a next step the numerical model could be extended with ion migration enabling the simulation of the full transient experiments. With this approach we are on the way to a complete model describing charge transport in perovskite solar cells from microseconds to minutes after excitation.

\chapter{Outlook}\label{ch:outlook}

\marginpar{\newline The major challenge for perovskite solar cells will be the device stability. So far $1000$ hours of operation have been demonstrated.}

The development of perovskite solar cells is still at the beginning. In order to become commercially relevant the device stability needs to be improved. Dye Sol recently presented a perovskite solar cell being stable over $1000$ hours \cite{dye_sol_stability_2015}.
Silicon solar module manufacturers offer a warranty over 20 years which corresponds to more than $20'000$ hours of operation. Reaching comparable device stability will be one of the major challenges for perovskite devices.\\

The toxicity needs to be investigated in more detail and environmental compatibility of modules needs to be proven. Perovskite containing lead or tin will probably not play a role in consumer electronics due to its toxicity.\\

There are two main applications where perovskites could play a role in the future. Either in cheap perovskite modules or in \gls{tandem} configurations with conventional solar cells. \\

\marginpar{\newline With silicon perovskite \glspl{tandem} efficiencies of $30\%$ could be realized.}

The current PV market is dominated by silicon wafer-based modules. As silicon solar cells are approaching their maximum theoretical efficiency the room for improvements with the current technology gets smaller. Using a \gls{tandem} configuration with a perovskite top cell and a silicon bottom cell could reach efficiencies up to $30\%$ \cite{filipic_ch_3nh_3pbi_3_2015}.\\

So far 4-terminal \glspl{tandem} with perovskite top cell and crystalline silicon bottom cell reaching $17\%$ have been demonstrated \cite{bailie_semi-transparent_2015}. By using perovskite on top of a \ac{CIGS} cell in 4-terminal configuration $19.5\%$ \ac{PCE} were reached \cite{kranz_high-efficiency_2015}. In perovskite-silicon 2-terminal cell configuration $18.1\%$ were demonstrated \cite{albrecht_monolithic_2015}. These results are promising as they could be easily added in a silicon solar cell production line.\\

\marginpar{\newline Further insight into device physics will be necessary to improve efficiency and lifetime. To characterise devices with mobile ions very systematic experiments are required.}

Whatever application will become relevant for perovskite devices, further insight into the physical device operating mechanism will be essential to improve efficiency and lifetime.
New measurement techniques like the \acf{TPC} presented in this paper can help to understand physical effects. To characterize devices with mobile ions, very systematic data acquisition is a necessity and all experiments need to be done with preconditioning.
Yet, there is no physical model that describes the physics of a perovskite solar cell completely. Models incorporating mobile ions in a drift-diffusion calculation will be a step in that direction.\\
To improve absorption light trapping structures suitable for perovskite need to be developed and optimized by numerical simulation. Fluxim has developed a prototype solver to calculate light scattering properties combining ray-tracing and thin-film optics leading towards comprehensive opto-electronic simulation of silicon-perovskite \glspl{tandem}.
On the long run module modelling can help optimising the grid and module structure.\\

Much research effort will be required that perovskite once generates electricity on our rooftops.

\chapter{Acknowledgement}\label{ch:acknowledgement}

I would like to acknowledge the following individuals that supported this work in one or the other way.

\begin{itemize}
\item Beat Ruhstaller (Fluxim AG, ZHAW) for supervising

\item Uli W\"urfel, Birger Zimmermann (Fraunhofer ISE), Stephane Altazin, Lieven Penninck (Fluxim AG) and Oskar Sandberg (Abo Akademi University) for fruitful discussions

\item Simon Z\"ufle (ZHAW) for proof-reading and discussions

\item Adrian Gentsch (Fluxim AG) for collaboration on the \paios development

\item Kurt Pernstich (ZHAW) for providing access to the glove-box at ZHAW

\item Philipp L\"oper, Jeremie Werner, Bjoern Niesen and Christophe Ballif (EPFL IMT) for the fabrication of the perovskite solar cells and for discussions

\item Cyrill Bolliger for providing the LaTeX template

\end{itemize}

\part{Appendix}

\chapter{Appendix}

%\section{Glossary}\label{ch:glossary}
%\printglossaries
%\clearpage

\section{Abbreviations}\label{ch:abbrievations}
%\chapter{Abbrievations}\label{ch:abbrievations}
%\input{FrontBackmatter/AbbrievationsBib}

\begin{acronym}[WWWWWWWWW]
	\acro{CdTe}{cadmium telluride}
	\acro{CIGS}{copper indium gallium diselenide}
	\acro{MOCVD}{metal organic chemical vapor deposition}
	\acro{DSSC}{dye sensitized solar cell}
	\acro{OSC}{organic solar cell}
	\acro{PCE}{power conversion efficiency}
	\acrodefplural{PCE}[PCE]{power conversion efficiencies}
	
	\acro{MALI}{methylammonium lead iodide}
	\acro{PSC}{perovskite solar cell}
	\acro{TEL}{transient electroluminescence}
	\acro{TPV}{transient photo-voltage}
	\acro{CELIV}{charge extraction by linearly increasing voltage}
	\acro{photo-CELIV}{photo-generated charge extraction by linearly increasing voltage}
	\acro{TPC}{transient photocurrent}
	\acro{SCLC}{space-charge limited current}
	\acro{SRH}{Shockley-Read-Hall}
	\acro{NREL}{National Renewable Energy Laboratory}
	\acro{HOMO}{highest occupied molecular orbit}
	\acro{LUMO}{lowest unoccupied molecular orbit}
	\acro{DFT}{density functional theory}
	\acro{TCO}{transparent conducting oxide}
	\acro{ETL}{electron transport layer}
	\acro{HTL}{hole transport layer}
	\acro{IV curve}{Current-voltage curve}
	\acro{TSC}{thermally stimulated current}
	\acro{DIT}{dark injection transients}
	\acro{EQE}{external quantum efficiency}
	\acro{MPP}{maximum power point}
	\acro{EL}{electroluminescence}
	\acro{IPCC}{Intergovernmental Panel Climate Change}
	\acro{RCP}{representative concentration pathway}
	%\acrodefplural{IV curve}[IV curves]{Current-voltage curves}
\end{acronym}

% --------------------- Abkürzungsverzeichnis -------------------
%
%

% --------------------- Symbolverzeichnis -------------------
%\section{Symbolverzeichnis}\label{sub:Symbolverzeichnis}
%\input{FrontBackmatter/Symbols}
%\clearpage

% ------------------- Abbildungsverzeichnis -----------------------
%\listoffigures
%\clearpage

% -------------------- Tabellenverzeichnis -----------------------
%\listoftables
%\clearpage

% -------------------- Literaturverzeichnis -----------------------
%\chapter{References}\label{ch:references}

\clearpage

\bibliography{Bibliography}

\clearpage

\pagestyle{plain}
\cleardoublepage

%  "/usr/texbin/makeindex" %.idx
% "/usr/texbin/makeindex" -s %.ist -o %.gls %.glo

\end{document}